\documentclass[a4paper,11pt]{article}
\usepackage{pos}
\usepackage{amsmath}

\numberwithin{equation}{section}

\title{From ten-flavor tests of the $\beta$-function  to  $\alpha_s$ at the Z-pole}

\author[a,b,c,d]{Zolt\'an Fodor}
\author*[e]{Kieran Holland}
\author*[f]{Julius Kuti}
\author[a]{Chik Him Wong}

\affiliation[a]{Department of Physics, University of Wuppertal\\
Wuppertal D-42097, Germany}

\affiliation[b]{ Juelich Supercomputing Center, Forschungszentrum Juelich\\
Juelich D-52425, Germany}

\affiliation[c]{ Department of Physics, Penn State University\\
	 University Park, 16802, USA}

\affiliation[d]{Department of Theoretical Physics, E\"otv\"os University\\
	P\'azm\'any P\'eter s\'et\'any 1, 1117 Budapest, Hungary}

\affiliation[e]{Department of Physics, University of the Pacific\\
3601 Pacific Ave, Stockton CA 95211, USA}

\affiliation[f]{Department of Physics, University of California, San Diego\\
9500 Gilman Drive, La Jolla, CA 92093, USA}

\emailAdd{fodor@bodri.elte.hu}
\emailAdd{kholland@pacific.edu}
\emailAdd{jkuti@ucsd.edu}
\emailAdd{cwong@uni-wuppertal.de}

\abstract{       
	
New tests are applied to  two $\beta$-functions of the much-discussed BSM model with ten
massless fermion flavors in the fundamental representation of the SU(3) color gauge group.
The renormalization scheme of the two $\beta$-functions  is defined on the gauge field gradient flow 
in respective finite or infinite physical volumes at zero lattice spacing. Recently published results in the ten-flavor theory
led to indicators of  an 
infrared fixed point (IRFP) in the finite-volume step $\beta$-function 
in the strong coupling regime of the theory~\cite{Hasenfratz:2020ess}. 
We analyze our substantially extended  set of ten-flavor lattice ensembles at 
strong renormalized gauge couplings and find no evidence
or hint for  IRFP in the finite-volume step $\beta$-function within controlled lattice reach. 
We also discuss new ten-flavor tests of the recently introduced lattice definition and
algorithmic implementation of the $\beta$-function defined
on the gradient flow of the gauge field over infinite Euclidean space-time in the continuum. 
Originally we introduced this new algorithm to match 
finite-volume step $\beta$-functions in massless near-conformal gauge theories 
with the infinite-volume $\beta$-function reached in the chiral limit from small fermion mass deformations
of spontaneous chiral symmetry breaking.
Results from the lattice analysis  of the  ten-flavor infinite-volume $\beta$-function 
are consistent with the absence of  IRFP from our step $\beta$-function based analysis. 
We make important contact at weak coupling in infinite volume
with gradient flow based three-loop perturbation theory, serving
as a first pilot study toward the long-term goal of developing 
alternate approach to the determination  of the strong coupling $\alpha_s$ at the Z-boson pole in QCD.
Without reporting here, our tests of this long-term goal continue in QCD with three massless fermion flavors 
and  in the SU(3) Yang-Mills limit of quenched QCD. 
}

\FullConference{%
 The 38th International Symposium on Lattice Field Theory, LATTICE2021
  26th-30th July, 2021
  Zoom/Gather@Massachusetts Institute of Technology
}

\begin{document}
\maketitle

\section{Introduction and outline}

We report new test results  here  for  two complementary $\beta$-functions of the much-discussed BSM model with ten
massless fermion flavors in the fundamental representation of the SU(3) color gauge group.
Details of the tests are presented for  the gauge field gradient flow  based  ten-flavor step $\beta$-function
in finite volumes, complemented by shorter discussion of the recently introduced  lattice based  algorithm for the ten-flavor $\beta$-function
over infinite four-dimensional Euclidean space-time.
The ten-flavor model 
is particularly relevant for its known BSM popularity when analyzed  as a mass-deformed conformal theory, built on
the hypothesis of conformal IRFP in the massless fermion limit~\cite{LatticeStrongDynamics:2020uwo}. 
The analysis of the mass-deformed conformal ten-flavor theory 
would be different under the alternate hypothesis of 
mass-deformed near-conformal behavior from spontaneous chiral symmetry breaking without IRFP in the 
massless limit. 
The anticipation of conformal behavior with IRFP at strong coupling in the massless fermion limit of the ten-flavor theory
has been motivated by published lattice analyses  in~\cite{Chiu:2017kza,Chiu:2018edw,Hasenfratz:2020ess}. 
In contrast,
our previous results~\cite{Fodor:2018tdg,Fodor:2019ypi}
and the extended new analysis reported here  
do not show any  hints or indicators of IRFP in the two $\beta$-functions  of the model within controlled lattice reach of the strong coupling region.

We will compare our new analysis with results from~\cite{Hasenfratz:2020ess} where the most 
recent systematic effort was presented on emergent  IRFP in the theory.  
Earlier results in~\cite{Chiu:2017kza,Chiu:2018edw} on the location of the
ten-flavor IRFP in the $g^2 \approx 6-8$ range were ruled out in~\cite{Fodor:2018tdg, Fodor:2019ypi},
in agreement with~\cite{Hasenfratz:2020ess}.
However, we differ with the analysis from~\cite{Hasenfratz:2020ess} where indicators were
presented for the
shift of the IRFP to the $g^2\approx 11-12$ range.
The authors  in~\cite{Hasenfratz:2020ess}  are careful to interpret their lattice evidence 
for the emergent IRFP presenting it as  a likely scenario
with added cautious warning on the control of lattice effects. 
We will show in Section 2 that the likely source of the disagreement is the statistical analysis 
in~\cite{Hasenfratz:2020ess} without  noticing the unresolved ambiguity in extrapolating from small lattice volumes to the continuum limit using  domain wall fermion (DWF) lattice implementation.
We did not find indicators for  IRFP in the theory from our own analysis of this ambiguity in fitting  the published DWF data sets of~\cite{Hasenfratz:2020ess}.
This is consistent with our staggered fermion based analysis not supporting IRFP
in significantly larger volumes than allowed by DWF based lattice resources.
We also discuss new ten-flavor tests of the recently introduced lattice definition and
algorithmic implementation of the $\beta$-function defined
on the gradient flow of the gauge field over infinite Euclidean space-time in the continuum. 
Originally we introduced this new algorithm to match 
finite-volume step $\beta$-functions in massless near-conformal gauge theories 
with the infinite-volume $\beta$-function in the chiral limit, reached from small fermion mass deformations  of spontaneous chiral symmetry breaking~\cite{Fodor:2017die}.
New results from the lattice analysis  of the  ten-flavor infinite-volume $\beta$-function 
are consistent with the absence of  IRFP from our finite-volume step $\beta$-function based analysis. 
We make important contact at weak coupling with infinite volume based
three-loop perturbation theory using the renormalization scheme as defined by the gradient flow~\cite{Narayanan:2006rf, Luscher:2010iy, Luscher:2010we,Harlander:2016vzb}.
This serves
as a first pilot study toward the long-term goal of developing 
alternate approach to the determination  of the strong coupling $\alpha_s$ at the Z-boson pole in QCD.
Our tests of this long-term goal continue in QCD with three massless fermion flavors 
and  in the SU(3) Yang-Mills limit of quenched QCD~\cite{Borsanyi}.
In Section 2 we present our new analysis of the finite physical volume based ten-flavor step $\beta$-function in the continuum limit. 
In Section 3 we report  new test results on the infinite volume based ten-flavor 
$\beta$-function with conclusions.

\section{New analysis of the finite-volume step $\beta$-function with ten flavors }

	\subsection{Renormalization on the gauge field gradient flow}
	
	The gradient flow in field theory was originally introduced as a method to regularize divergences
	and ultraviolet fluctuations in lattice
	calculations~\cite{Narayanan:2006rf,Luscher:2009eq}. 
	The gradient flow based diffusion of the gauge fields on  lattice configurations from 
	 Hybrid Monte Carlo (HMC) simulations became the method of choice  for studying renormalization 
	effects with great accuracy on the lattice~\cite{Narayanan:2006rf, Luscher:2010iy, Luscher:2010we,Luscher:2011bx, Lohmayer:2011si}.
	Scale setting in lattice QCD was the 
	immediate first application~\cite{Luscher:2010iy,Luscher:2013vga,Borsanyi:2012zs}  before it 
	became an important nonperturbative tool to calculate $\beta$-functions and scale dependent running couplings
	in gauge theories with the scale set by the finite physical volume. 

In particular, we introduced earlier the gradient flow based
scale-dependent renormalized gauge coupling $g^2(L)$ where the scale is 
set by the linear size $L$ of the finite physical volume~\cite{Fodor:2012td}. This implementation is based on
the gauge invariant trace of the non-Abelian quadratic field strength,
$E(t) = -\frac{1}{2} {\rm Tr} F_{\mu\nu} F_{\mu\nu}(t)$,
renormalized as a composite operator at gradient flow time $t$ on the gauge configurations,
\begin{equation}
g^2(L) = \frac{128 \pi^2 \langle t^2 E(t) \rangle}{3(N^2-1)(1 + \delta(c))}, \hspace{5mm} 
E(t) = -\frac{1}{2} {\rm Tr}~F_{\mu \nu}(t) F_{\mu \nu}(t), 
\end{equation}
where $t$ is the flow time parameter with SU(N) color  group of the non-Abelian gauge field
from  discretized lattice implementation of the flow time evolution. 
The one-parameter finite-volume renormalization scheme and the related gradient flow time $t(L)$  are set by the choice 
$c = \sqrt{8t}/L$ in the $L^4$ physical volume  and  with $\delta(c)$ defined in terms of 
Jacobi elliptic functions in~\cite{Fodor:2012td}. The definition is designed to match the gradient flow based gauge coupling 
with the $\overline{\rm MS}$ scheme at leading 
order $g^2 (L) = g^2_{\overline{\rm MS}}$ for any chosen value of $c$ from this one-parameter family.

\subsection{The ten-flavor lattice ensembles of our analysis}
The renormalization schemes  $c=0.25,$  $c=0.275$, and   $c=0.30$, used  in our work, are identical 
to the ones used in~\cite{Hasenfratz:2020ess} including periodic boundary conditions 
on gauge fields and anti-periodic boundary conditions on fermion fields in all four directions of the lattice.
A general method for the scale-dependent renormalized gauge coupling $g^2(L)$  was introduced earlier to probe the
step $\beta$-function, defined as $( g^2(sL) - g^2(L) ) / \log( s^2 )$ for some preset finite scale change $s$
in the linear physical size $L$
of the  four-dimensional volume in the continuum limit of 
lattice discretization~\cite{Luscher:1992an}. 
To avoid any confusion, the sign convention of the lattice based step $\beta$-function is the opposite of the conventional off-lattice literature.
We use step sizes $s=2,~s=3/2,~s=4/3$ in our analysis of the gradient flow based renormalization scheme.
Restricted to  limited volume sizes by the DWF  implementation, only step size $s=2$ was analyzed in~\cite{Hasenfratz:2020ess}.
In our implementation of the step $\beta$-function analysis, staggered lattice fermions are used with stout smearing in the fermion Dirac operator. 
Large volumes in staggered fermion implementation lead to important cross-checks from multiple step choices of $s$.

The Markov Chain Monte Carlo (MCMC) based gauge field generation of the lattice ensembles use
the Rational Hybrid Monte Carlo (RHMC) evolution code as
described in~\cite{Fodor:2016zil} together with further details on the lattice implementation of the 
step $\beta$-function and its continuum limit. Similar but not identical procedures are followed here. 
We generated lattice ensembles with $L^4$ lattice volumes in the range $L=8,10,12,14,16,18,20,24,28,30,32,36,40,48$ at 21 bare gauge couplings $g_0^2$ with
$6/g_0^2=2.6,2.7,2.8,2.9,3.0,$ $3.1,3.2,3.3,3.4,3.5,3.6,3.7,3.8,3.9,4.0,4.1,4.5,5.0,6.0,7.0,8.0.$ 
In the new infinite volume $\beta$-function analysis of Section 3 only the large volumes with $L=32,36,40,48$ were used. 
For comparison, in~\cite{Hasenfratz:2020ess} lattice ensembles at 17 bare gauge couplings were analyzed with linear lattice sizes $L=8,10,12,14,16,20,24,28,32$ 
restricted by the cost of the DWF implementation for improved  chiral fermion properties  in comparison with staggered fermions we use in large volumes.

Implementing the gradient flow on the lattice requires the discretization of the action density $E = -\frac{1}{2} {\rm Tr}~F_{\mu \nu} F_{\mu \nu}$. 
This appears in three places: the weight in MCMC simulation, the flow of the gauge field, and the observable $\langle t^2 E(t) \rangle$ at flow time $t$. 
Varying the discretization scheme is a critical test for controlled cutoff effects, as all discretization schemes must agree in the continuum limit. 
Our RHMC simulations use  Symanzik-improved action and along the gradient flow independently both the Symanzik and Wilson gauge actions, 
and for the observable $\langle t^2E(t) \rangle$  both the clover and Symanzik versions are used. 
We do not use the Wilson plaquette action for the observable 
$\langle t^2 E(t) \rangle$  with the clover operator showing improved cutoff effects~\cite{Luscher:2010iy}.
This gives four combinations, e.g.~$WSC$ for Wilson flow on Symanzik RHMC gauge configuration generation and with clover observable
for the renormalized coupling.  The other three schemes $SSC$, $SSS$, and $WSS$ are designated accordingly. 
Consistency for results from these scheme on the gradient flow were tested with results presented in what follows. 

Each of the four schemes can be implemented with their original lattice definition (unimproved),
or with tree-level  improvement to reduce cutoff effects. We introduced the  method of tree-level improvement for the gradient flow earlier~\cite{Fodor:2014cpa}. 
These improvements are expected to be most effective at weak couplings. 
Since tree-level improvements only effect the gauge field flow, 
it is expected to work best in perturbation theory when the  one-loop $\beta$-function dominates. 
Large fermion loop contributions, 
which would be improved only in next order beyond tree-level, will begin to dominate 
for large flavor numbers, like the model here with nf=10.  
The large influence of fermion dynamics   drives the nf=10 theory toward the conformal window.
Tree-level improved operators are more ad hoc modifications at strong coupling 
and  it is difficult to predict their cutoff-reducing effects. 
Their deployment requires care because  in some cases it can lead to counter-intuitive effects.
We tested tree-level improvement 
of  the four schemes on the gradient flow at every targeted coupling $g^2$ for both $\beta$-functions with consistent results,
not all of it shown below in limited space.
\subsection {Analysis of the finite volume based step $\beta$-function at c=0.25}

We applied the same algorithmic procedure as in~\cite{Hasenfratz:2020ess} to compare results.  
At fixed $L$, we measure the renormalized gauge coupling $g^2$ on the gradient flow 
at each bare  gauge coupling  $g_0^2$ appearing in the $6/g_0^2$ form in the lattice gauge action. 
The flow time is set by the choice $c=0.25$ in the procedure. For a given choice of the scaled step, like $s=2$, or $s=3/2$, we pair $L$ and $sL$
lattices giving lattice step functions  $(g^2(sL)-g^2(L))/{\rm log}(s^2)$  at each bare lattice gauge coupling, with  $g^2$ 
measured  for a given $L$ on the gradient flow at $c=0.25$ for each $g_0^2$.
In the next step we target preselected $g^2$ values, held to be identical 
for all $L$-pairs to be able to take the continuum limit $a^2/L^2 \rightarrow 0$ at each targeted renormalized coupling $g^2$.
One way to implement this synchronization is to fit a 4th order polynomial first with good statistical confidence at each L to $(g^2(sL)-g^2(L))/{\rm log}(s^2)$ 
as a function of  $g^2$ measured at 21 bare gauge couplings for fixed $L$. 
Using  standard interpolations procedure from the polynomial fits, we calculate the lattice step $\beta$-functions $(g^2(sL)-g^2(L))/{\rm log}(s^2)$   at any given $L$-pair for each of the 
targeted  $g^2$ values. 
The targeted $g^2$-sequence seems to be arbitrary but it is simply designed to sample the continuous range of $g^2(L)$
where the lattice spacing can be removed safely, based on fits to data from the selected lattice ensembles. Every preselected $g^2$ implicitly defines 
a physical scale $L$ in the continuum. 
For illustration we show these fits and interpolations at $c=0.25$ in Fig.~\ref{c250App} of the Appendix for  lattice step functions $ (g^2(sL)-g^2(L))/{\rm log}(s^2)$  with and without tree-level 
improvement for the $ L=24 \rightarrow L=48$ pair with steps $s=2, s=3/2, s=4/3$. For each $L$-pair we have 24 combinations of the $SSC,SSS,WSC,WSS$ schemes 
with three step choices of $s$, unimproved, or improved. We have eight $L$-pairs with 24 fits each
for the lattice ensembles listed above. Red circles mark the lattice step $\beta$-functions at targeted $g^2$ locations in the unimproved schemes and cyan circles show the 
analysis with tree-level improvement in Fig.~\ref{c250App}. Similar procedures are followed for $c=0.275$ and $c=0.30$ with $c=0.30$ shown in Fig.~\ref{c300App}. 

In the last step of the fitting procedure 
the interpolated lattice step $\beta$-functions, $(g^2(sL)-g^2(L))/{\rm log}(s^2)$,  at all  targeted $g^2$ values are 
extrapolated to the $a^2/L^2 \rightarrow 0$ continuum limit. 
Typically six or seven pairs were used with linear or quadratic fits in $a^2/L^2$ at step $s=2$ and  four or five pairs at $s=3/2$ and $s=4/3$. The linear and quadratic fits 
in $a^2/L^2$ were accepted if they produced statistically consistent results.
\begin{figure}[h!]	
	\begin{center}
		\begin{tabular}{cc}
			\includegraphics[width=0.4\textwidth]{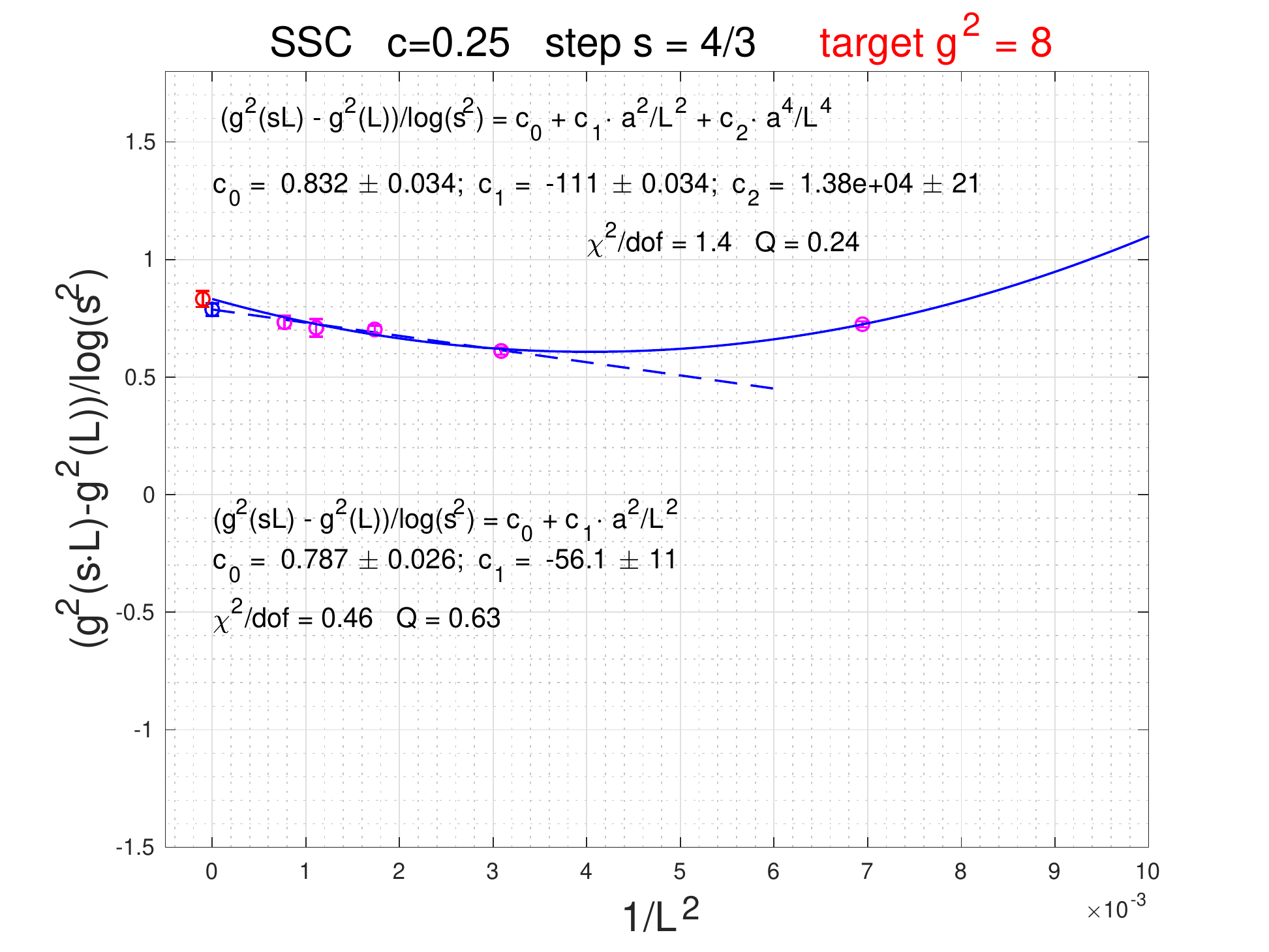}&
			\includegraphics[width=0.4\textwidth]{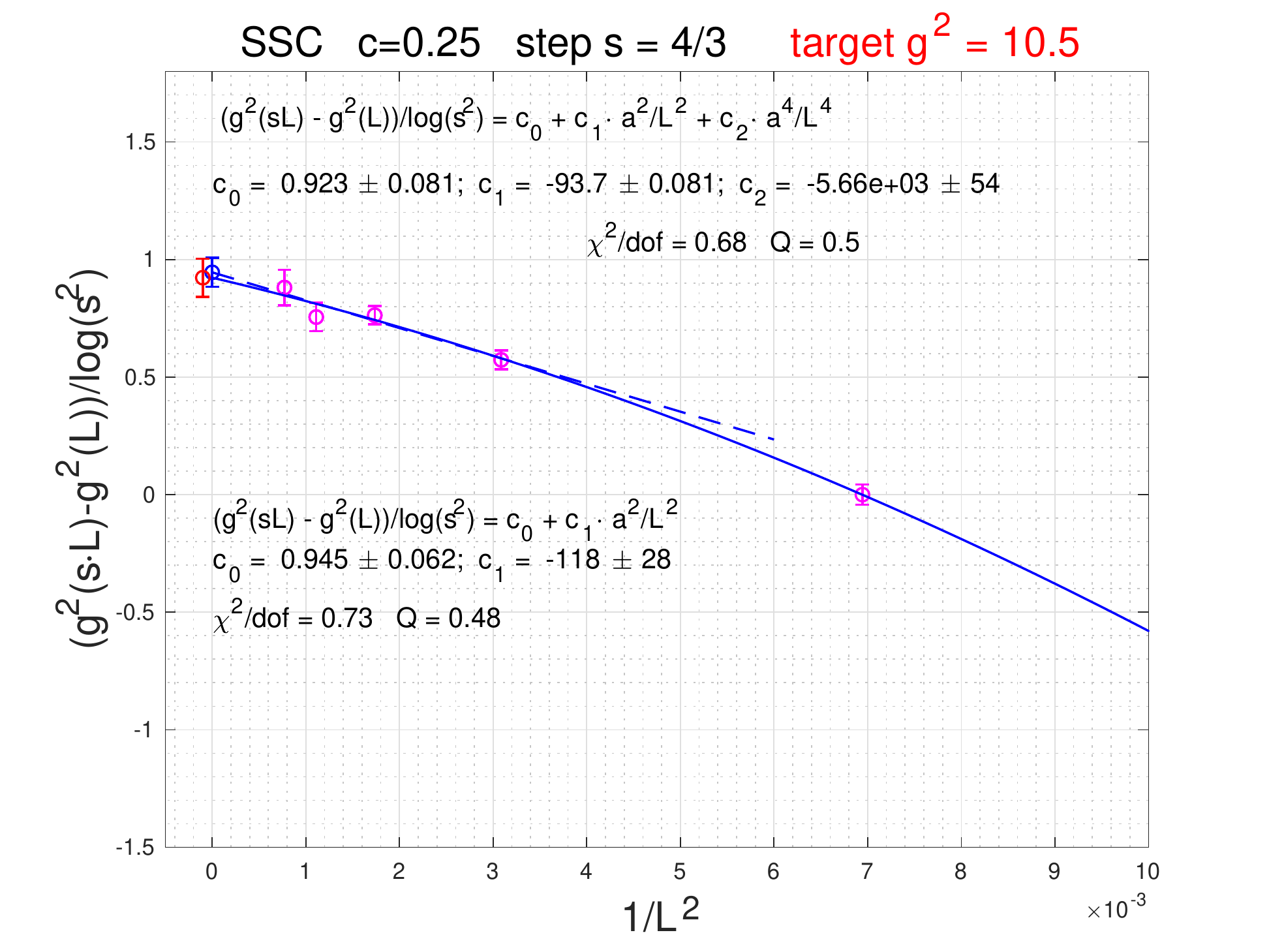}\\
		   \includegraphics[width=0.4\textwidth]{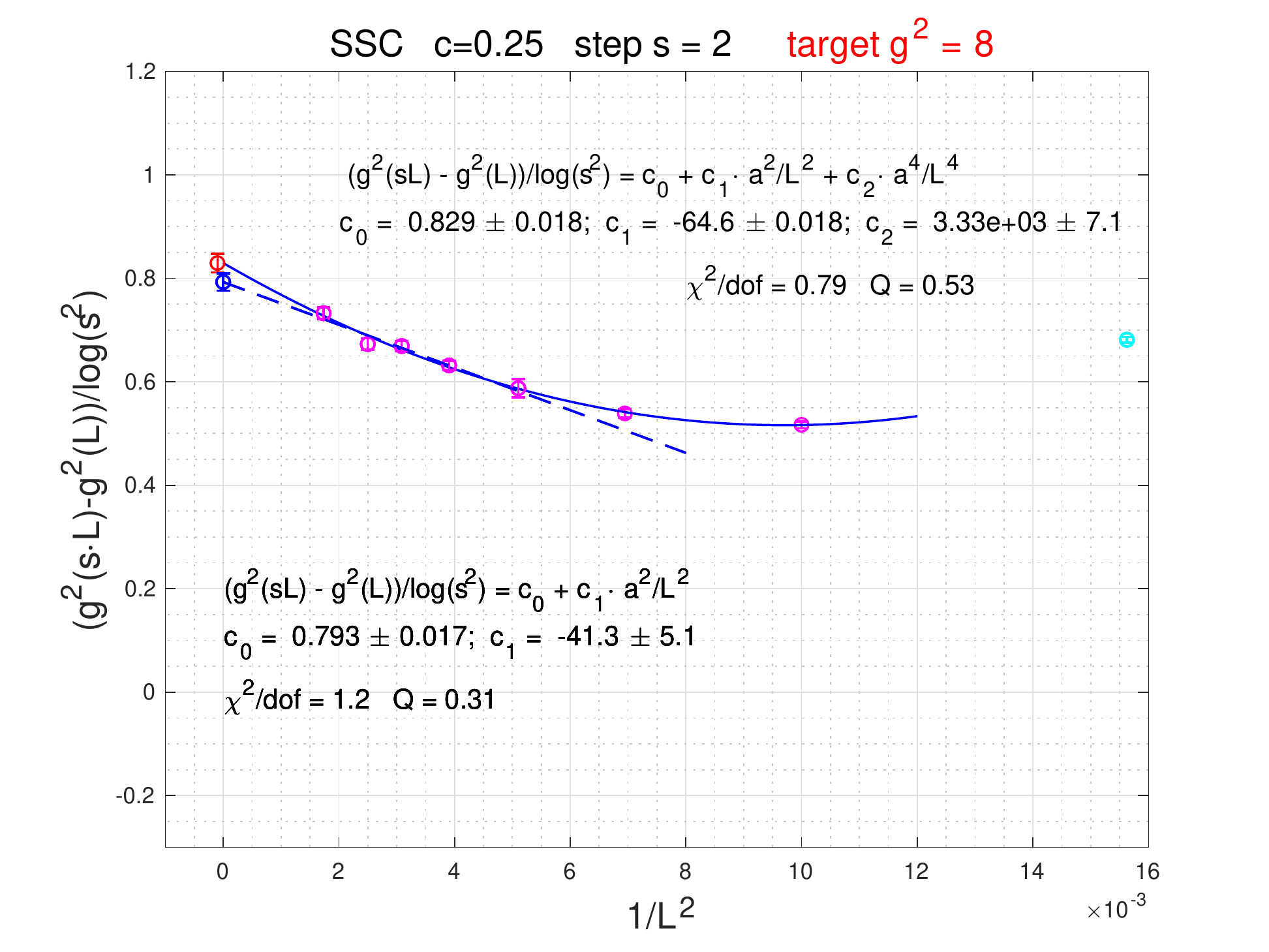}&
			\includegraphics[width=0.4\textwidth]{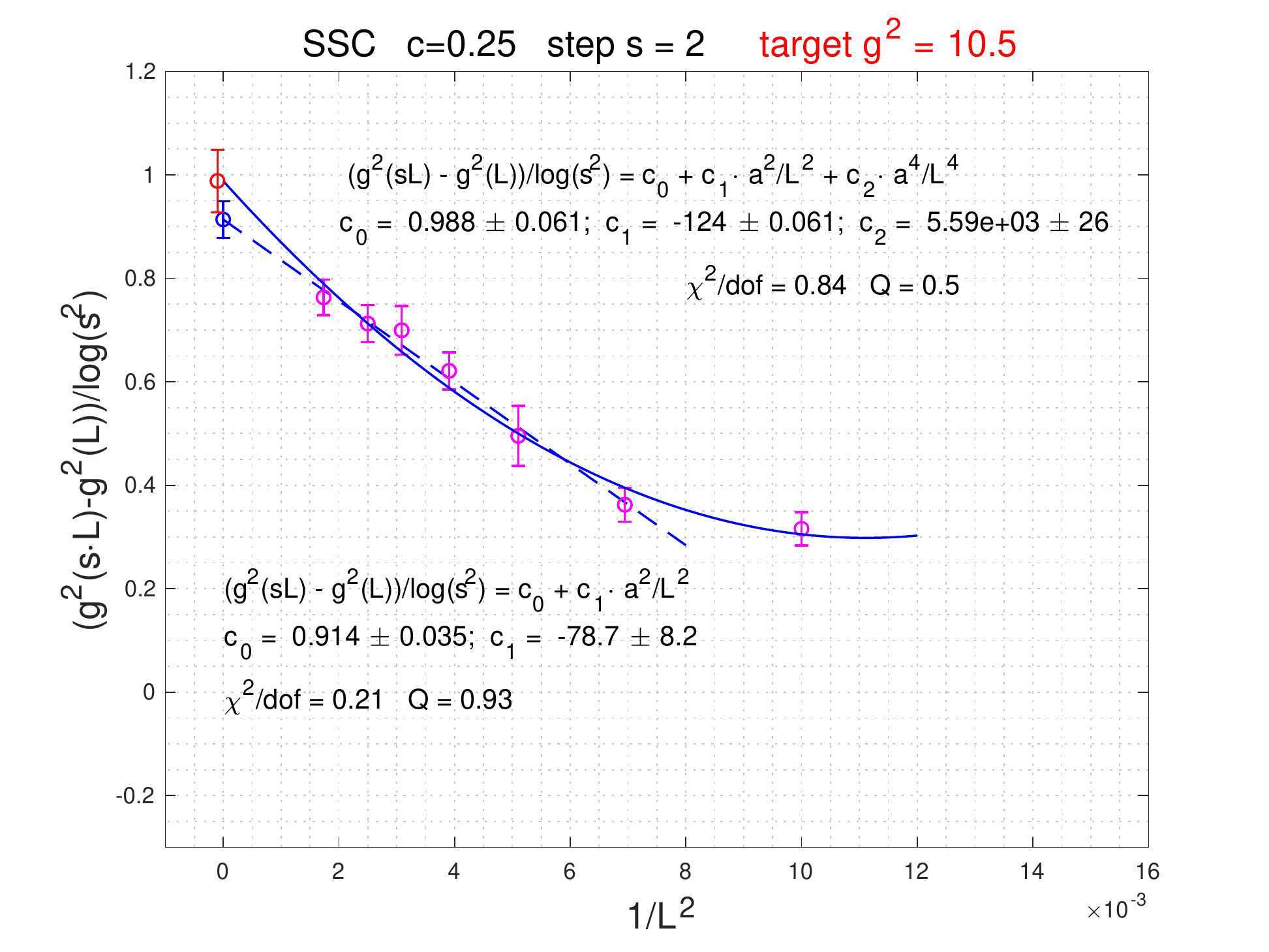}
		\end{tabular}
	\end{center}		
    \vskip -25pt
	\caption{\label{c250a}{\small For $g^2=8$ in the continuum limit, the upper left panel shows linear fit in $a^2/L^2$ 
			with lower  L-values of the pairs at $L=18,24,30,36$ for step $s=4/3$ and quadratic fit when the $L=12\rightarrow 16$ pair is added to the fit.  
			The upper right panel shows similar fits when $g^2=10.5$.  The two lower panels are fits at $s=2$.
	       The Q-values of the fits are discussed  as P-value hypothesis tests in the notation of~\cite{Hasenfratz:2020ess}}.}
\end{figure}
Fig.~\ref{c250a} shows fits in the $SSC$ scheme at targeted couplings $g^2=8$ and $g^2=10.5$  with scaled step sizes $s=4/3$ and $s=2$ and with statistical consistency between linear and quadratic fitting 
in $a^2/L^2$ for  extrapolation to the continuum limit. 
The  consistency of the continuum limit with linear and quadratic fitting in the $g^2=8-10.5$ range is particularly 
important in comparison with the $c=0.25$ analysis of~\cite{Hasenfratz:2020ess} in the same range, as further discussed below.

\begin{figure}[h!]	
	\begin{center}
		\begin{tabular}{cc}
			\includegraphics[width=0.4\textwidth]{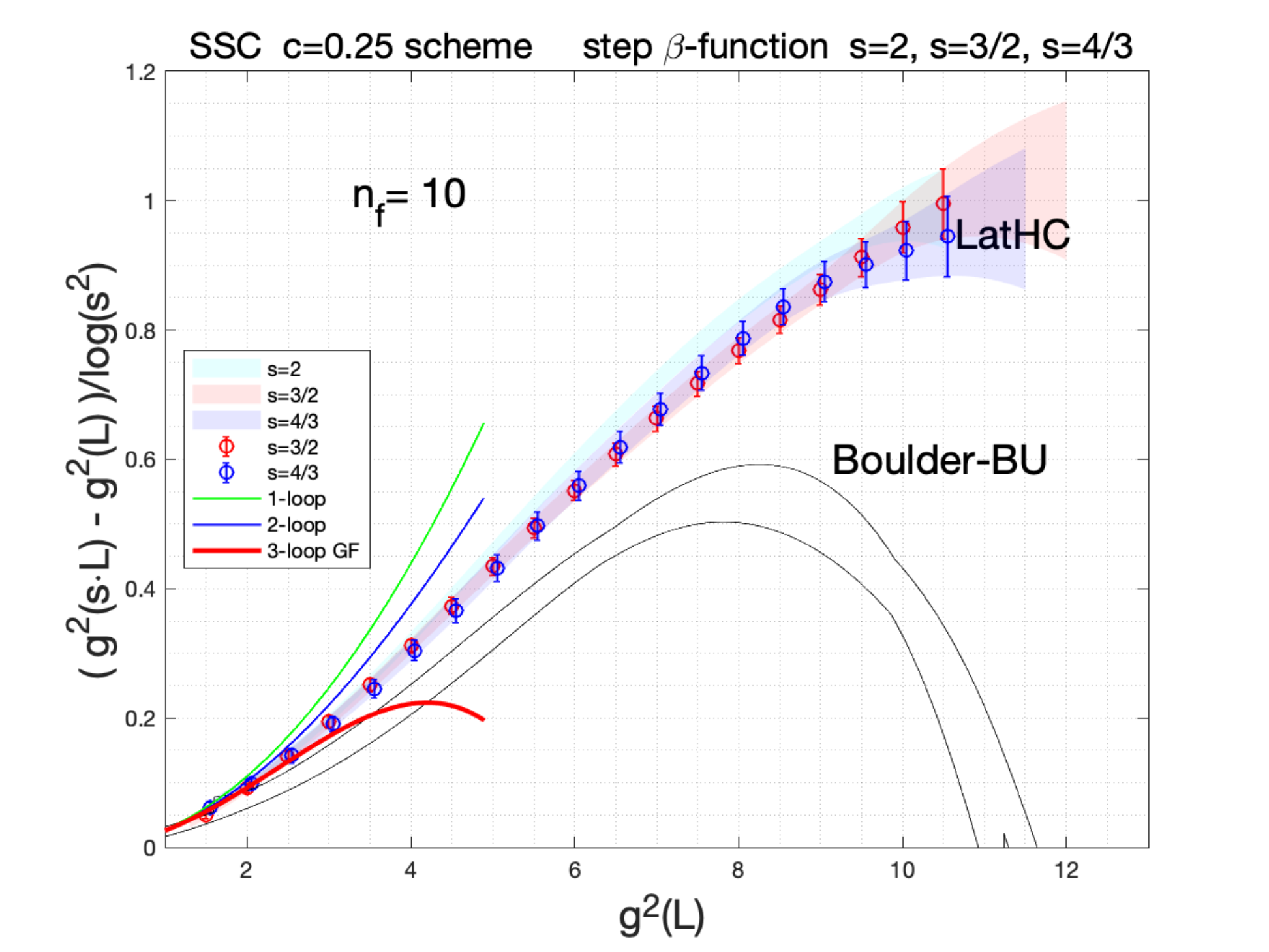}&
			\includegraphics[width=0.4\textwidth]{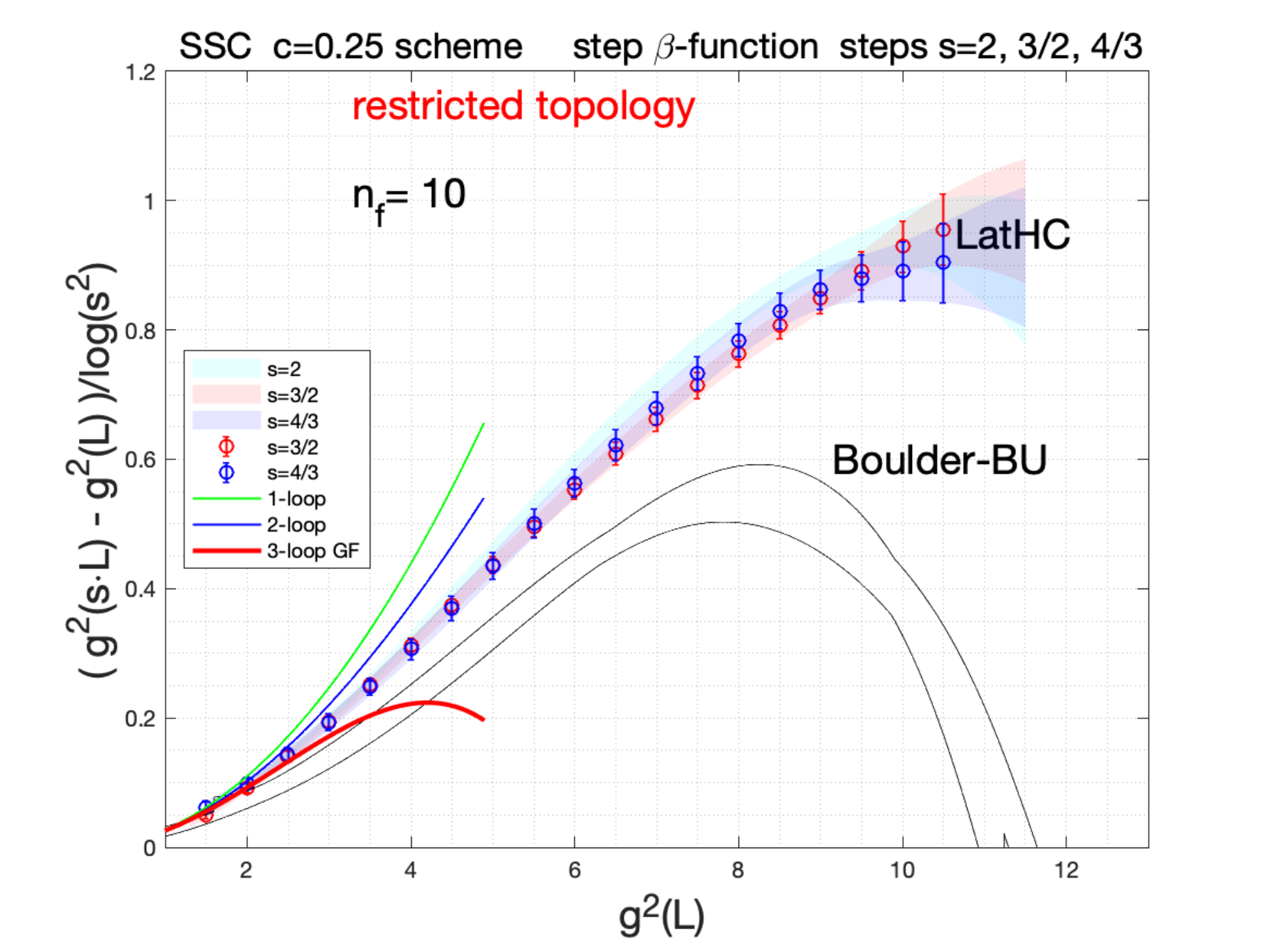}\\
			\includegraphics[width=0.4\textwidth]{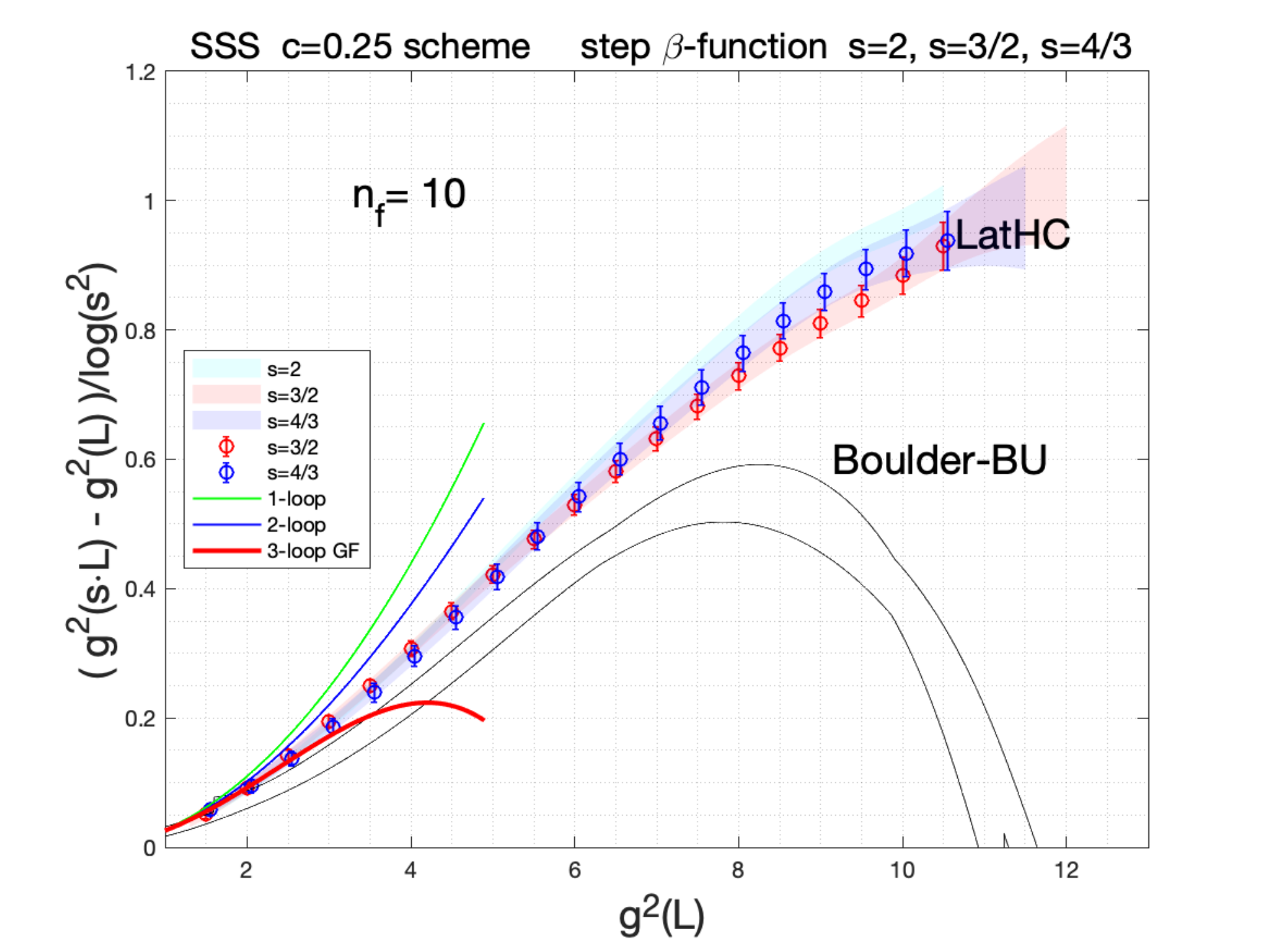}&
			\includegraphics[width=0.4\textwidth]{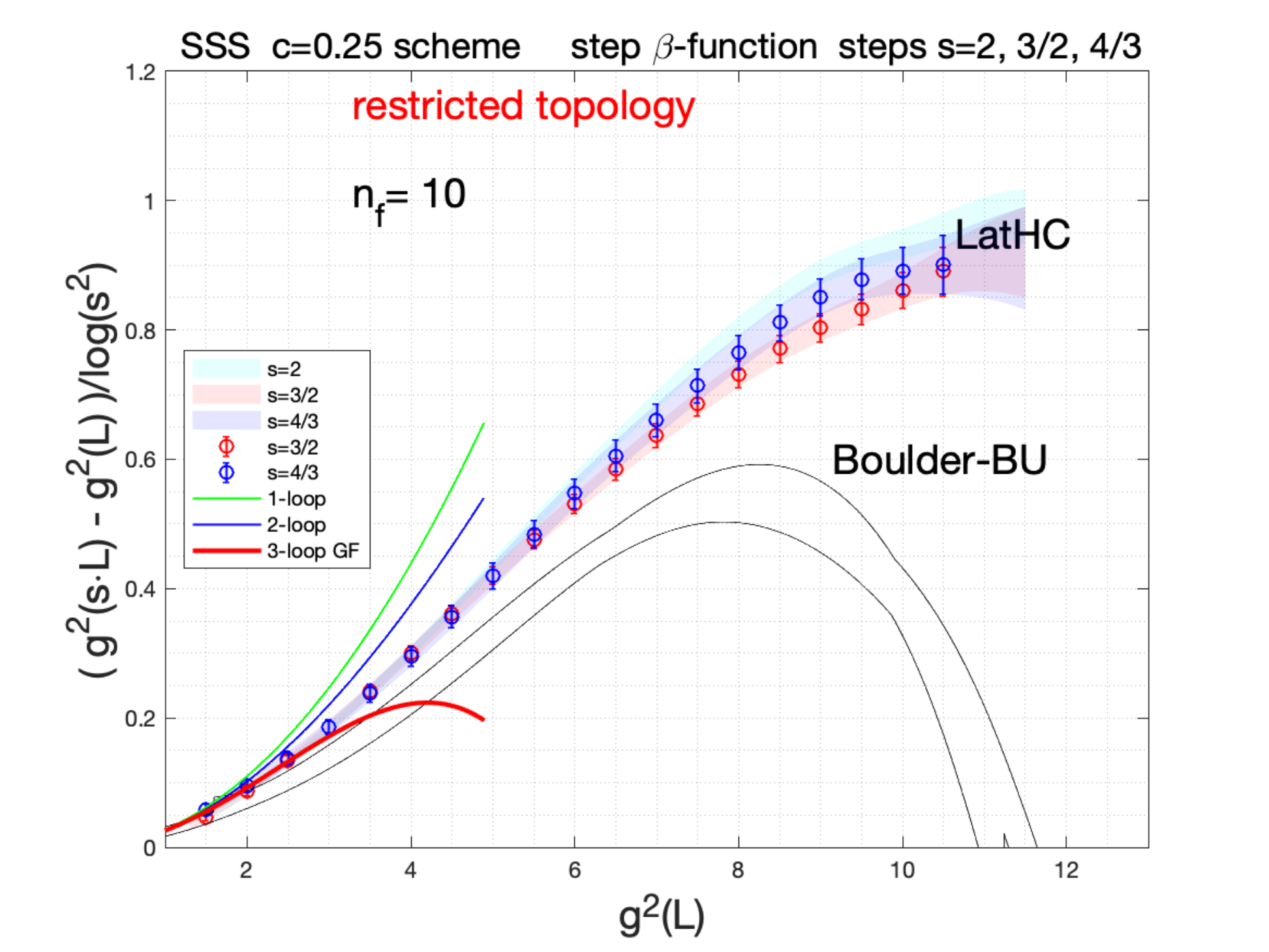}			
		\end{tabular}
	\end{center}		
     \vskip -25pt
	\caption{\label{c250b} {\small  Fits of our data sets without tree-improvement are marked with LatHC tag in the plots and fit results from the authors of~\cite{Hasenfratz:2020ess} are marked with  Boulder-BU tag.
			We will keep these tags in several  plots of the report. 
			The upper left panel shows LatHC fit results  for the continuum 
			step $\beta$-function,  $(g^2(sL)-g^2(L))/{\rm log}(s^2)$, in the $SSC$ scheme  at three different step sizes in the $g^2=1.5-12$ range of the continuum theory. The $s=2$ fits were quadratic in $a^2/L^2$,
			the $s=3/2$ and $s=4/3$ fits were linear in $a^2/L^2$.
			Two black lines represent the upper and lower errors of the band from~\cite{Hasenfratz:2020ess} in the tree-level improved nWSS scheme of the paper.
	     	The upper right panel shows the same fitting procedure for gauge configurations with the topological charge restricted 
			to the $Q_{\rm top} \leq 0.3$ range without any significant effect, as discussed in the text. The two lower panels show the tests for the  $SSS$ scheme with the same fitting procedure as the $SSC$ scheme. }}
\end{figure}

Moving now to the targeted continuous $g^2$ range where continuum limits can be taken with statistical consistency,
in Fig.~\ref{c250b} we show  fitted results of the continuum step $\beta$-function, $(g^2(sL)-g^2(L))/{\rm log}(s^2)$.  
Sampling from the continuous (color-shaded) range of $g^2$ values are shown with fits  in the $\chi^2/dof\approx 1$ range.
The size of the finite physical volume in the continuum limit is implicitly set by the  value of $g^2(L)$ which is held fixed for all $L$-pairs while the cutoff is being removed.
The incompatible results in Fig.~\ref{c250b} are striking with overwhelming statistical  significance between the LatHC-tagged analysis with monotonic step $\beta$-function 
and the BoulderBU-tagged analysis of~\cite{Hasenfratz:2020ess} hitting an IRFP around $g^2\approx 11$. We will identify the most likely source of this discrepancy in what follows below.
Cutoff effects from topological charge fluctuations do not explain the discrepancy. On the two right-side panels of Fig.~\ref{c250b} we show that restricting the topological charge 
in the LatHC fits to the $Q_{\rm top} \leq 0.3$ range has no significant effect. We only did this in response to~\cite{Hasenfratz:2020ess}  where the issue was raised, although 
with small effects found in the DWF based analysis as well. We do not see any justification for imposing topological charge based cuts on the analysis with added remark
in Section 2.5 on the Dirac spectrum of staggered lattice fermions.

There are two added features of the LatHC fits in Fig.~\ref{c250b}. With $SSC$ and $SSS$ fits shown,  consistency is as expected  at fixed choices of $c$ and  $s$ across 
different schemes on the gradient flow leading to the same continuum step $\beta$-function.  We also find similar consistency fitting the $WSC$ and $WSS$ schemes. 
The three step sizes with $s=2, 3/2, 4/3$ will  have slightly  different shapes which we cannot resolve with the limited accuracy of our data.
They all would be calculable from the $s\rightarrow 1$ limit of the finite-volume step function and they all should flow 
into the same IRFP in the $c=0.25$ finite-volume setting, so they are directly relevant for our tests.
The first  contact with perturbation theory is also encouraging.
We show in Fig.~\ref{c250b} the infinite-volume based perturbative loop expansion in the gradient flow based renormalization scheme which will be more directly 
relevant in the $s\rightarrow 1$ limit of Section 3 for infinite physical volume analysis 
of the gradient flow based $\beta$-function~\cite{Harlander:2016vzb,Artz:2019bpr}. 
The finite-volume step $\beta$-functions are expected to be close to the perturbative loop expansion of the plot
at weak coupling without strictly tracking it. 

\begin{figure}[h!]	
	\begin{center}
		\begin{tabular}{cc}
			\includegraphics[width=0.4\textwidth]{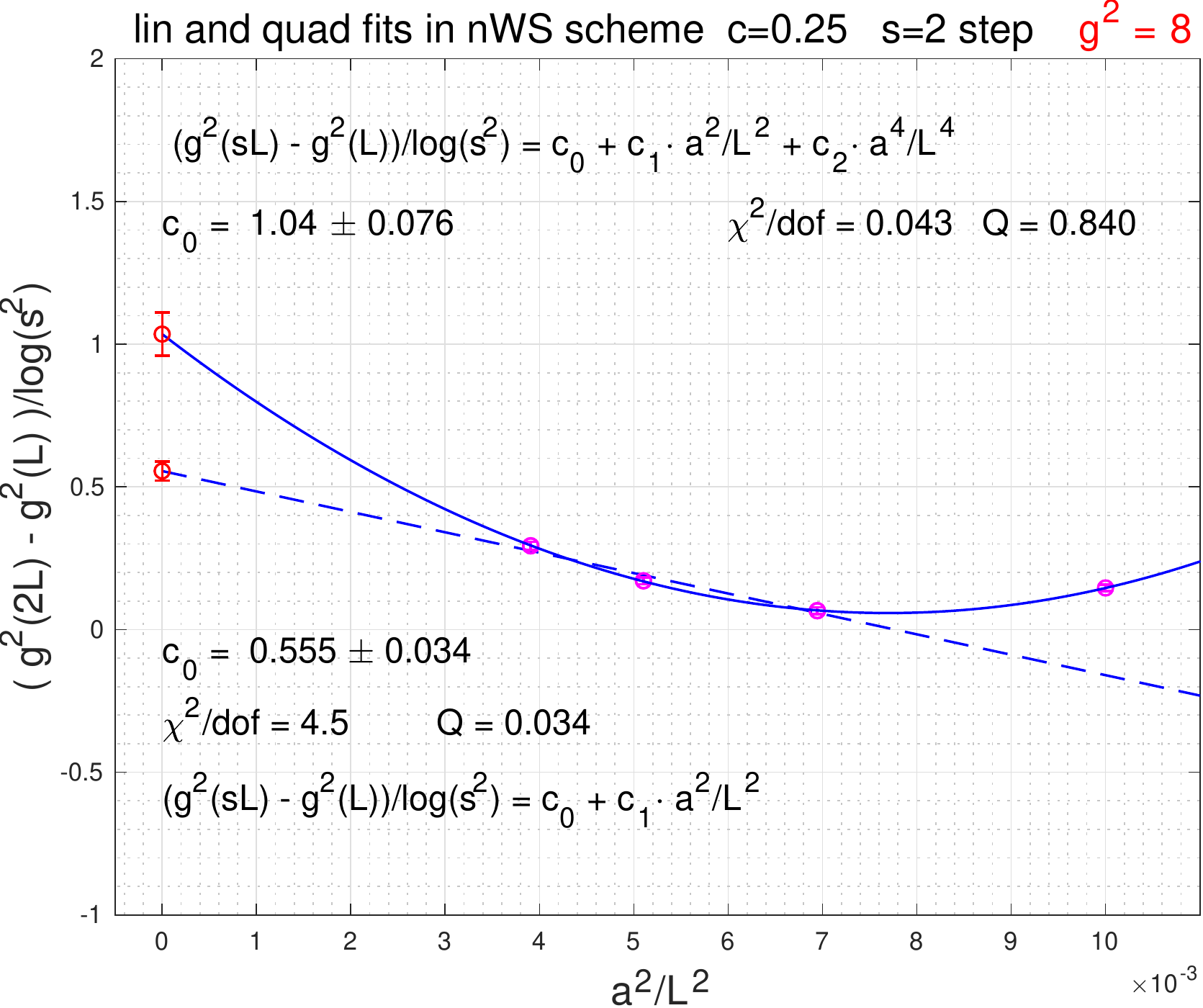}&
			\includegraphics[width=0.4\textwidth]{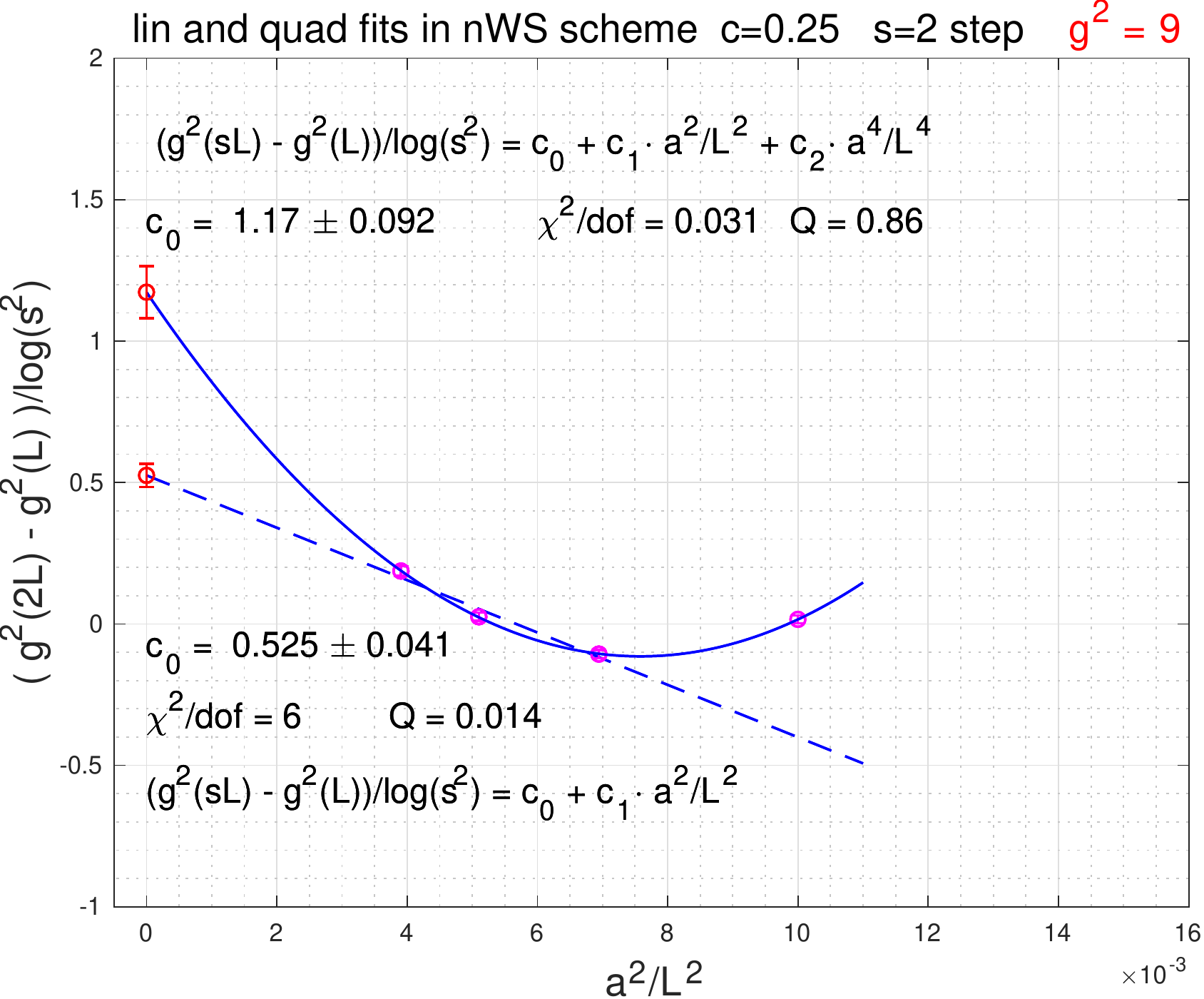}\\
			\includegraphics[width=0.4\textwidth]{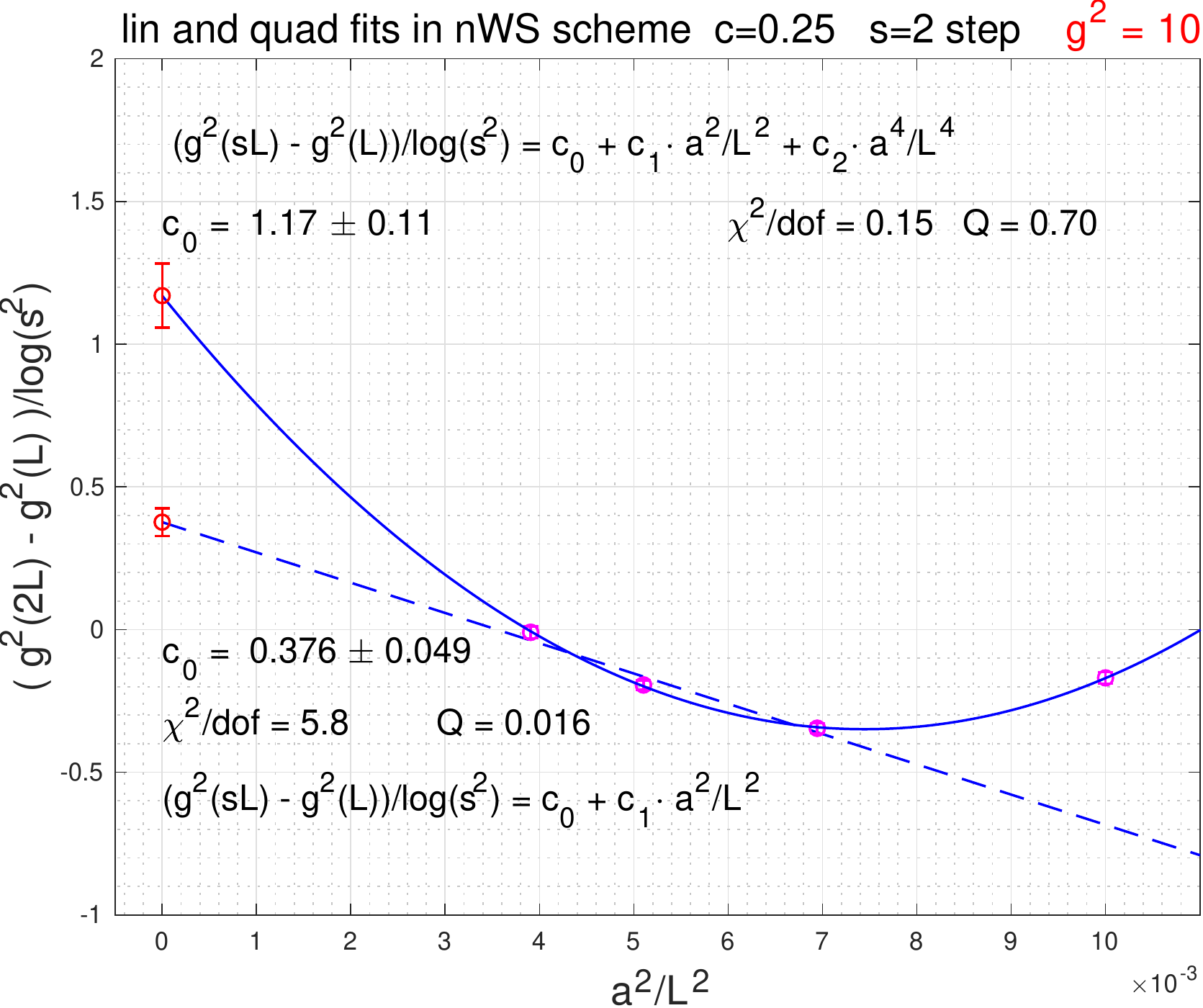}&
			\includegraphics[width=0.45\textwidth]{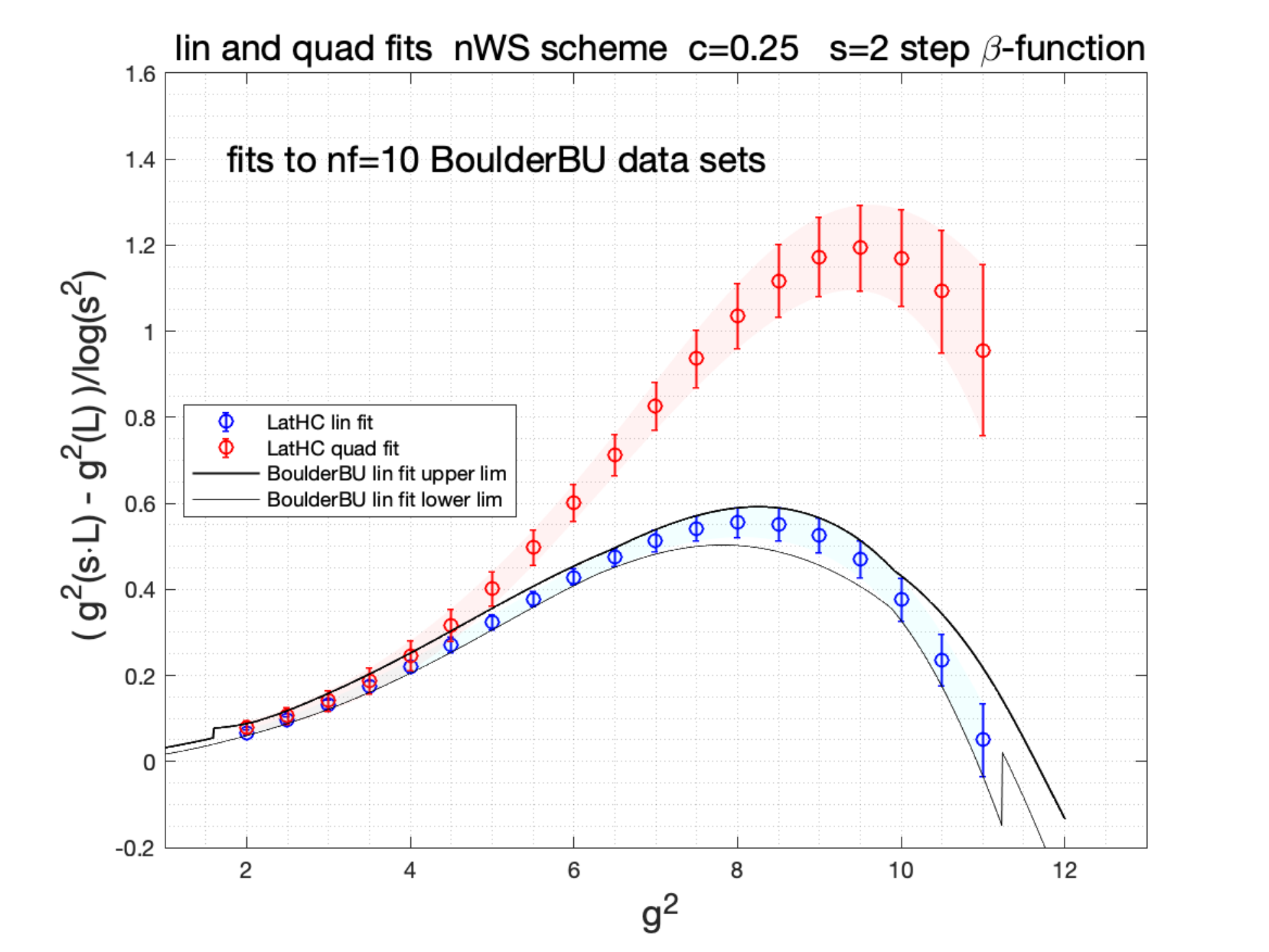}
		\end{tabular}
	\end{center}		
	\vskip -25pt
	\caption{\label{c250c}  {\small Our own  fits are shown in four panels to the published $c=0.25$ data of~\cite{Hasenfratz:2020ess}.
	}}   
\end{figure}

Based on Fig.~\ref{c250c} we will  investigate  now  what we believe to be the source of the controversy at $c=0.25$.
We show in Fig.~\ref{c250c}  our own fits to the published $c=0.25$ data from~\cite{Hasenfratz:2020ess}. 
We selected  the preferred nWS scheme of the authors
from data in Table III from~\cite{Hasenfratz:2020ess}. Similarly to tests of our own lattice ensembles 
which we presented in Fig.~\ref{c250a} we are probing here the consistency 
of linear and quadratic fitting in $a^2/L^2$ for the DWF  based step $\beta$-function in the continuum limit. 
Our linear fits in $a^2/L^2$ are in good agreement  with the published fits of~\cite{Hasenfratz:2020ess} for all three choices of $c$ with  $c=0.25, 0.275, 0.30$ at each coupling. This is not the issue. 
Differing from~\cite{Hasenfratz:2020ess}, we also investigate quadratic fits in the $a^2/L^2$ variable over the entire $g^2=2-11$ range under consideration. 
Comparing linear and quadratic fits to the data of~\cite{Hasenfratz:2020ess} at  $c=0.25$ leads to rather peculiar results.
The linear fitting hypothesis  turns out to be unacceptable at strong couplings in the $g^2 \approx 8-10.5$ range. 
The authors of~\cite{Hasenfratz:2020ess} know that.
Linear fitting works better at weaker couplings in the $g^2 \approx 3-7$ range but rapidly deteriorates
to unacceptable level toward $g^2\approx 8$ and remains unacceptable out to $g^2\approx 10$. 
At very strong coupling in the $g^2\approx  11 - 12$ range linear fits begin to work again in the nWS scheme but with rapidly increasing errors in the fits. 
Accepting the linear fits would lead to the IRFP, reached around $g^2 \approx 11$. 
In sharp contrast,  quadratic fitting in $a^2/L^2$ works  well in the entire $g^2\approx 3-12$ range 
deteriorating below $g^2\approx 3$ which is irrelevant for the issue here.
The nWS based quadratic  fit to DWF data in Fig.~\ref{c250c} overshoots the results from our staggered fermion based fits peaking around
$g^2\approx 9.5$ with   $ (g^2(sL)-g^2(L))/{\rm log}(s^2)\approx1.2$ of the step $\beta$-function and  dropping back to $\beta\approx 0.95$ at $g^2=11$.

Based on our analysis, it is not credible to rely on the hypothesis of linear fits leading to the highly suspect
IRFP at $c=0.25$.  This would point then to the hypothesis of quadratic fits with much better statistical significance. 
However, the two hypotheses,  both showing
statistically acceptable fits to the continuum $\beta$-function outside the $g^2 \approx 8-10$ range, should be compatible with each other for meaningful continuum fits. 
Unfortunately, they are not compatible, leading to 
paradoxical outcome in their predictions of the continuum limit, statistically far separated.
Perhaps further scrutiny of the statistical analysis might resolve this contradiction. 
For example, quadratic DWF fits, extrapolating from small  volumes to $L/a\rightarrow \infty$,  are suspect 
with less controlled  cutoff effects from smaller $L$ at shorter flow time. 
We speculate that  larger volumes are needed  for more consistent DWF analysis at $c=0.25$ for consistent  linear and quadratic fits
to the continuum limit.

\subsection {Analysis of the finite volume based step $\beta$-function at c=0.275 and c=0.30}
\begin{figure}[h!]	
	\begin{center}
		\begin{tabular}{cc}
			\includegraphics[width=0.4\textwidth]{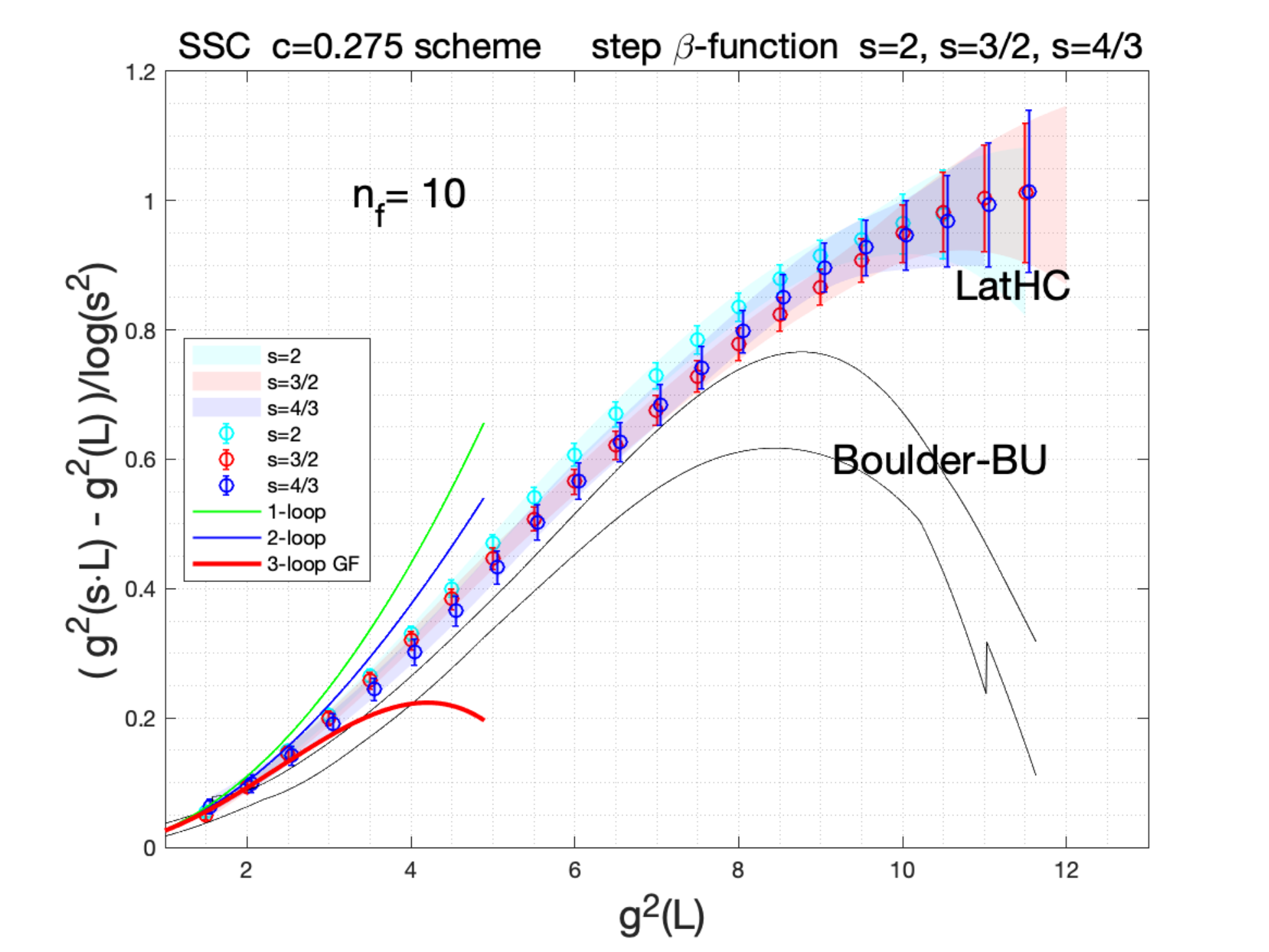}&
			\includegraphics[width=0.4\textwidth]{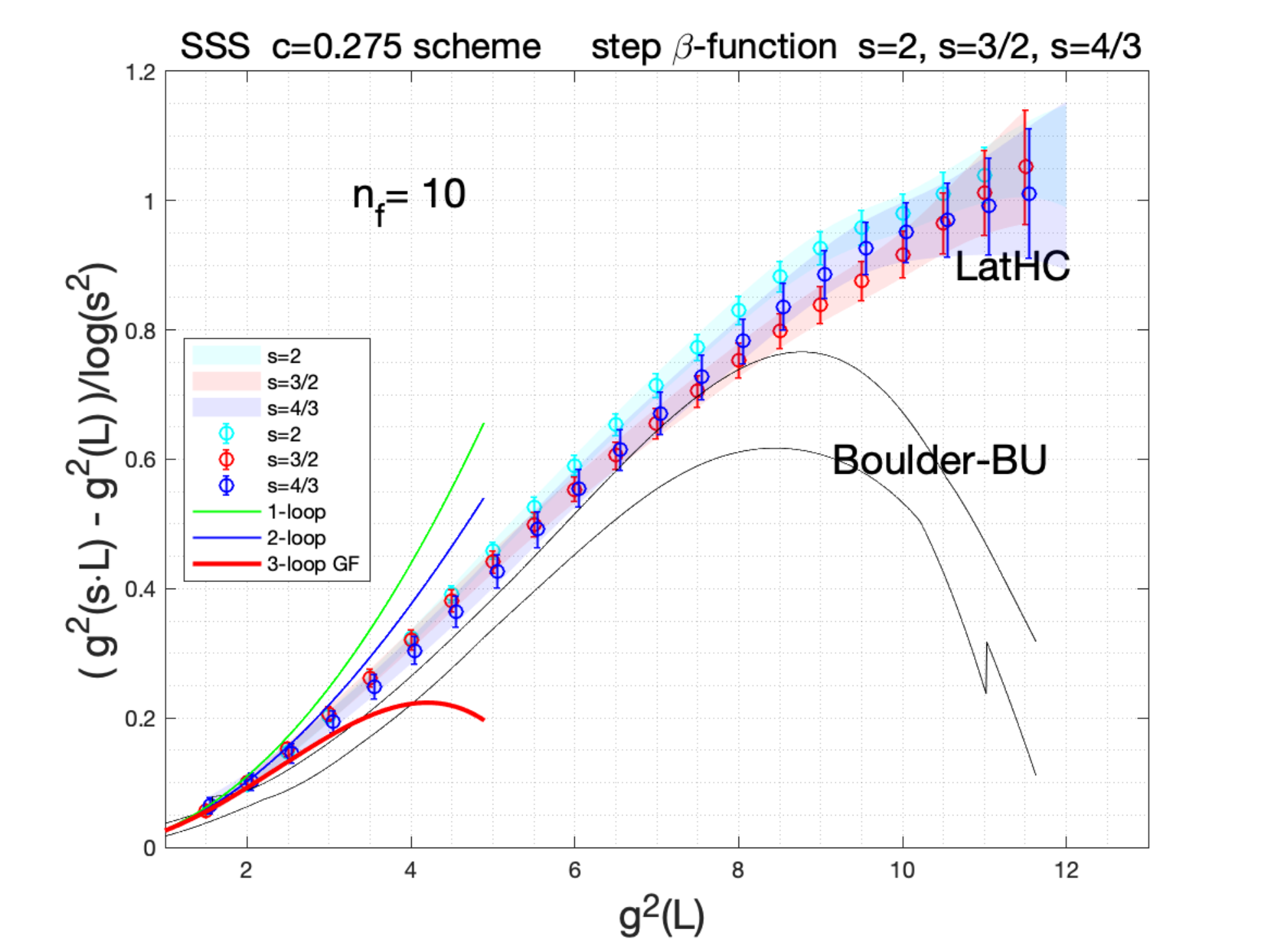}\\
			\includegraphics[width=0.4\textwidth]{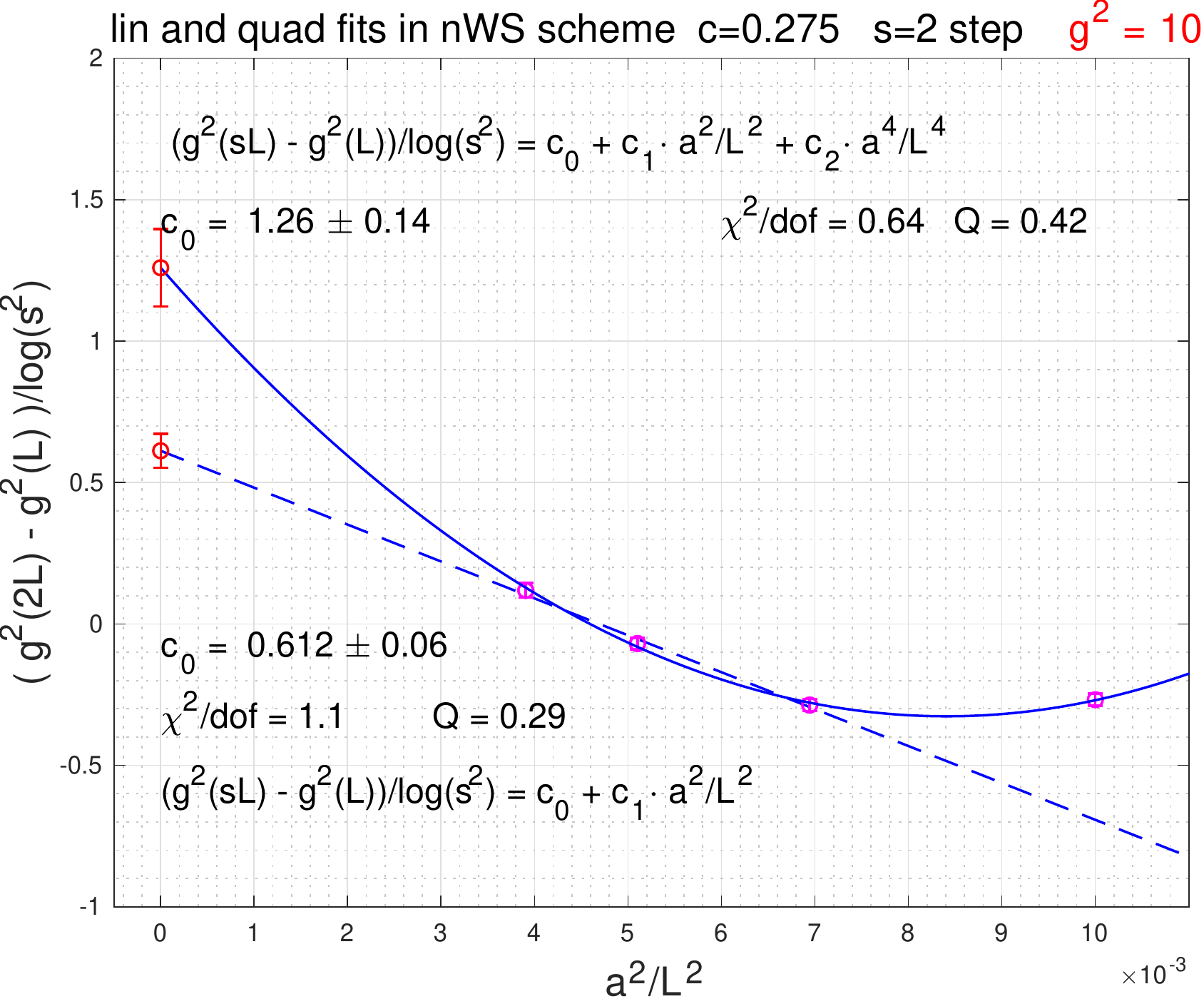}&
			\includegraphics[width=0.45\textwidth]{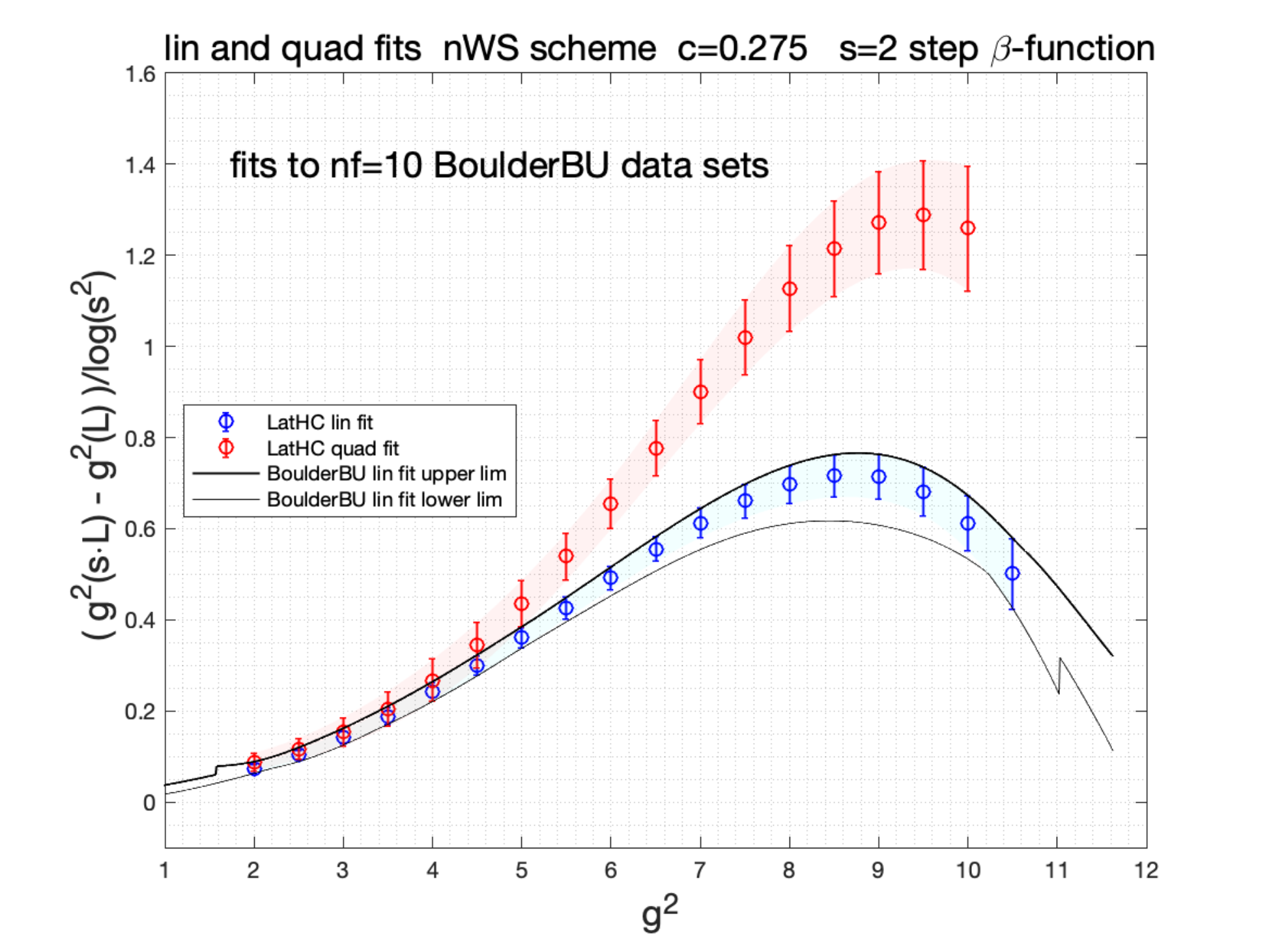}
		\end{tabular}
	\end{center}		
    \vskip -25pt
	\caption{\label{c275a}{\small  Results from LatHC staggered fermions at $c=0.275$ are shown in the two upper panels with fits from~\cite{Hasenfratz:2020ess} also marked.
	                                                   Our own fits at $c=0.275$  to DWF data from~\cite{Hasenfratz:2020ess} are shown  in the two lower panels.}}
\end{figure}

The two upper panels of the fits in Fig.~\ref{c275a}  show our reasonably consistent fits with staggered fermions at $c=0.275$ in the $SSC$ and $SSS$ schemes at steps $s=2, 3/2, 4/3$,
again  without any indicator of an IRFP developing in the $g^2\approx 11-12$ range. 
Our own fits to the DWF data of ~\cite{Hasenfratz:2020ess} changed somewhat. There are now statistically reasonable  linear and quadratic fits to the data of~\cite{Hasenfratz:2020ess}
in the entire $g^2$ range but they are still incompatible with each other, separating around $g^2\approx 5$. 
The linear fits are closer to our fits in a broader $g^2$ range but separate around $g^2\approx 9-10$ with some significance.  Our quadratic fits to DWF data from~\cite{Hasenfratz:2020ess} 
significantly overshoot the fits to our staggered fermion based ensembles. 
We speculate again that the strange results the fitted DWF data exhibit is due to systematic difficulties when extrapolating to infinite 
lattice volumes from small DWF lattice. 

\begin{figure}[h!]	
	\begin{center}
		\begin{tabular}{cc}
			\includegraphics[width=0.40\textwidth]{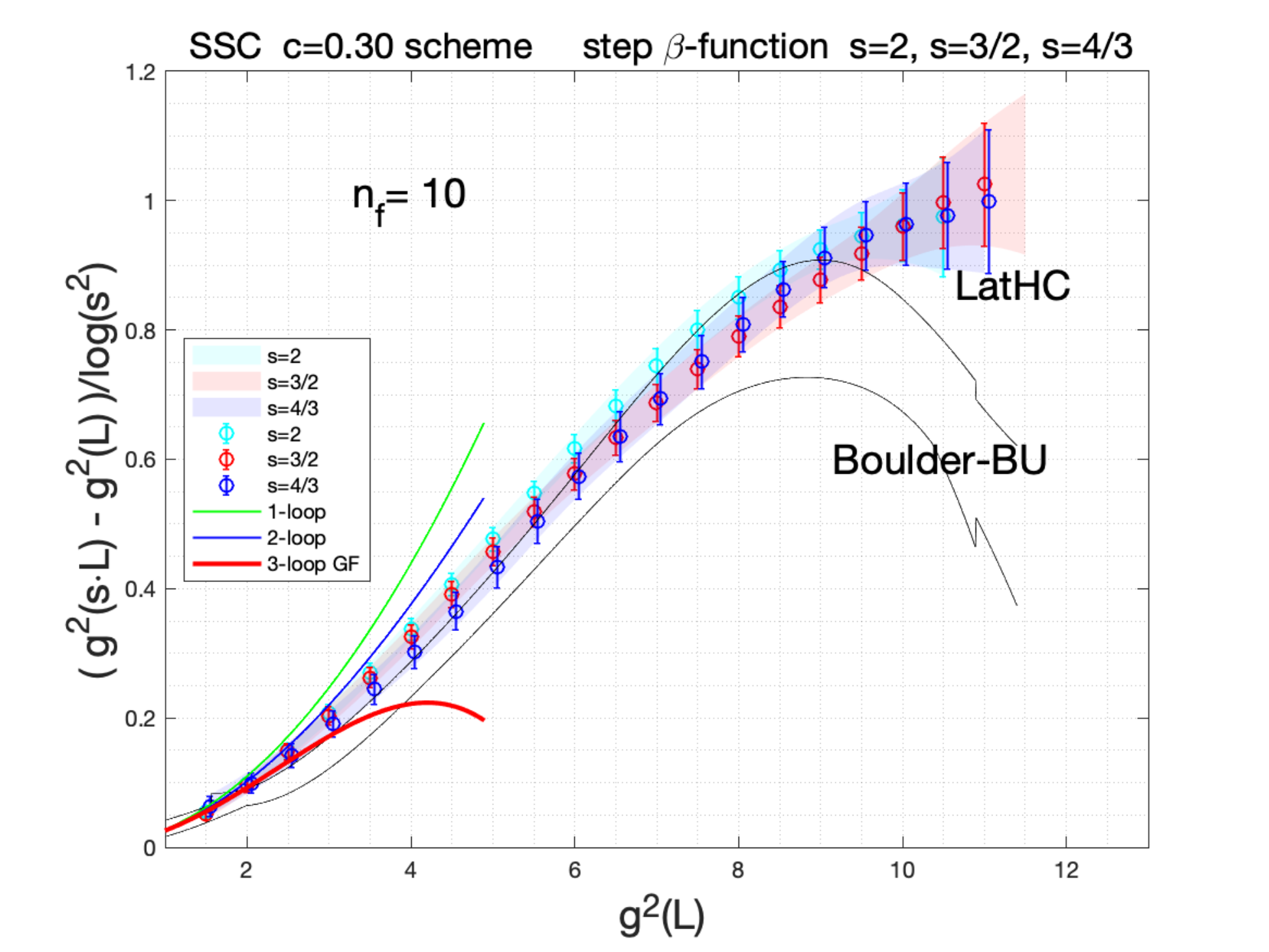}&
			\includegraphics[width=0.40\textwidth]{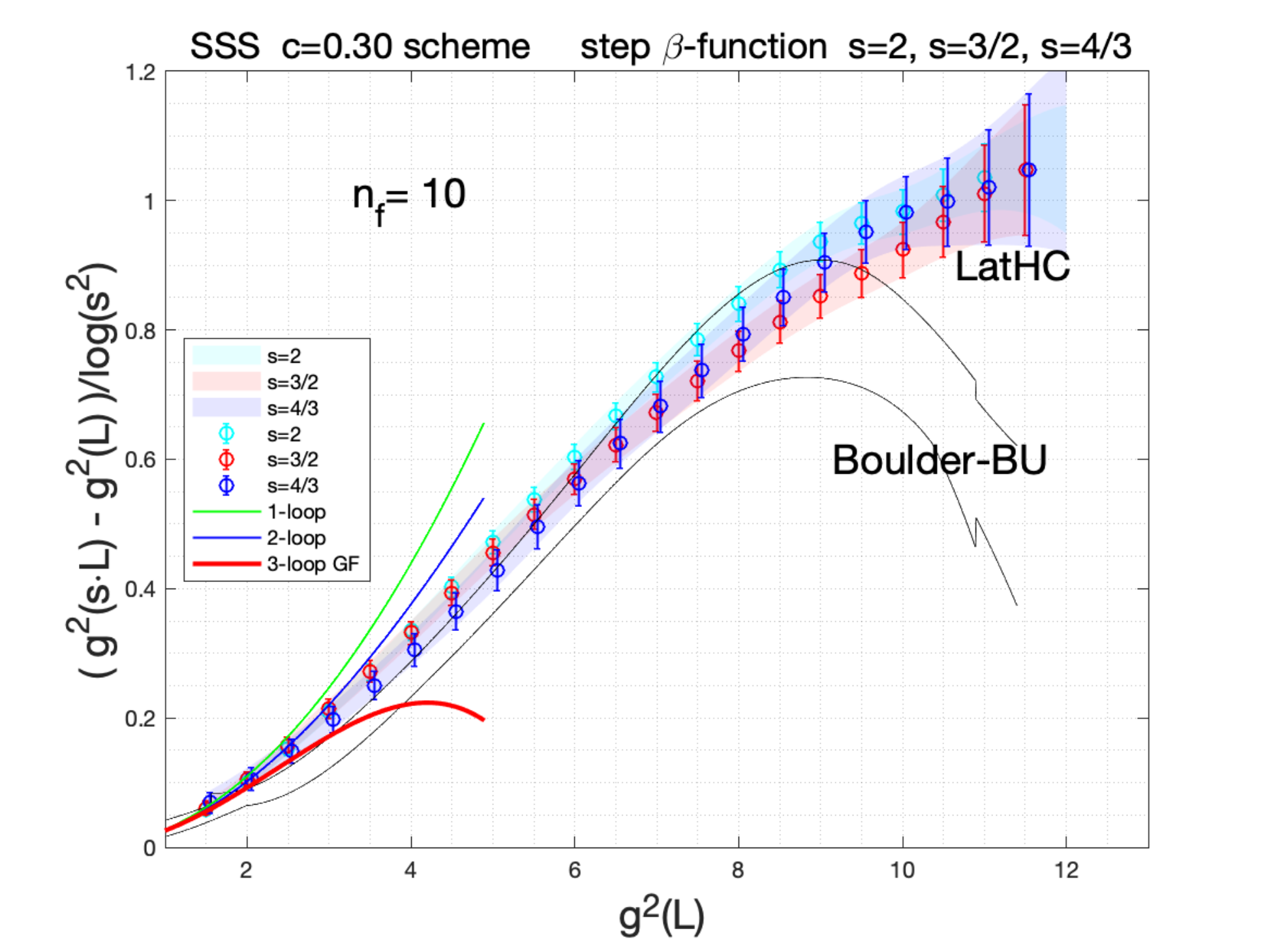}\\
			\includegraphics[width=0.4\textwidth]{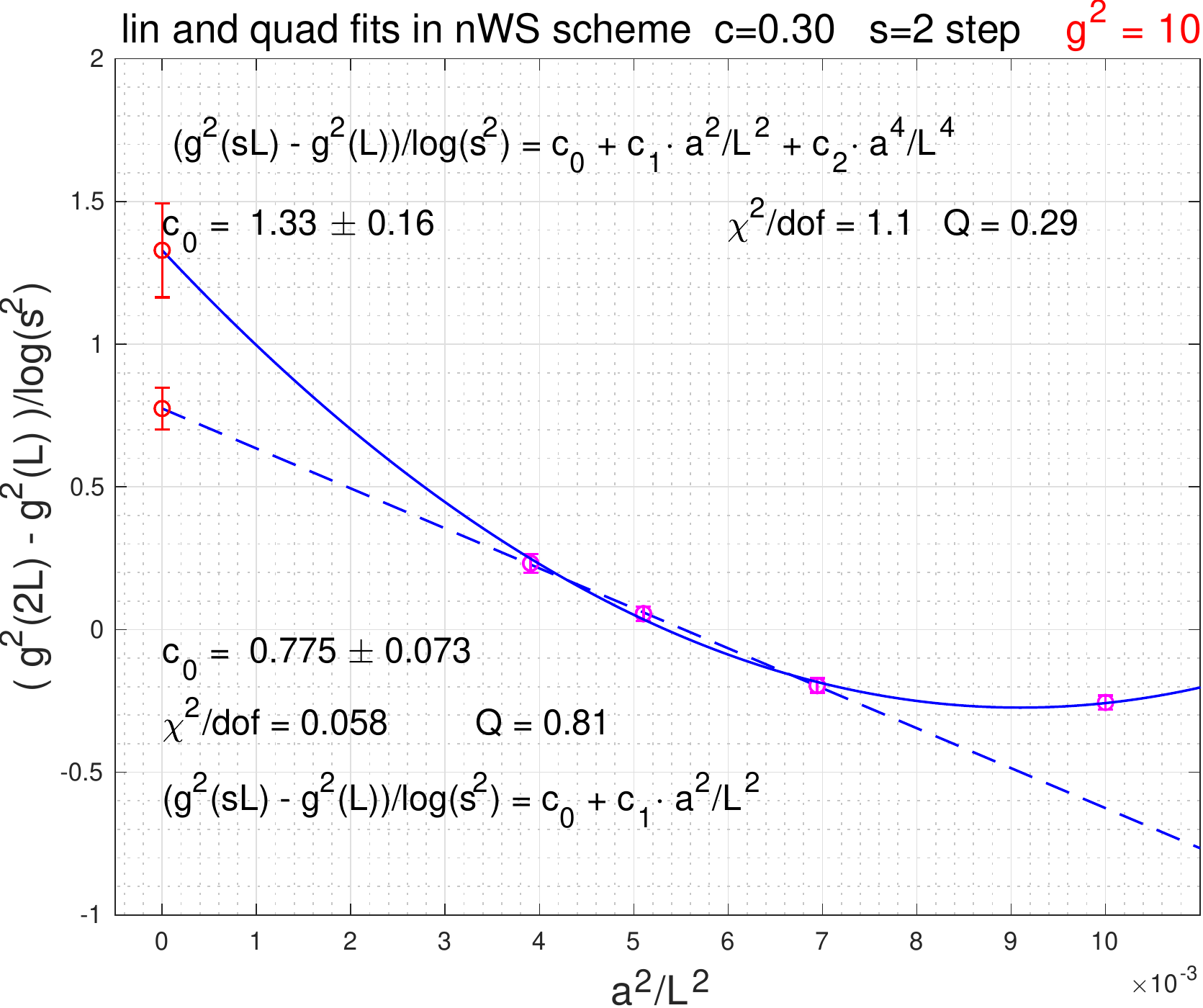}&
			\includegraphics[width=0.45\textwidth]{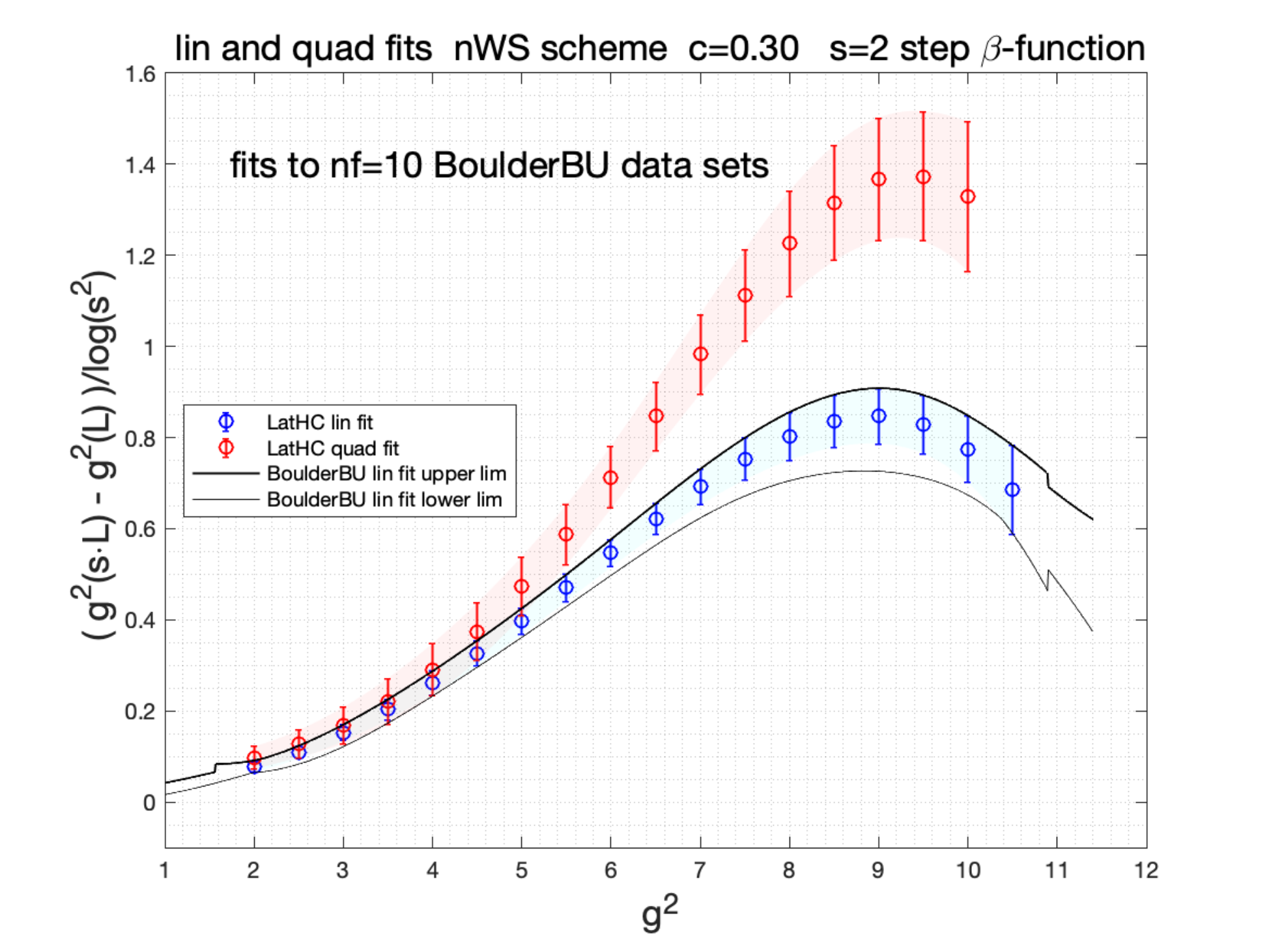}
		\end{tabular}
	\end{center}		
    \vskip -25pt
	\caption{\label{fig1h} {\small  Fits at $c=0.30$ with respective upper and lower panels explained in the caption of Fig.~\ref{c275a}.}}
\end{figure}

The trends we observe at $c=0.275$ continue  in $c=0.30$ based fits to the staggered fermion data and the DWF data as shown in Fig.~\ref{fig1h}. 
Our staggered fermion based analysis remains consistent and without any sign of an IRFP developing.
There are again statistically reasonable  linear and quadratic fits to the DWF data 
in the entire $g^2$ range but they are still incompatible with each other, separating around $g^2\approx 5$, similarly to the $c=0.275$ fits. 
The linear fits overlap with our staggered fermion fits below $g^2\approx 10$, with some separation setting in
beyond. 

In summary of all DWF fits analyzed for three choices of $c$,  no signs were found for any indicator of conformal behavior in the 
ten-flavor theory. 
This is a  significant disagreement from two independent lattice analyses of the same model and should warrant further discussion.

\newpage

\newpage
\subsection{The limits of lattice reach in the strong coupling regime}
\begin{figure}[h!]	
	\begin{center}
		\begin{tabular}{cc}
			\includegraphics[width=0.7\textwidth]{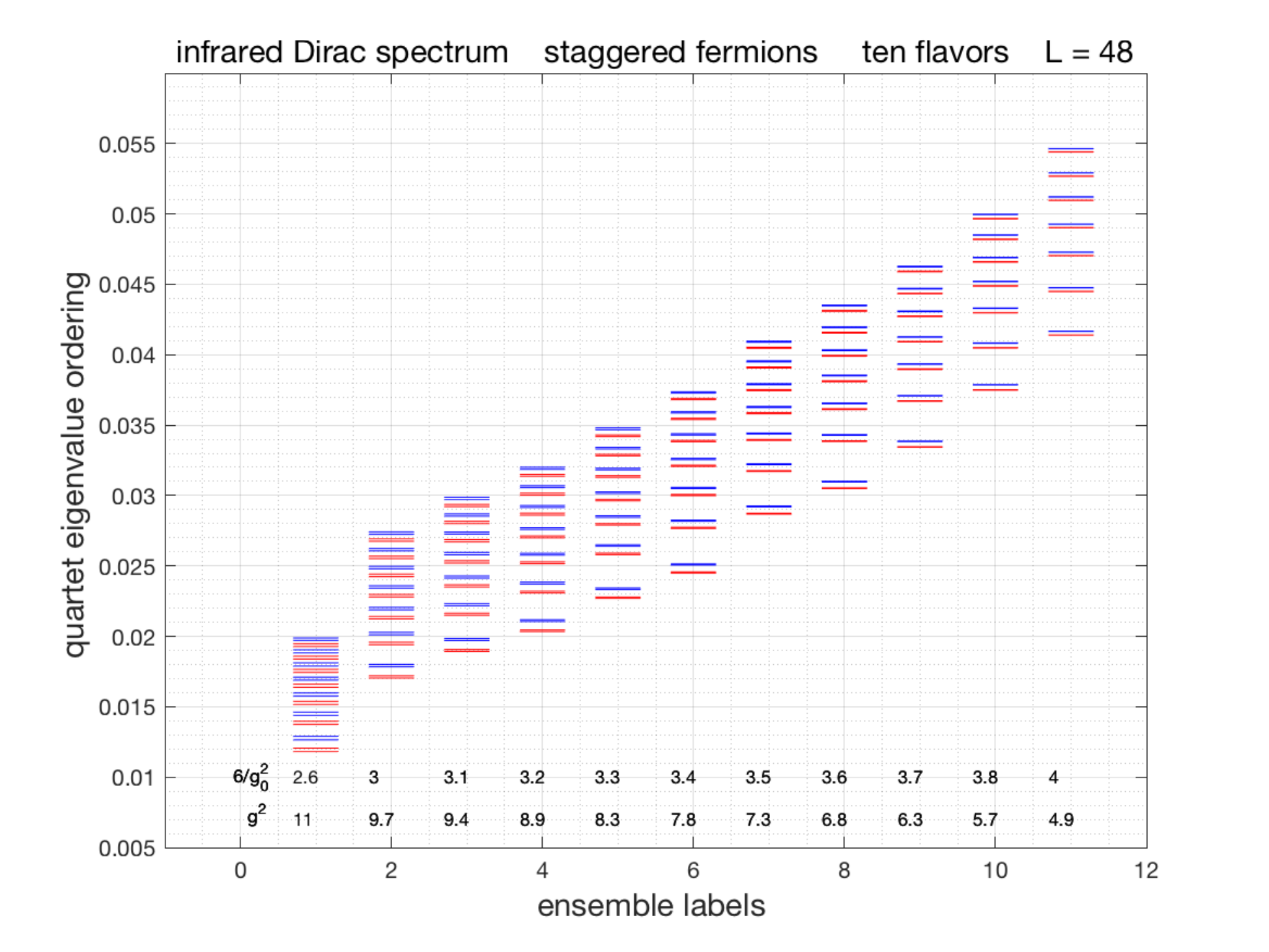}
		\end{tabular}
	\end{center}		
	\vskip -25pt
	\caption{\label{fig1e} {\small For illustration, the eigenvalue tower at the strongest bare coupling at $6/g_0^2 = 2.6$  of the extended work is shown in reaching toward the  renormalized $g^2\approx 11$ range. 
	 The visible change in quartet ordering is discussed in the text. 
 }}
\end{figure}

The nf = 10 $\beta$-functions of the staggered fermion formulation require the square root of the fermion determinant 
to obtain the correct flavor number in the continuum limit. We have previously provided a renormalization group based comprehensive theoretical argument
to validate this procedure~\cite{Fodor:2015zna}.  
Implicit in the proof is the restoration of quartet degeneracy in the Dirac spectrum in the continuum limit. 
Once the quartet degeneracy is established in the Dirac spectrum 
toward the $a\rightarrow 0$  continuum limit, the rooted theory is expected to become equivalent to the correct continuum fermion theory. 
Toward the continuum limit eigenvalue pairs within quartets would start forming narrow doublets with two doublets  split with  more spacing within the quartet.
Closer to the continuum limit  doublets start merging  into degenerate quartets.  
The emerging pattern is illustrated in Fig.~\ref{fig1e} with the infrared part of the Dirac spectrum shown at the largest $48^4$ volume of the $L=32,36,40,48$ sequence we used in the earlier analysis
with a set of bare gauge couplings in the $6/g_0^2 = 3 - 4$ range~\cite{Fodor:2018tdg}.
It was shown  that the infrared part of the Dirac determinant  approached in the $a \rightarrow 0$ limit the fourth power of the determinant built from degenerate quartets,
supporting the validation for  the use of rooted determinants of staggered fermions~\cite{Fodor:2018tdg}.  
The earlier analysis  was extended for this report to ensembles with bare gauge couplings 
added in the $6/g_0^2 = 2.6 - 2.9$ strong coupling range to reach renormalized couplings in the $g^2\approx 11$ range.
The added infrared eigenvalue tower at $6/g_0^2 = 2.6 $ in Fig.~\ref{fig1e} illustrates the challenge as we are pushing toward the limit of controlled lattice reach in the $g^2\approx 11$
range. 
The change in the spectrum at $6/g_0^2 = 2.6 $ with significantly less quartet ordering of doublets is a first indicator of increased cutoff effects with delayed quartet degeneracy. Added 
larger volumes, like $L=64$, would help the convergence in reaching the continuum limit from an $L$-sequence of lattices. 

More details will be provided about this analysis in the journal publication of this report.

\newpage 
\section{The infinite-volume based continuum $\beta$-function in multi-flavor QCD at nf=10}

\subsection{Brief history and implementation of the method}

We will discuss here new ten-flavor tests of the recently introduced lattice definition and
algorithmic implementation of the $\beta$-function defined
on the gradient flow of the gauge field over infinite Euclidean space-time in the continuum. 
 The infinite-volume $\beta$-function is based directly on the gradient flow coupling $g^2(t)$ defined in infinite volume as a function of  continuous flow time $t$. 
 This will allow the direct definition  $\beta(g^2(t))=t \cdot dg^2/dt$ from infinitesimal RG scale change using opposite sign convention from
 the conventional definition.
 The derivative will be approximated by  five-point discretization of the flow time for any observable $E(t)$,
 \begin{equation}
 	[-E(t + 2\epsilon) + 8E(t + \epsilon) - 8E(t - \epsilon) + E(t - 2 \epsilon)]/(12 \epsilon) = dE/dt + {\cal O}(\epsilon^4).
 \end{equation}
 with discrete step $\epsilon$ used in the integration of the gradient flow equations. 
We tested several implementations and applications of the method~\cite{Fodor:2017die,Fodor:2019ypi}.
Originally we introduced this new algorithm to match 
finite-volume step $\beta$-functions in massless near-conformal gauge theories 
with the infinite-volume $\beta$-function in the chiral limit of fermion mass deformations from the phase with spontaneous chiral symmetry breaking. 
This implementation of the algorithm was tested first in a study of the near-conformal two-flavor sextet model 
reaching the chiral limit from small fermion mass deformations $m$ in the chiral symmetry breaking phase~\cite{Fodor:2017die}. 

An alternative implementation of the infinite-volume $\beta$-function through $t \cdot dg^2/dt$, as applied here to the ten-flavor model,
is based on simulations directly at $m = 0$ and in the infinite-volume limit taken at fixed reference values of flow time $t/a^2$ in lattice units $a$.
This lattice algorithm was first tested in the two-flavor QCD model~\cite{Hasenfratz:2019hpg}  and in the multi-flavor nf=12 theory~\cite{Fodor:2019ypi}. 
We will show in the ten-flavor lattice implementation  how to make important contact at weak coupling 
with gradient flow based three-loop perturbation theory in infinite volume~\cite{Harlander:2016vzb,Artz:2019bpr}, serving
as a first pilot study toward the long-term goal of developing 
alternate approach to the determination  of the strong coupling $\alpha_s$ at the Z-boson pole in QCD.
It will be also shown that results from the lattice analysis of the  infinite-volume based ten-flavor $\beta$-function 
are consistent with the absence of  IRFP from our finite-volume ten-flavor step $\beta$-function in the range of renormalized couplings discussed in Section 2. 

\subsection{The ten-flavor lattice analysis}

\noindent{The algorithmic implementation of the  lattice analysis has three steps. }

\begin{figure}[h!]	
	\begin{center}
		\begin{tabular}{cc}
			\includegraphics[width=0.45\textwidth]{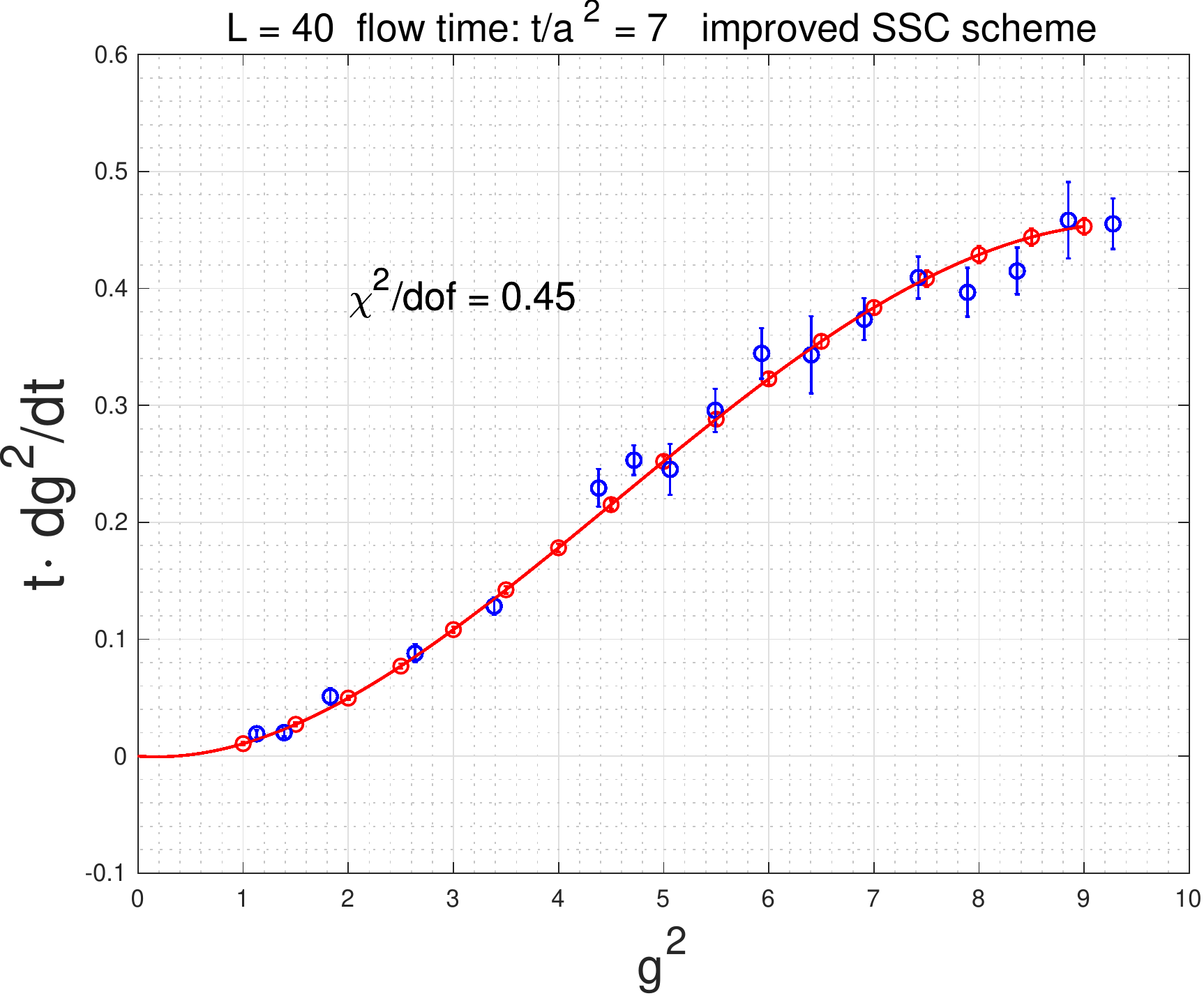}&
			\includegraphics[width=0.45\textwidth]{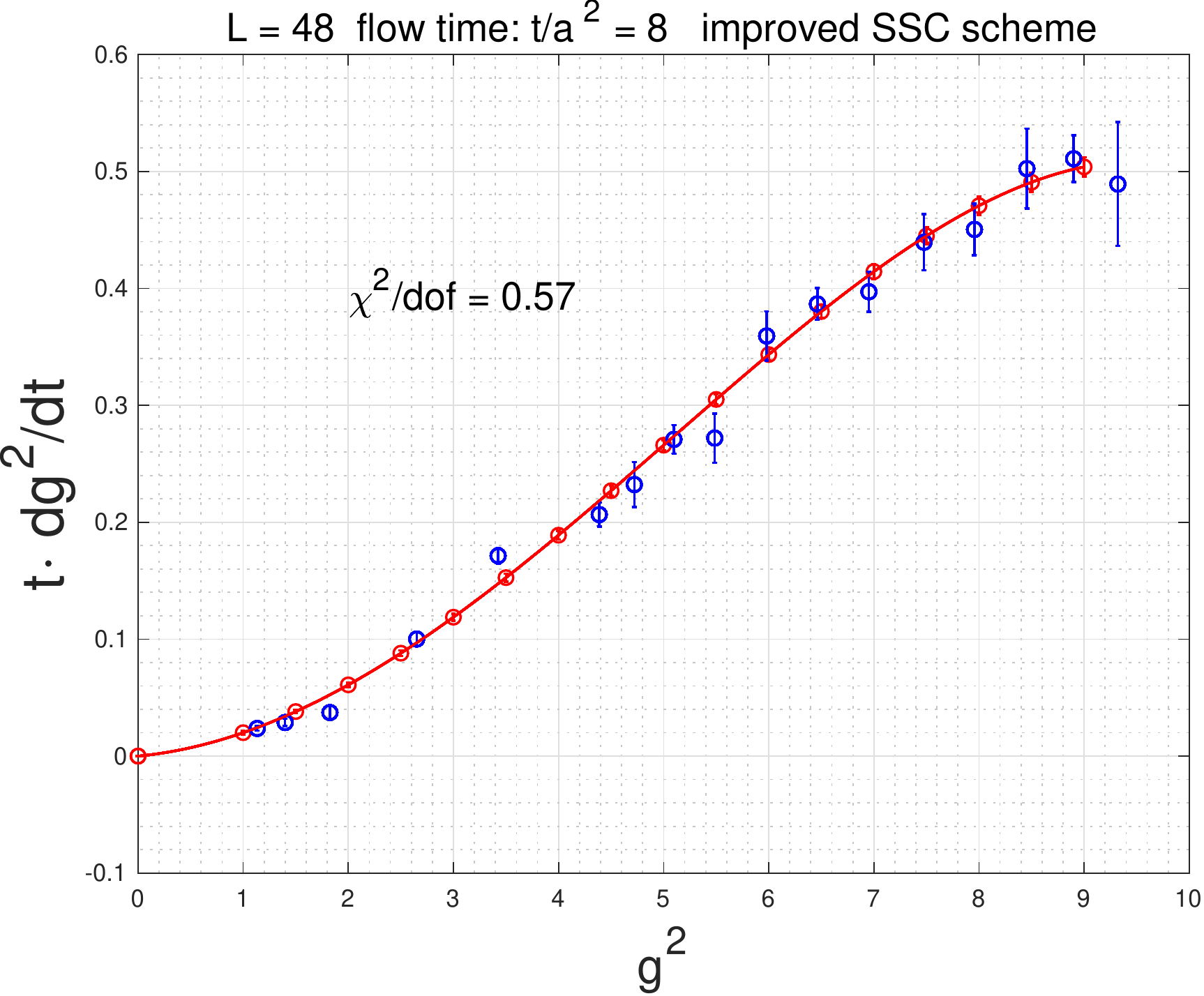}
		\end{tabular}
	\end{center}		
    \vskip -25pt
	\caption{\label{ContBeta1}  {\small Fourth order polynomial fits of $t\cdot g^2/dt$ are shown in the $SSC$ scheme  for $L=40$ and $L=48$
			with good statistical quality. On the left panel the fit is shown to the fitted blue points $t\cdot g^2/dt$ for $L=40$ at locations of $g^2$ determined at bare 
		    gauge couplings at flow time $t/a^2 = 7$.  The interpolated values of $t\cdot g^2/dt$  are marked with red symbols for  targeted $g^2$ values.  On the right
	       panel similar fit and interpolation are shown for $L=48$ and $t/a^2 = 8$.     }}
	
\end{figure}

\noindent{\bf Step 1:} the four  largest volumes $L^4 = 32^4, 36^4, 40^4, 48^4 $  are selected  for the analysis from the full set, discussed in Section 2,
at 21  bare gauge couplings $6/g_0^2$ for each $L$ with periodic gauge and antiperiodic fermion boundary conditions.
For the control of the continuum limit at zero lattice spacing, 12 preset values of the flow time $t$ were preselected for the agorithm at each $L$ and each $6/g_0^2$ 
starting at $t/a^2 = 2.5$ and extended to $t/a^2 = 8$ in increments of $0.5$ steps. We also preselect targeted $g^2$ values for the determination of the 
infinite volume based $\beta$-function in the limit of zero lattice spacing. We analyze here a choice of the preset range $g^2 = 1-10.5$ in increments of $0.5$.
The preset $t/a^2$-sequence and $g^2$-sequence seem to be arbitrary but they are designed to sample the range of $g^2(t)$ in the continuum theory 
where the lattice spacing can be removed safely, based on fits to data from the selected lattice ensembles. Every preselected $g^2$ implicitly defines 
the flow time in the continuum at scale $\mu=1/\sqrt{8t}$. At each $L$ we calculate  $g^2$ and its discretized 
$t\cdot dg^2/dt$ derivative on the gradient flow  at 21 different $6/g_0^2$ couplings and at the 12 preselected  $t/a^2$-values. 
At each $L$ we fit $t\cdot dg^2/dt$  as a function of $g^2$ to a fourth order polynomial at 21 points in $g^2$ and interpolate $t\cdot dg^2/dt$  from the polynomial fits
to the preselected $g^2$ values at each preset $t/a^2$ for fixed $L$. Samples of this fitting procedure are shown in Fig.~\ref{ContBeta1}.

\begin{figure}[h!]	
	\begin{center}
		\begin{tabular}{cc}
			\includegraphics[width=0.45\textwidth]{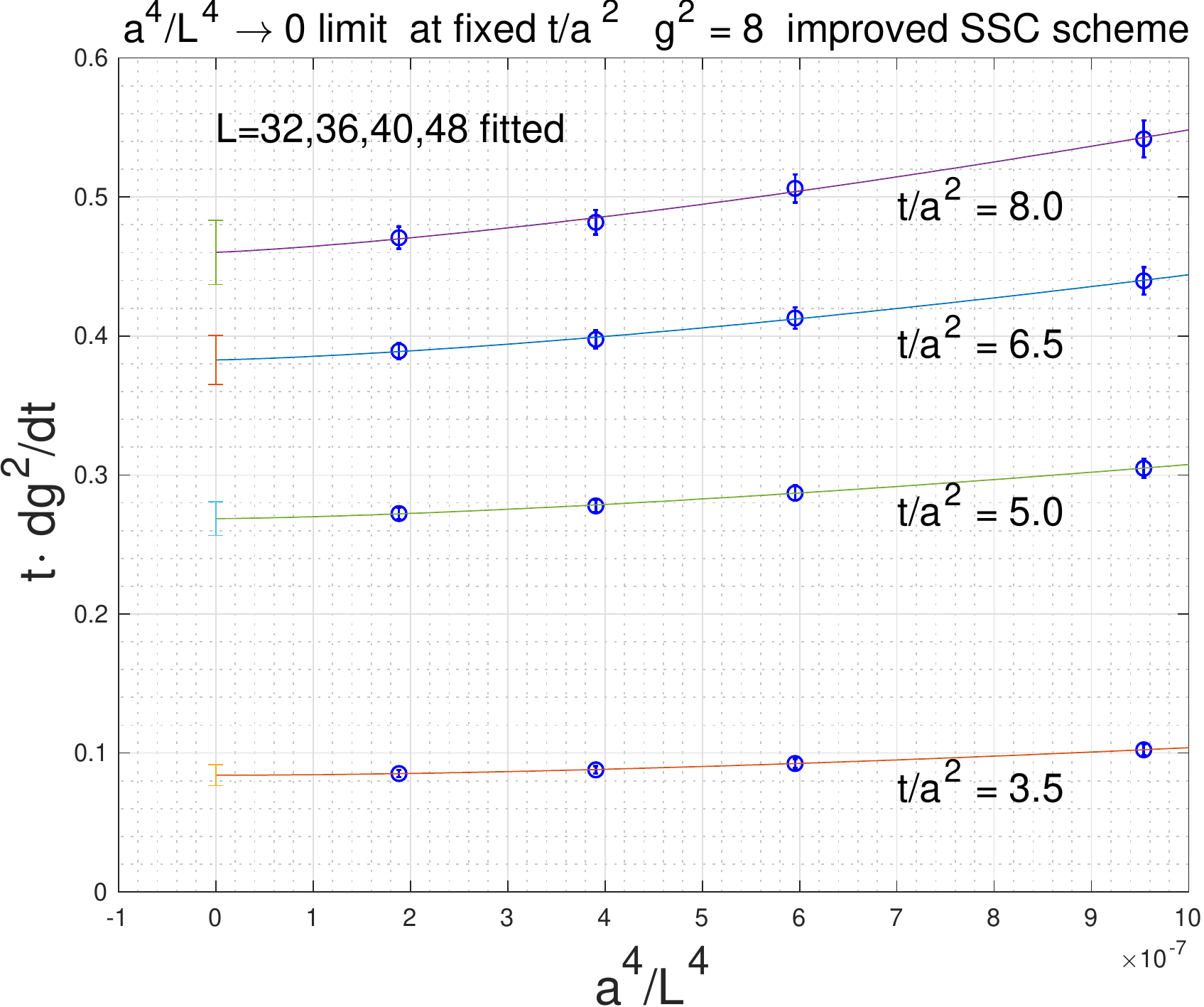}&
			\includegraphics[width=0.45\textwidth]{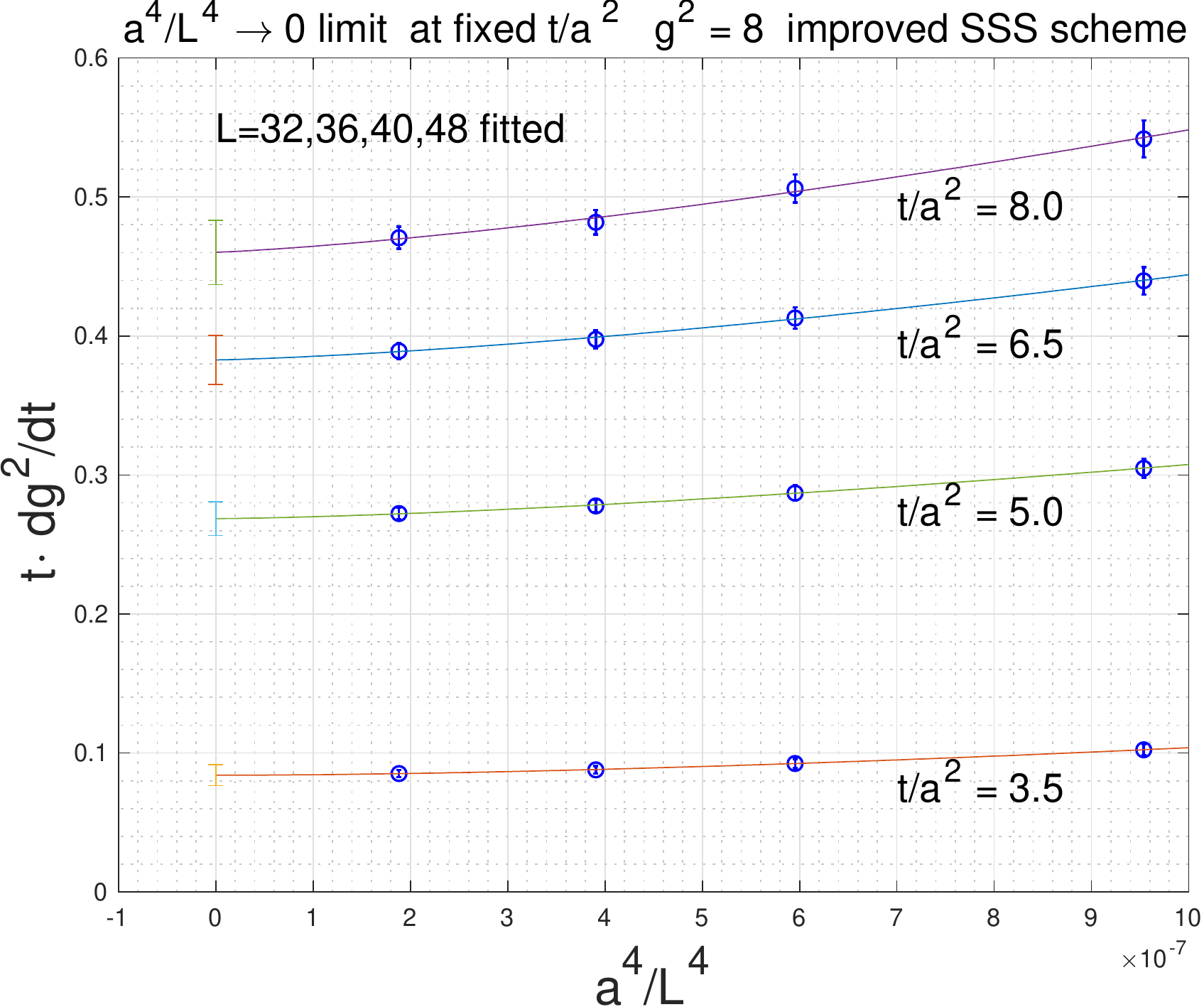}
		\end{tabular}
	\end{center}		
	\vskip -25pt
	\caption{\label{ContBeta2}  {\small  The interpolated $L$-dependent beta-functions of $\beta=t\cdot dg^2/dt$ are fitted  to the $a^2/L^2 \rightarrow 0$  infinite volume limit 
	at fixed lattice spacing set by the four values of  $t/a^2$ from the $2.5-8$ range at $g^2=8$.  The fits with very good statistics are polynomials in the $a^2/L^2$ 
	variable with a leading $a^4/L^4$ term and the higher order $a^6/L^6$ term with $L/a=32,36,40,48$ in the SSC scheme on the left and the SSS scheme on the right. }}
	
\end{figure}

\noindent{\bf Step 2:}  The interpolated $L$-dependent $\beta=t\cdot dg^2/dt$ beta-functions are extrapolated to the infinite volume limit for each $t/a^2$ and each $g^2$ from Step 1, 
with samples of these fits shown in Fig~\ref{ContBeta2}.

\begin{figure}[h!]	
	\begin{center}
		\begin{tabular}{cc}
			\includegraphics[width=0.45\textwidth]{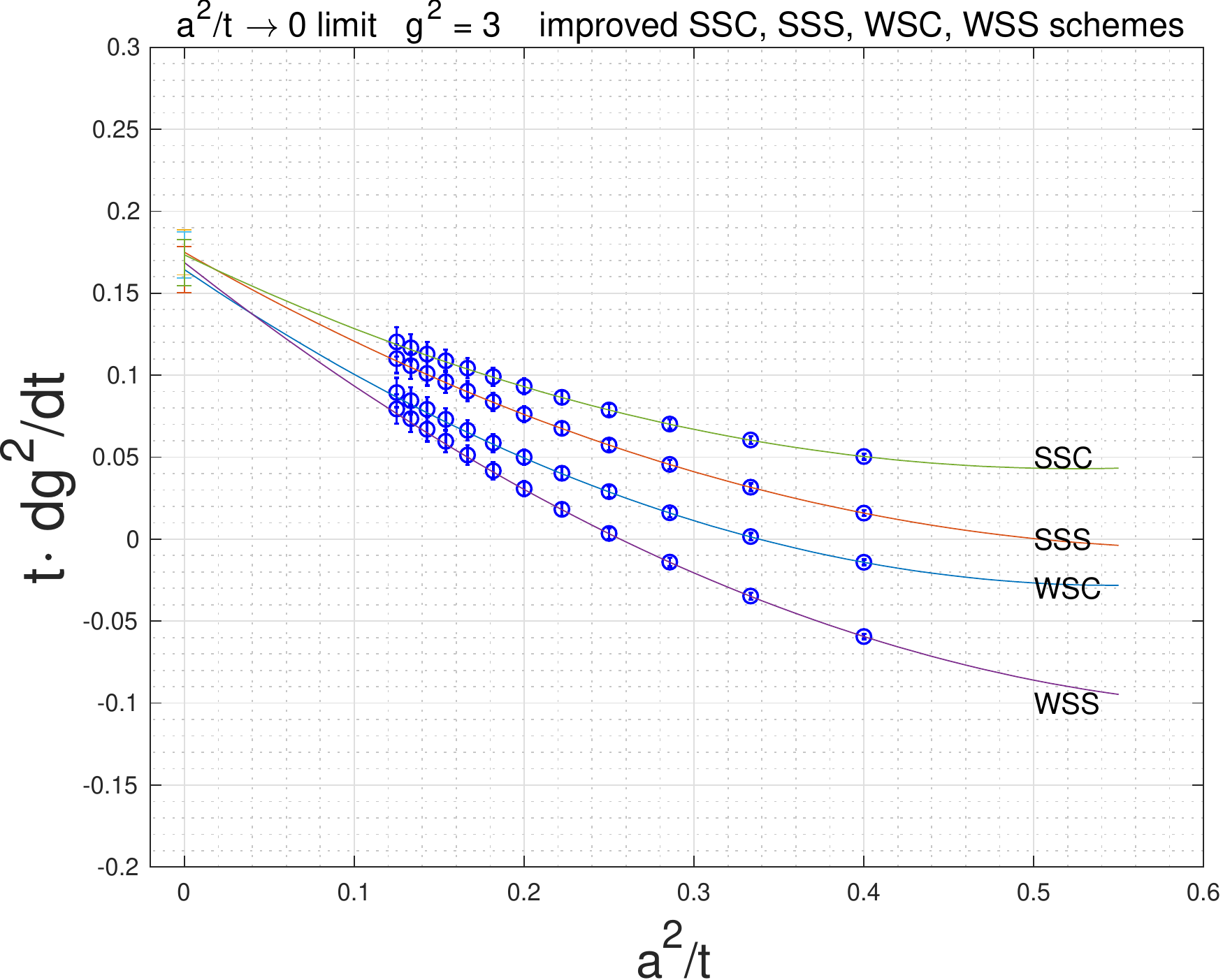}&
			\includegraphics[width=0.45\textwidth]{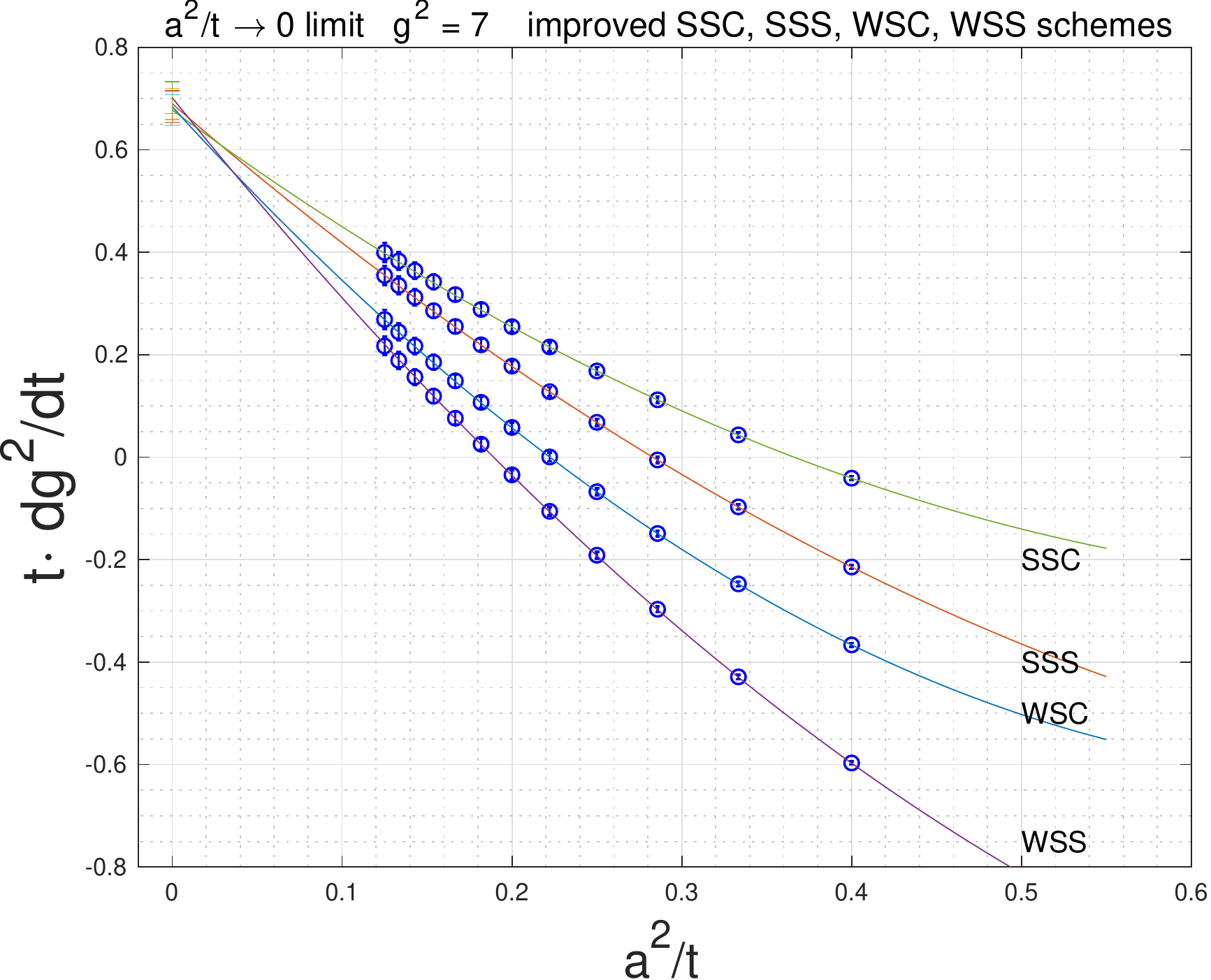}
		\end{tabular}
	\end{center}		
	\vskip -25pt
	\caption{\label{ContBeta3}  {\small Continuum fits of $\beta=t\cdot dg^2/dt$ are shown in all four schemes at two renormalized couplings with choices of $g^2=3$
			and $g^2=7$. The fits are quadratic in the $a^2/t$ variable.  }}
	
\end{figure}

\noindent{\bf Step 3:}  After Step2 we have the infinite volume based beta-functions of $\beta=t\cdot dg^2/dt$ at 12 values of $t/a^2$ for each targeted value of the 
gradient flow based renormalized coupling $g^2$ held fixed in the algorithm. In Step 3, we fit the cutoff dependence of   $\beta=t\cdot dg^2/dt$  to determine its
$a\rightarrow 0$ continuum limit at fixed $g^2$. This is obtained by quadratic fits in the $a^2/t$ variable with fit samples shown in Fig.~\ref{ContBeta3}. 
If the correlated fits of the three steps are statistically consistent and acceptable, we reached the goal in the covered and sampled  $g^2 = 1.5-10.5$ range 
which can be extended by added analysis. 

\subsection{Ten-flavor tests toward  the goal of the strong coupling $\alpha_s$ at the Z-pole in QCD}

\begin{figure}[h!]	
	\begin{center}
		\begin{tabular}{cc}
			\includegraphics[width=0.45\textwidth]{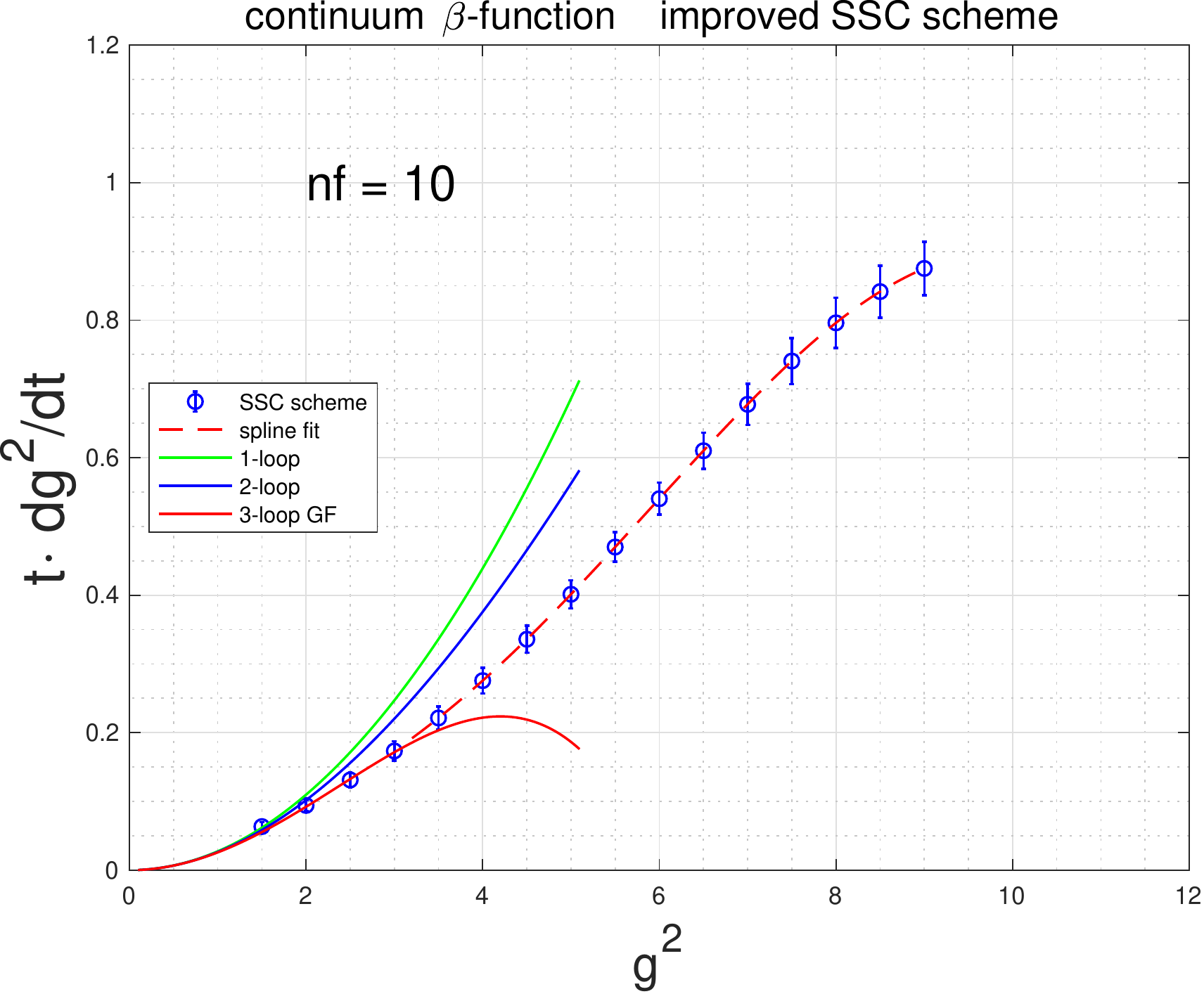}&
			\includegraphics[width=0.45\textwidth]{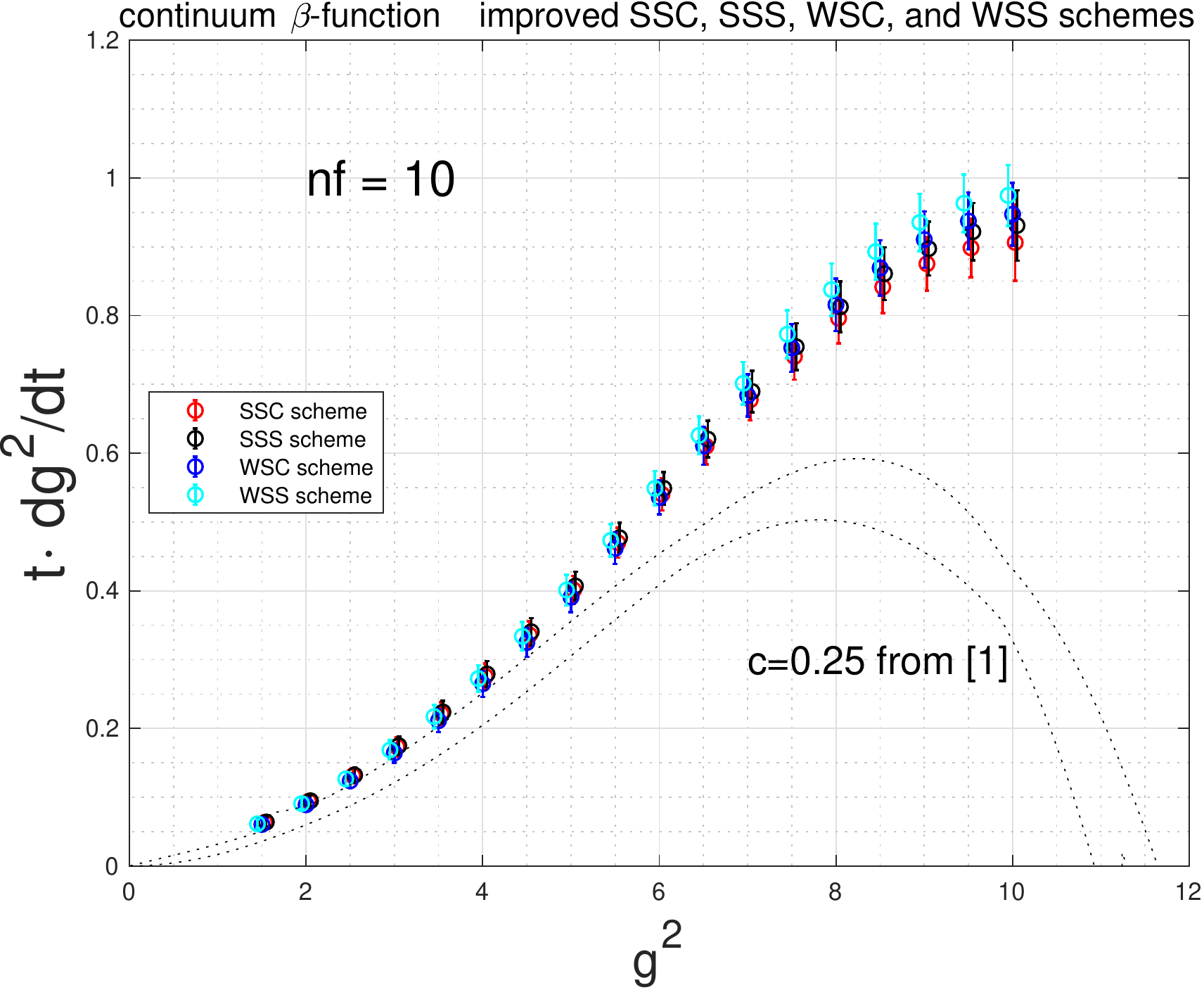}
		\end{tabular}
	\end{center}		
	\vskip -25pt
	\caption{\label{ContBeta4}  {\small The dashed line in the left panel shows spline based fitting in the  $g^2=3-9$  range of the SSC scheme. The fit would be identical 
			in the SSS scheme, or in the WSC and WSS  schemes,  added  in the right panel. 
			Below $g^2=3$ three-loop perturbation theory can be used for the $\beta$-function from~\cite{Harlander:2016vzb,Artz:2019bpr}. }}
\end{figure}

After the three-step analysis of the ten-flavor model we reached the set goal of $\beta(g^2(t)) = t\cdot dg^2/dt$ in the $g^2(t)  = 1.5-10.5$ range, as shown in Fig.~\ref{ContBeta4}.
The value of $g^2(t)$ implicitly defines the scale $\mu=1/\sqrt{8t}$ which ultimately would be connected to a nonperturbative parameter, like the decay constant $f=F_\pi$ 
of the Goldstone pion, if the theory turns out to be near-conformal with spontaneous chiral symmetry breaking.  
Similar to~\cite{DallaBrida:2018rfy}  in three-flavor QCD with massless fermions, we can connect now the $\Lambda_{GF}$ scale of the gradient flow scheme to other scales,
like the scale $\mu=1/\sqrt{8t}$ set by  the choice $g^2(t)$ in the ten-flavor theory with massless fermions. 
As an example, we will express the scale $L_0 = \sqrt{8t}$ set at ${\bar g}^2 = 5$ in $\Lambda_{GF}$ units of the ten-flavor theory,
\begin{equation}
L_0\cdot \Lambda_{GF} = (b_0{\bar g}^2)^{-b_1/2b_0^2}\cdot {\rm exp}(-1/2b_0{\bar g}^2)\cdot {\rm  exp}\Bigl (-\int_0^{\bar g} dx\bigl [1/\beta(x) + 1/b_0x^3 - b_1/b_0^2x\bigr] \Bigr ).
\end{equation}
The integral in Eq.(3.2) was broken up into two parts. In the $x=0-3$ range the three-loop value of the $\beta$-function was used and the $x=3-5$ range was evaluated with numerical 
\begin{figure}[h!]	
	\begin{center}
		\begin{tabular}{c}
			\includegraphics[width=0.60\textwidth]{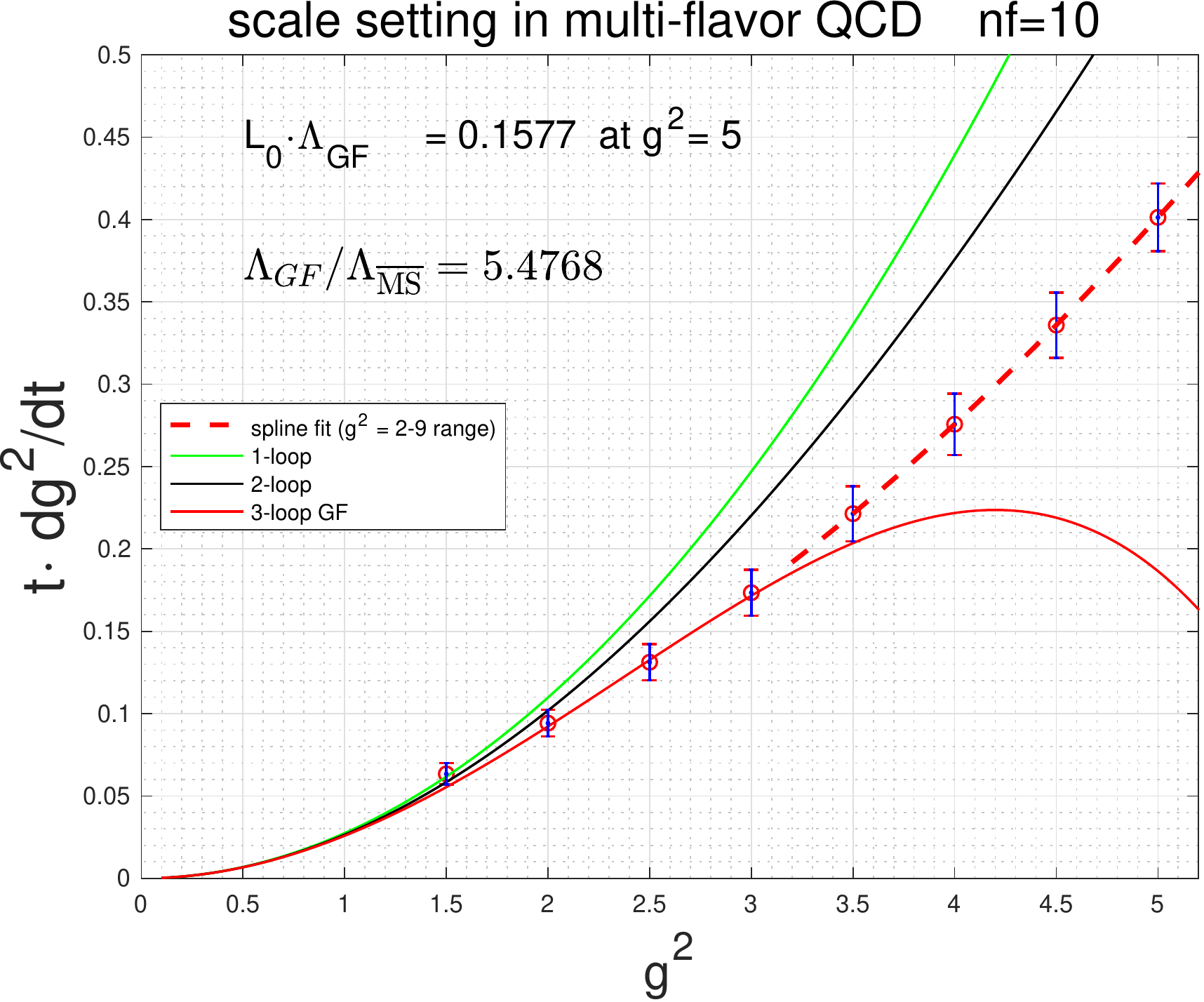}
		\end{tabular}
	\end{center}		
	\vskip -25pt
	\caption{\label{ContBeta5}  {\small The dashed line in the figure shows spline based fitting in the  $g^2=3-9$  range of the SSC scheme. 
			Below $g^2=3$ three-loop perturbation theory can be used for the $\beta$-function from~\cite{Harlander:2016vzb,Artz:2019bpr}. }}
\end{figure}	
integration, based on the spline fit made to the data. The result of $L_0\cdot\Lambda_{GF} = 0.1577$ can be converted to $\Lambda_{\overline{\rm MS}}$  units of the ten-flavor theory,
$L_0\cdot  \Lambda_{\overline{\rm MS}} = 0.02879$, using the conversion 
factor $\Lambda_{GF}/\Lambda_{\overline{\rm MS}}=5.4768$ from a well-know one-loop calculation~\cite{Harlander:2016vzb,Artz:2019bpr}.
The precision of the calculation is on the few percent level with combined systematic and statistical uncertainties, not far from the goal of performing similar analysis in 
three-flavor QCD with massless fermions.
\vskip 0.1in
\noindent{\bf\large Conclusions:}
We have presented strong evidence from our finite volume based step $\beta$-function analysis that massless multi-flavor QCD  with ten flavors shows no IRFP, or any hint for it within controlled lattice reach
with added support from the infinite volume based $\beta$-function.
Scale setting of the gradient flow time at selected $g^2(t)$ values, defined over unlimited Euclidean space-time,
was successfully demonstrated in $\Lambda_{\overline{\rm MS}}$ units of the theory using the infinite volume based ten-flavor $\beta$-function. 
This holds promise for alternate determination of the strong coupling $\alpha_s$ in QCD.

\section*{Acknowledgments}
\vskip -0.1in
We acknowledge support by the DOE under grant DE-SC0009919, by the NSF under grant 1620845,  and by the Deutsche Forschungsgemeinschaft grant SFB-TR 55. 
Computations for this work were carried out in part on facilities of the USQCD Collaboration, which are funded by the Office of Science of the U.S. Department of Energy.
Computational resources were also provided by the DOE INCITE program  on the SUMMIT gpu platform at ORNL, by the University of Wuppertal, 
and by the Juelich Supercomputing Center.

\newpage
\section*{Appendix}
\appendix
\renewcommand\thefigure{\thesection\arabic{figure}} 
\counterwithin{figure}{section}
\numberwithin{equation}{section}
\section{Polynomial interpolations in the  c=0.25 and c=0.30 step $\beta$-function schemes}
\begin{figure}[h!]	
	\begin{center}
    	\begin{tabular}{ccc}
			\includegraphics[width=0.33\textwidth]{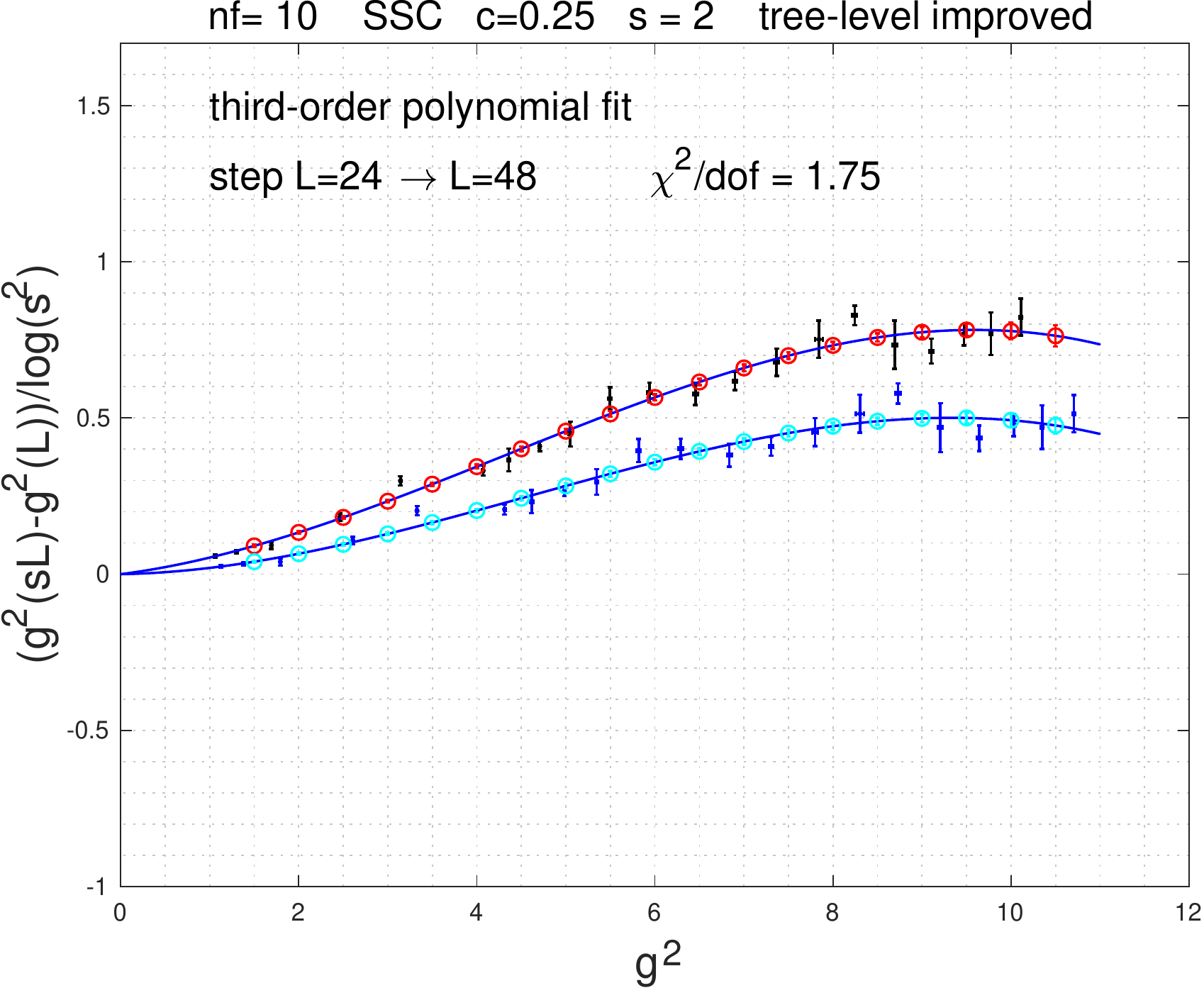}&
			\includegraphics[width=0.33\textwidth]{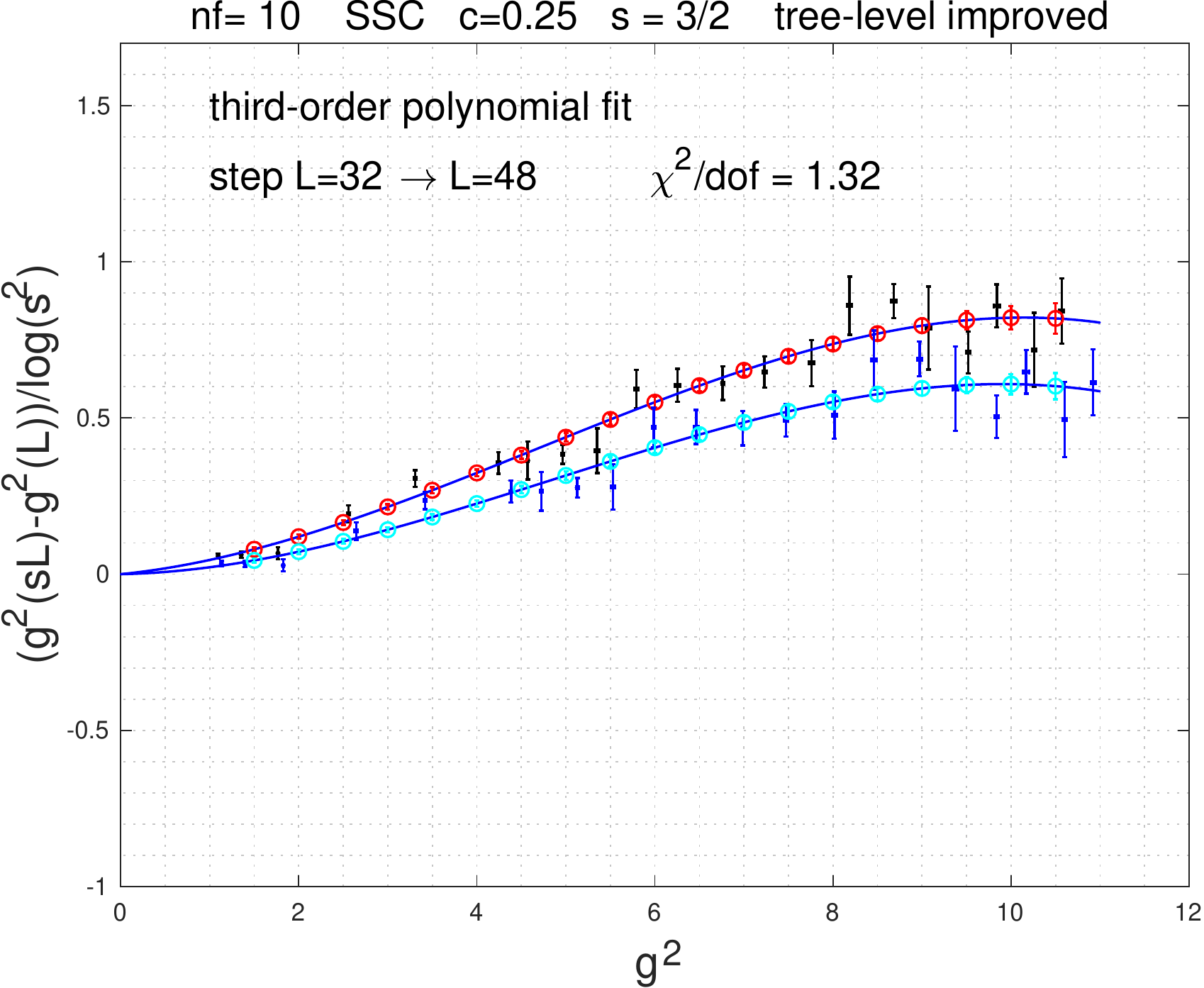}&
			\includegraphics[width=0.33\textwidth]{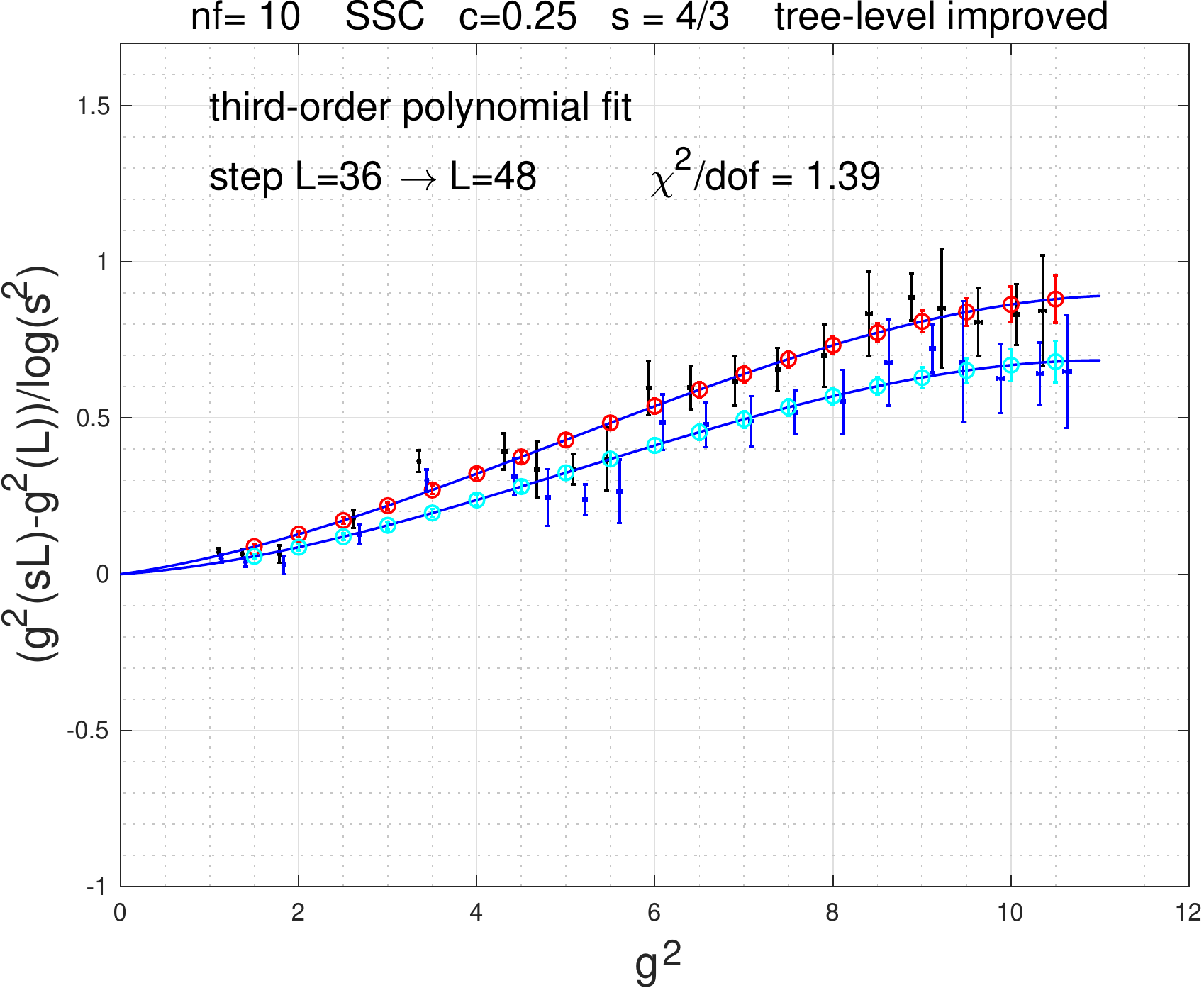}\\
			\includegraphics[width=0.33\textwidth]{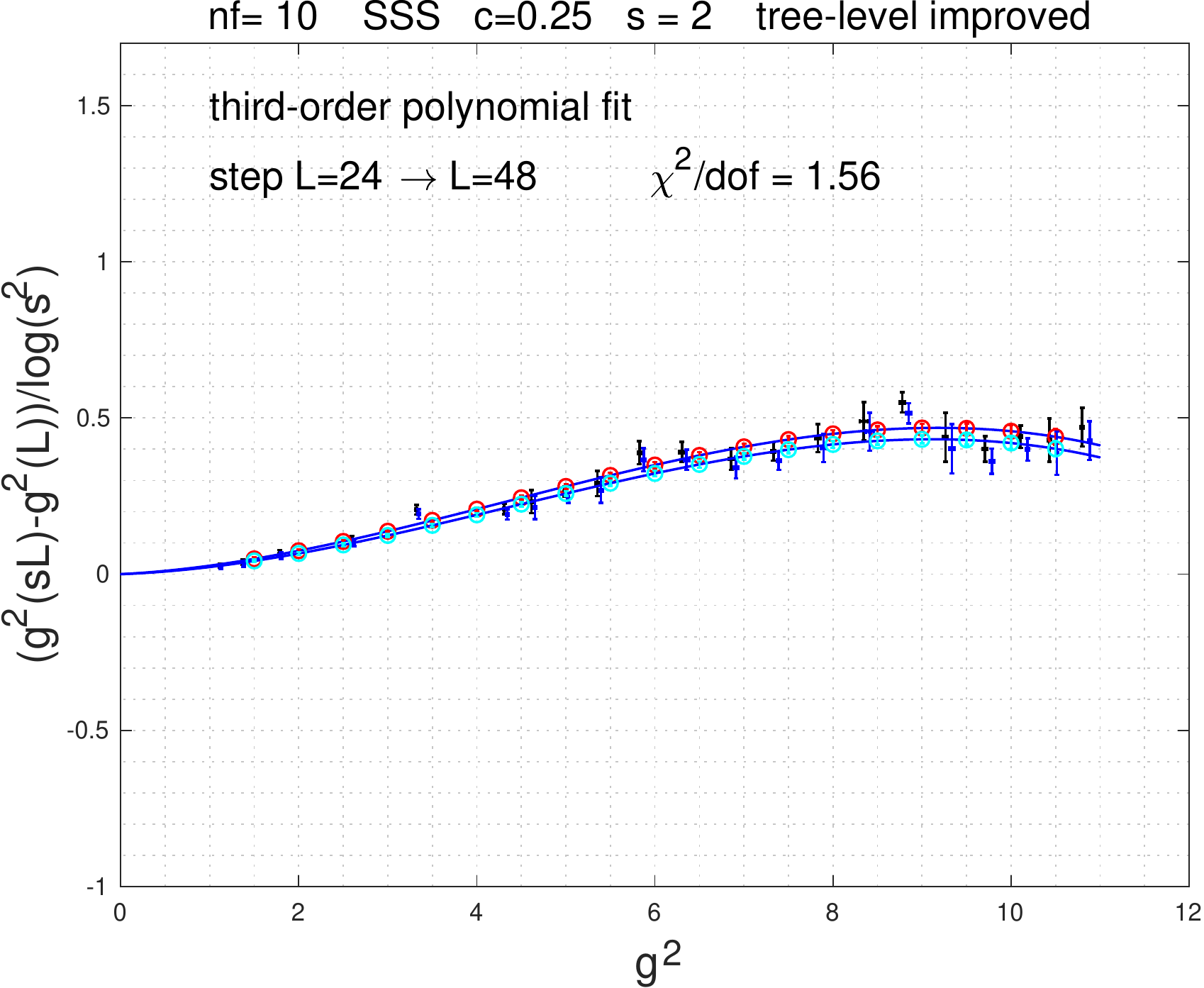}&
			\includegraphics[width=0.33\textwidth]{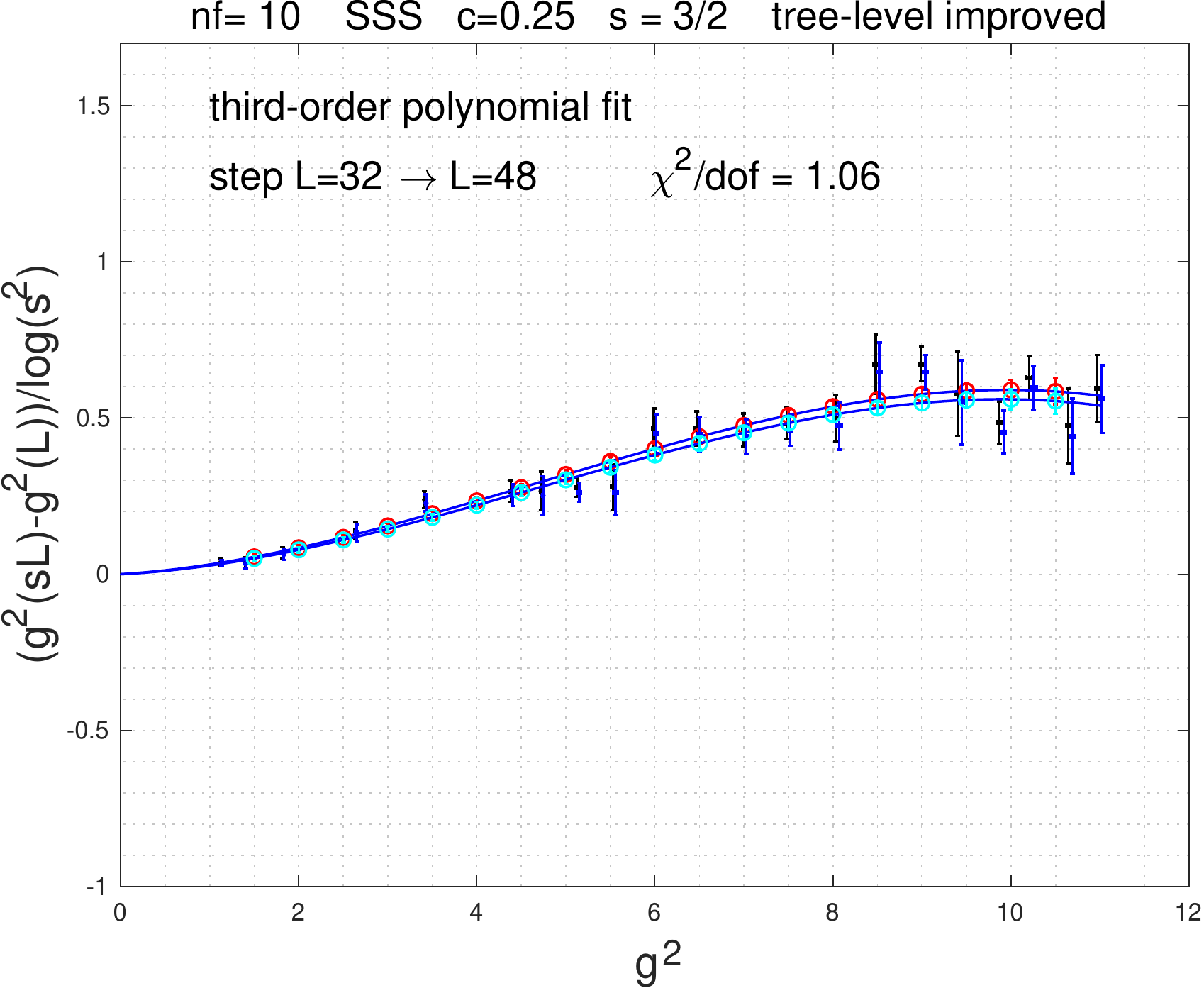}&
			\includegraphics[width=0.33\textwidth]{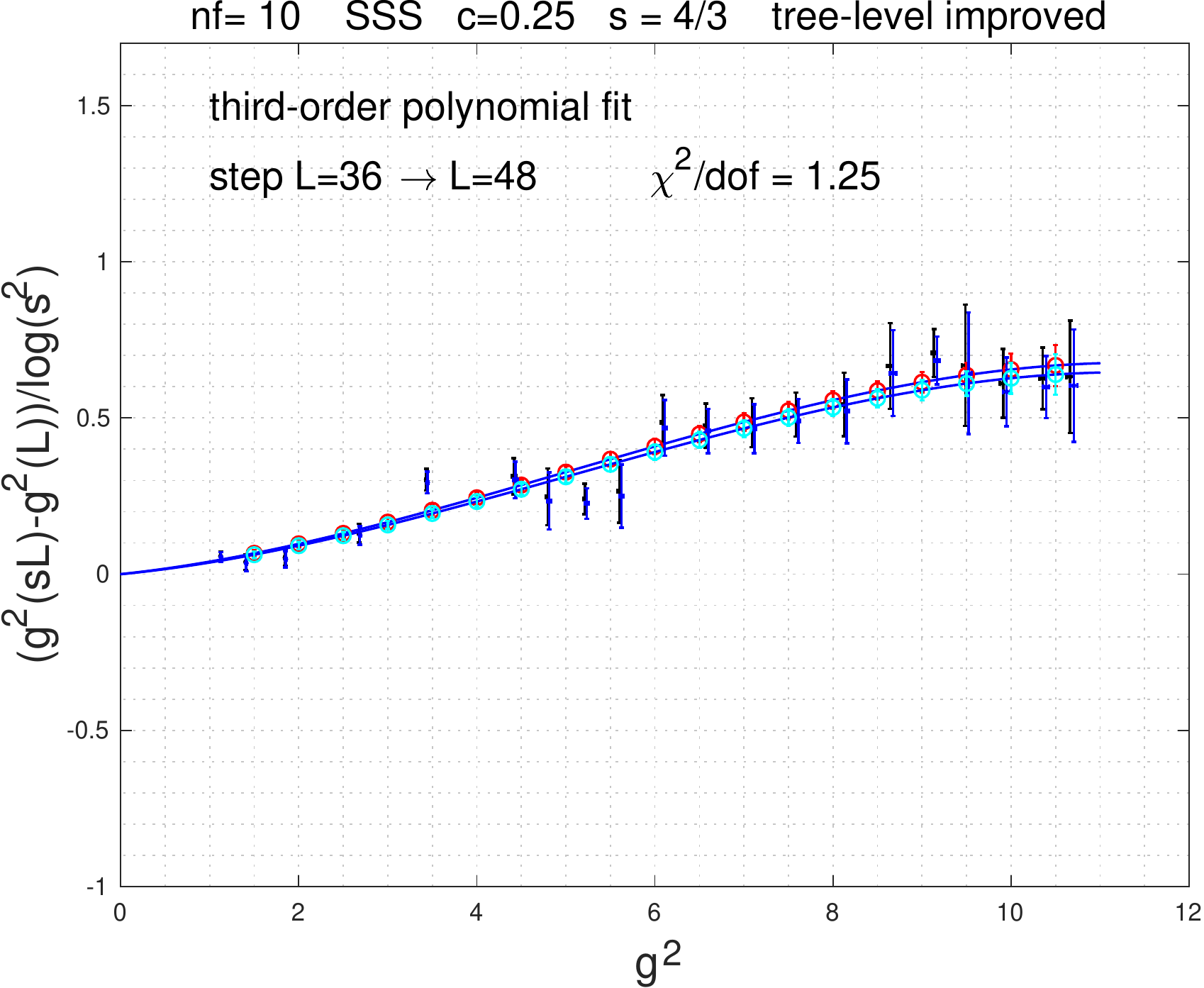}\\			
			\includegraphics[width=0.33\textwidth]{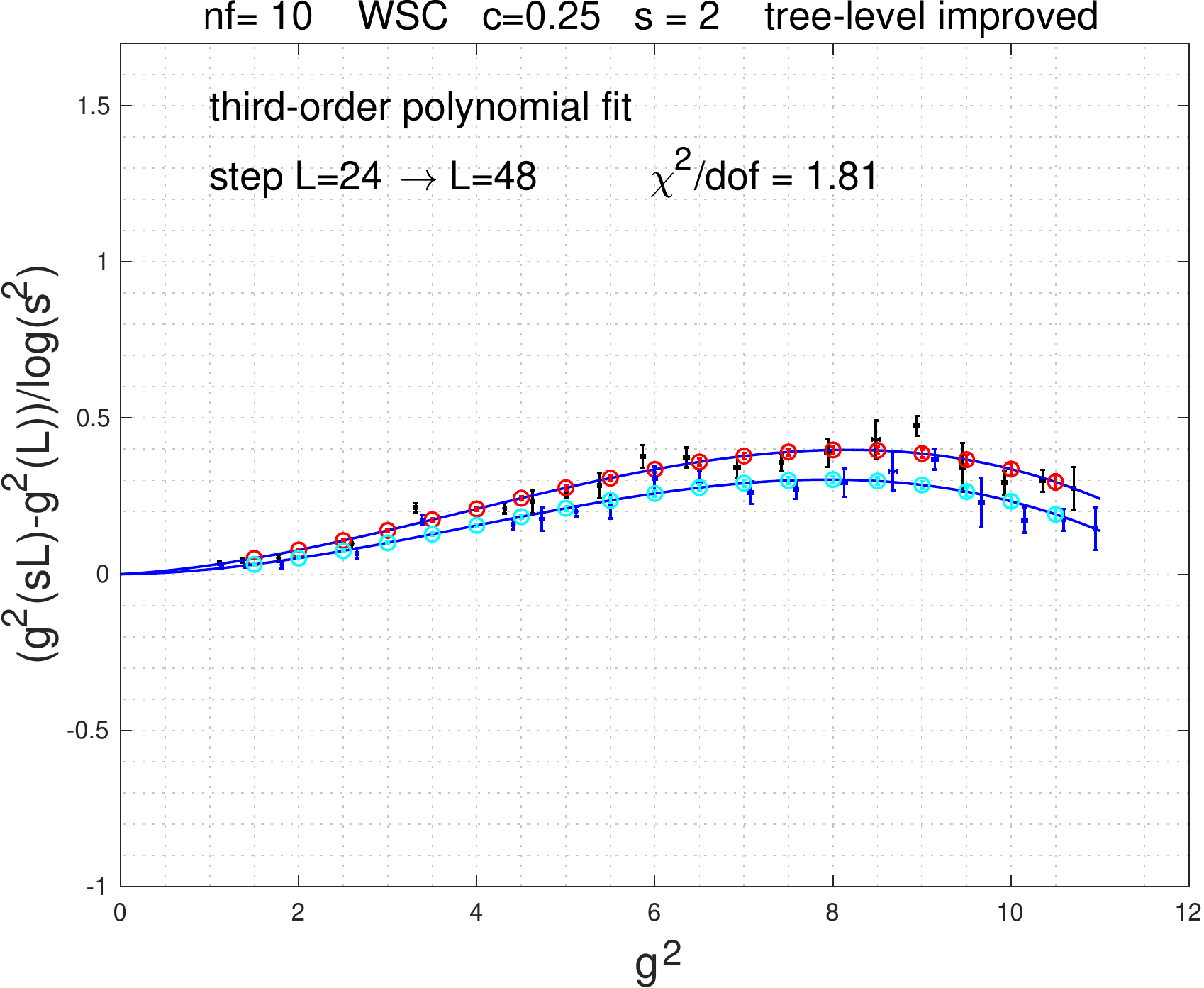}&
			\includegraphics[width=0.33\textwidth]{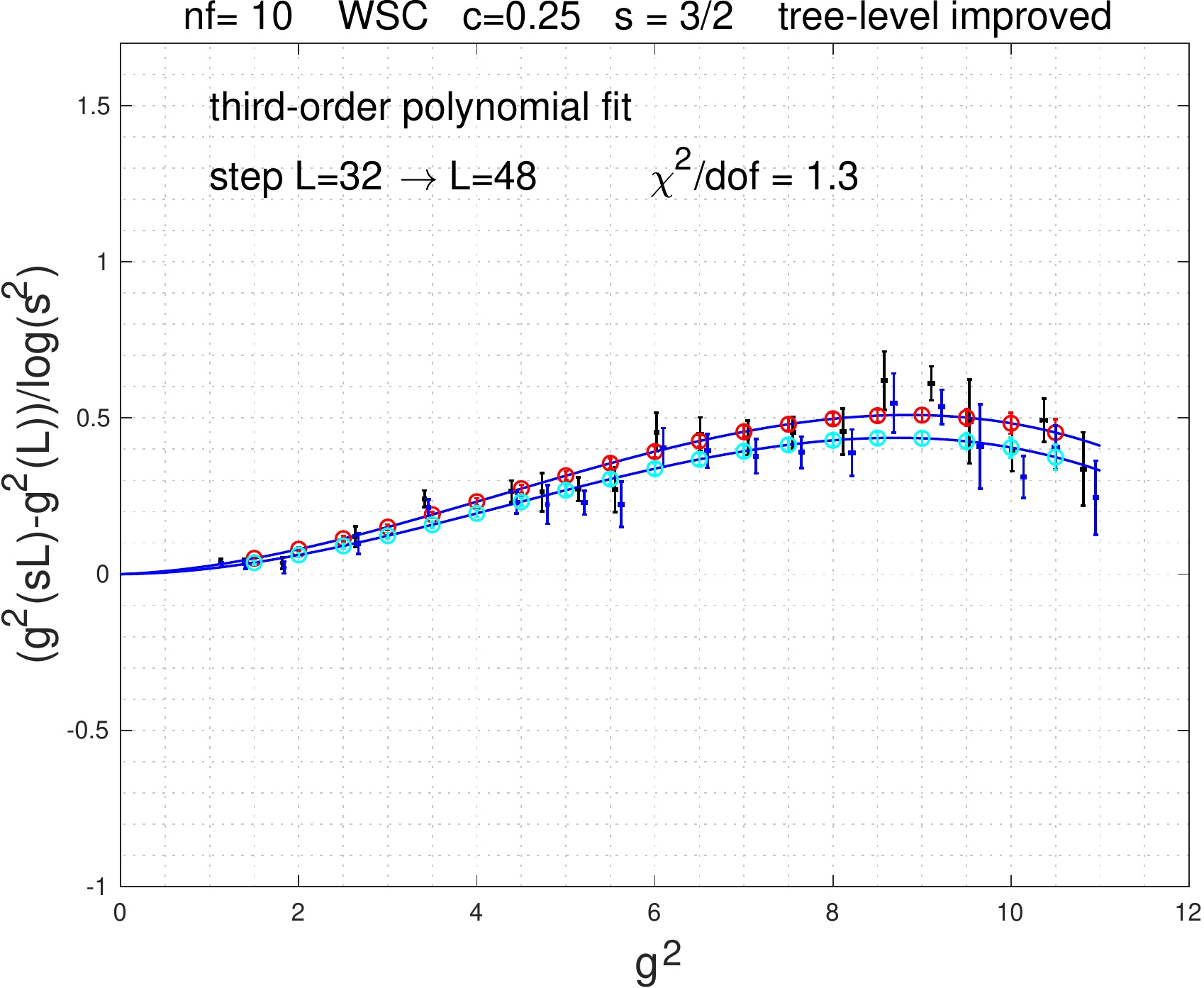}&
			\includegraphics[width=0.33\textwidth]{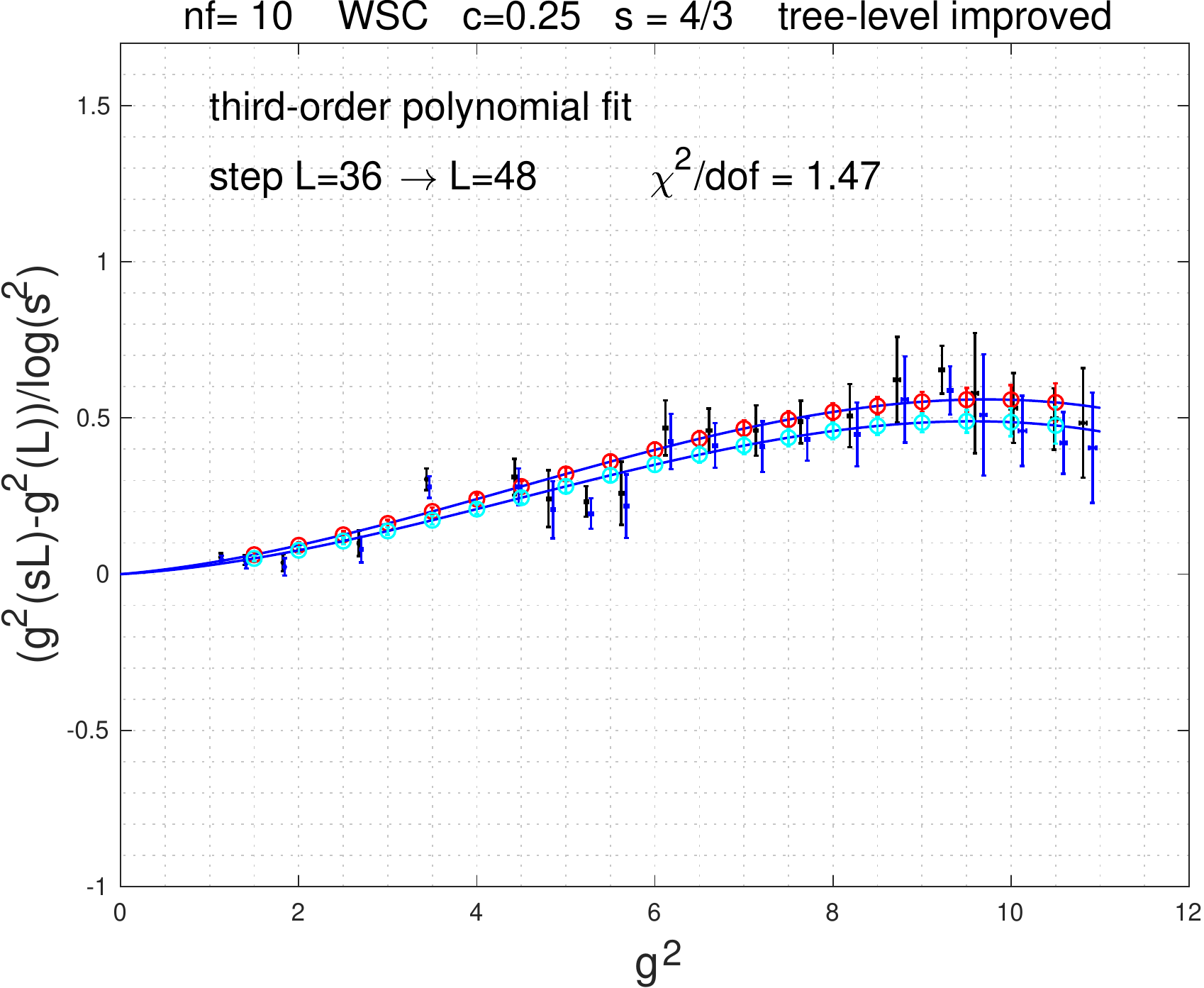}\\
			\includegraphics[width=0.33\textwidth]{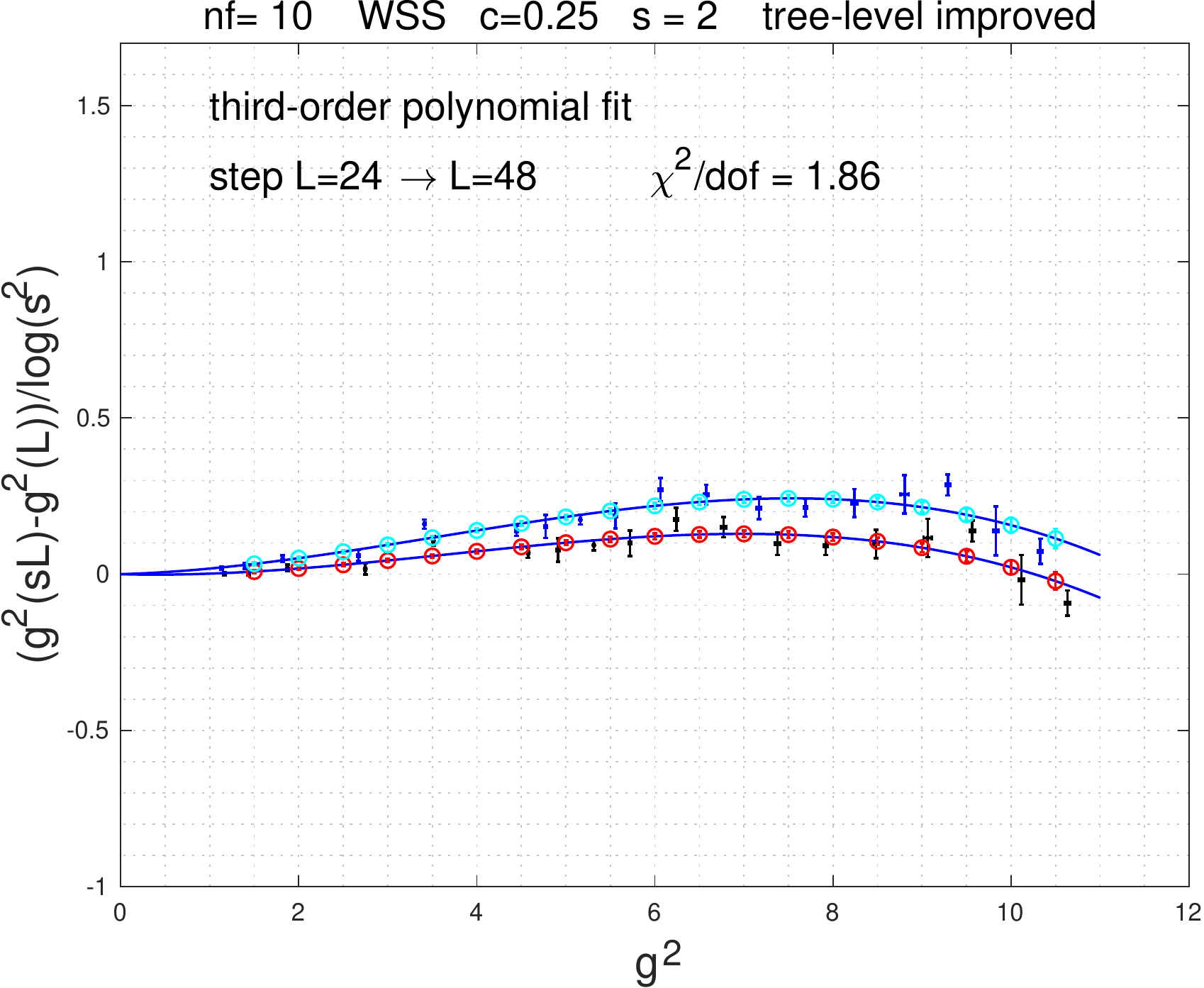}&
			\includegraphics[width=0.33\textwidth]{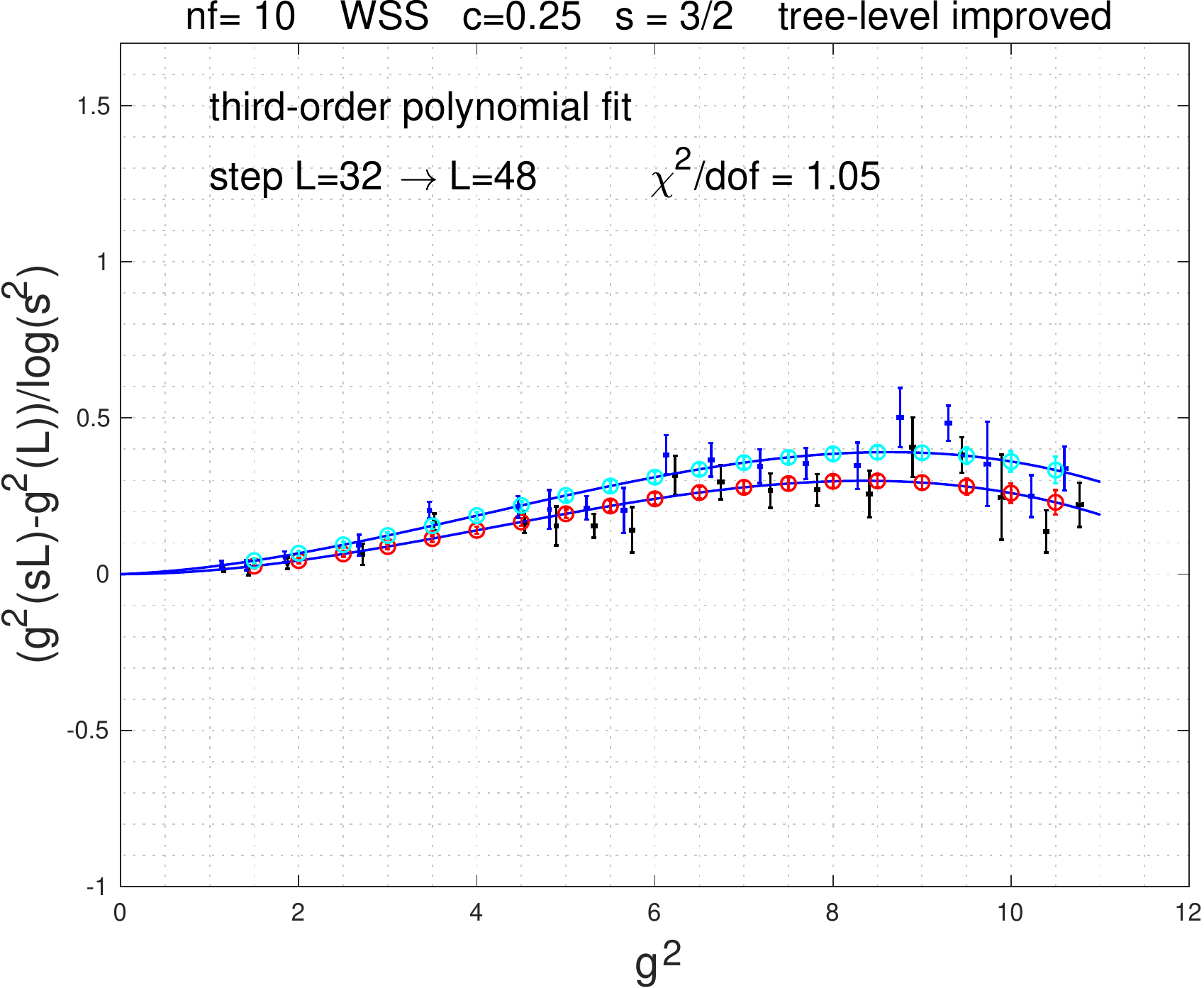}&
			\includegraphics[width=0.33\textwidth]{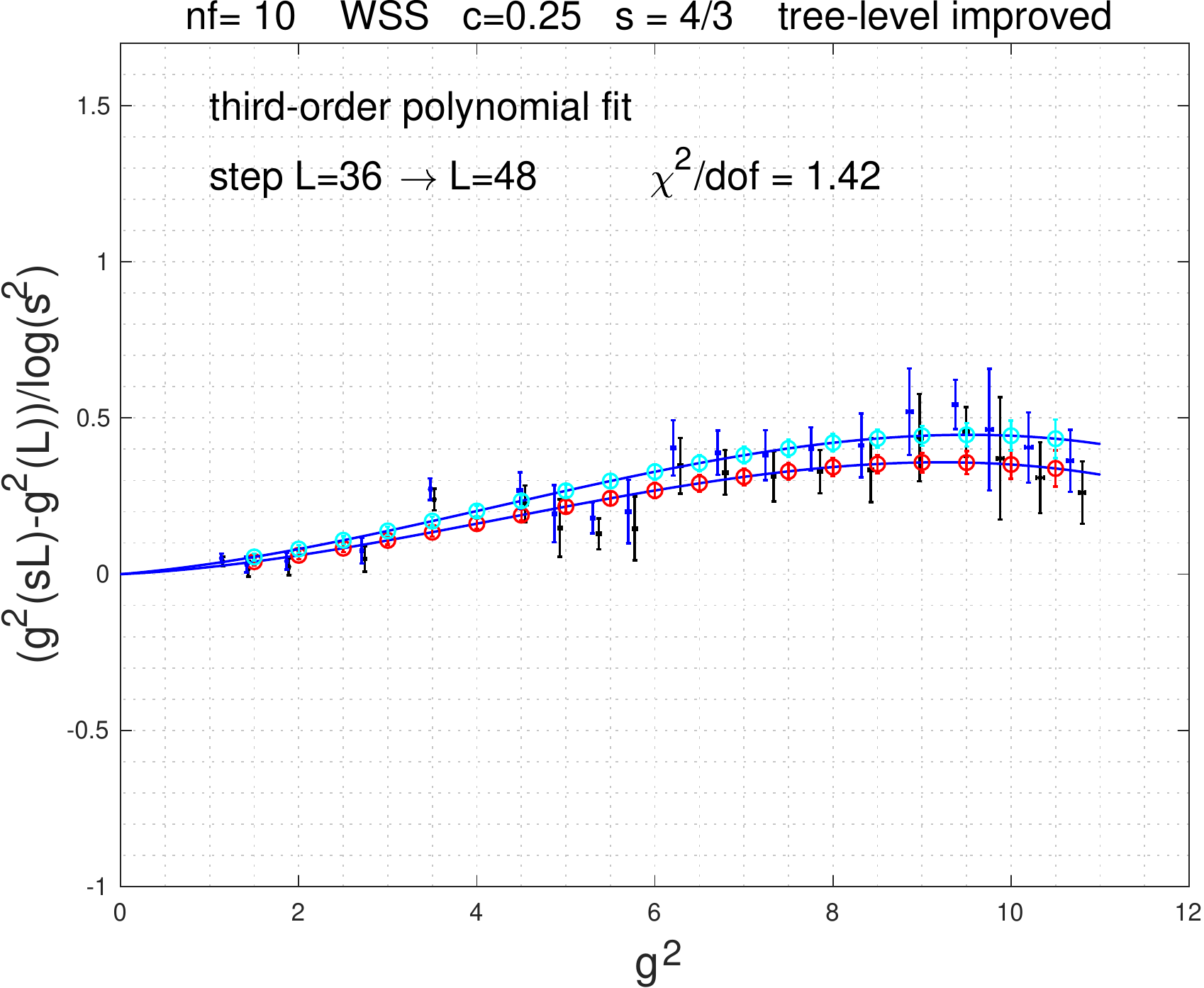}
		\end{tabular}
	\end{center}		
	\vspace{-20pt}
	\caption{\label{c250App} {\small Polynomial fits and interpolations are shown at $c=0.25$  for  lattice step functions $ (g^2(sL)-g^2(L))/{\rm log}(s^2)$  with and without tree-level 
		improvement for the $ L=24 \rightarrow L=48$ pair with step $s=2$,  $L=32\rightarrow L=48$ at $s=3/2$,
		and  $L=36\rightarrow L=48$ at $s=4/3$.  4th order polynomials are used in $g^2$ and constrained to vanish at $g^2=0$.
		Red circles mark the lattice step $\beta$-functions at targeted $g^2$ locations in the unimproved schemes and cyan circles are the targets 
		with tree-level improvement.  Fits to the measured data points at bare couplings $g_0^2$ are shown in black for unimproved data and in blue for tree-improved data. }}
\end{figure}

\begin{figure}[h!]	
	\begin{center}
		\begin{tabular}{ccc}
\includegraphics[width=0.33\textwidth]{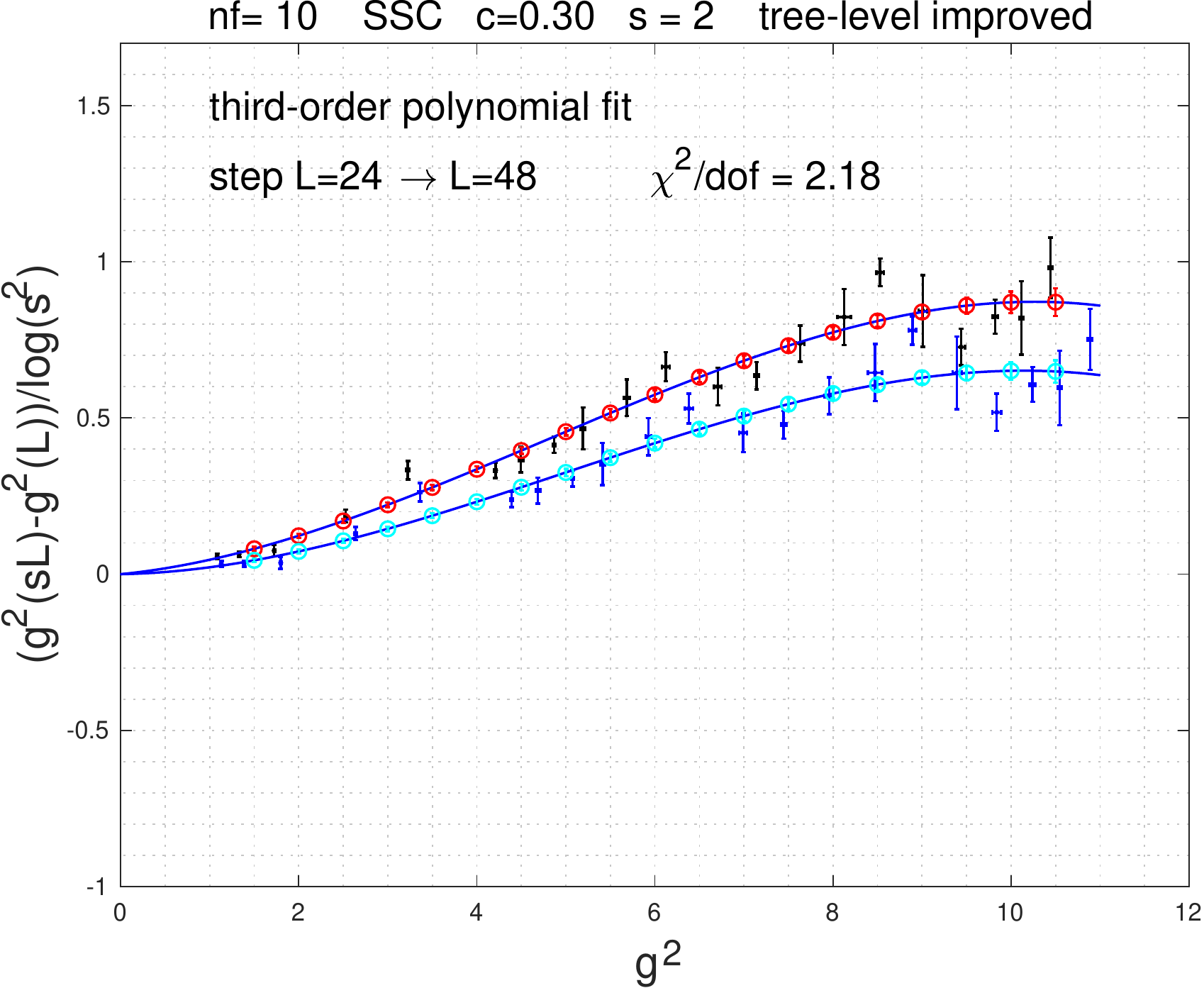}&
\includegraphics[width=0.33\textwidth]{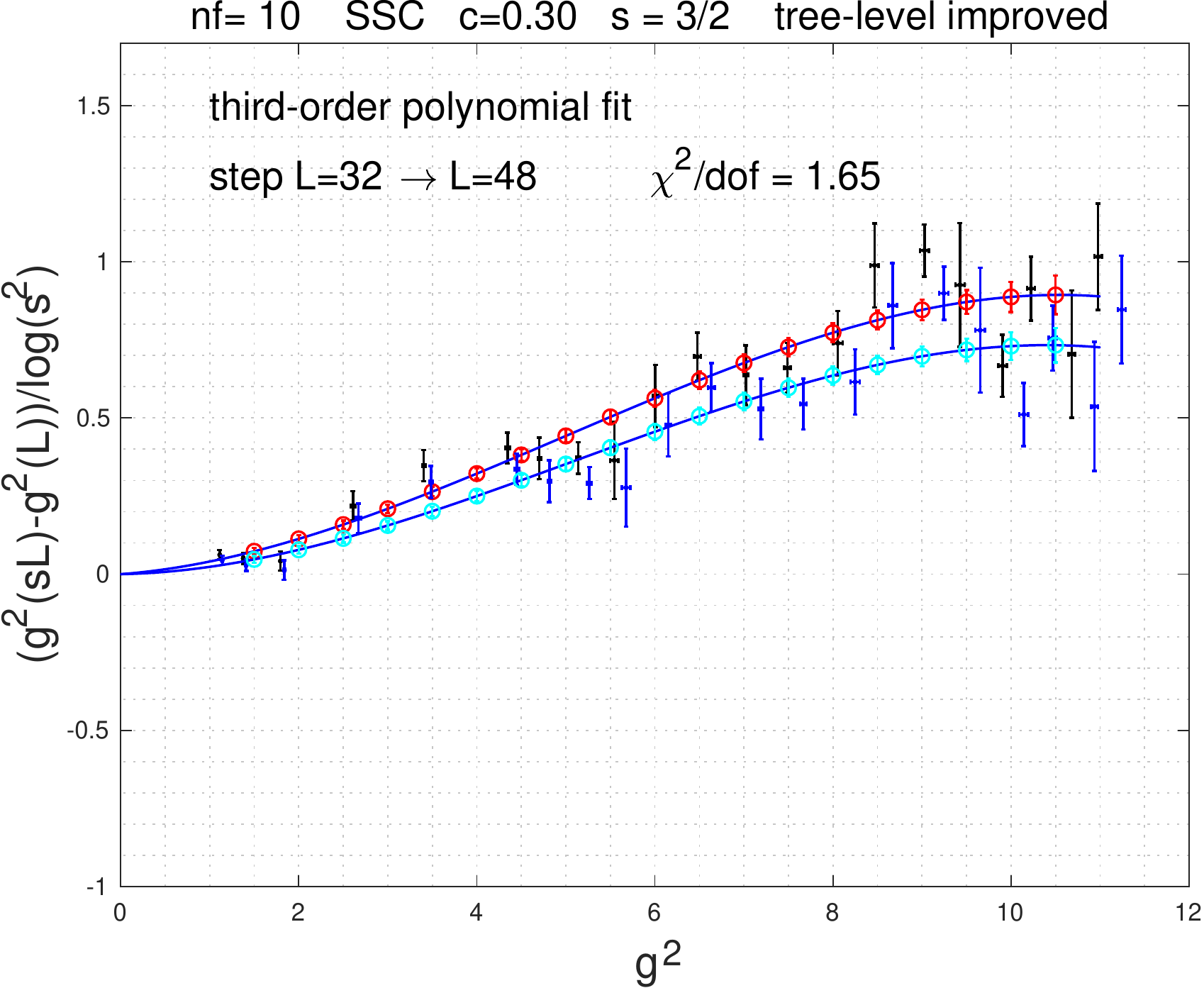}&
\includegraphics[width=0.33\textwidth]{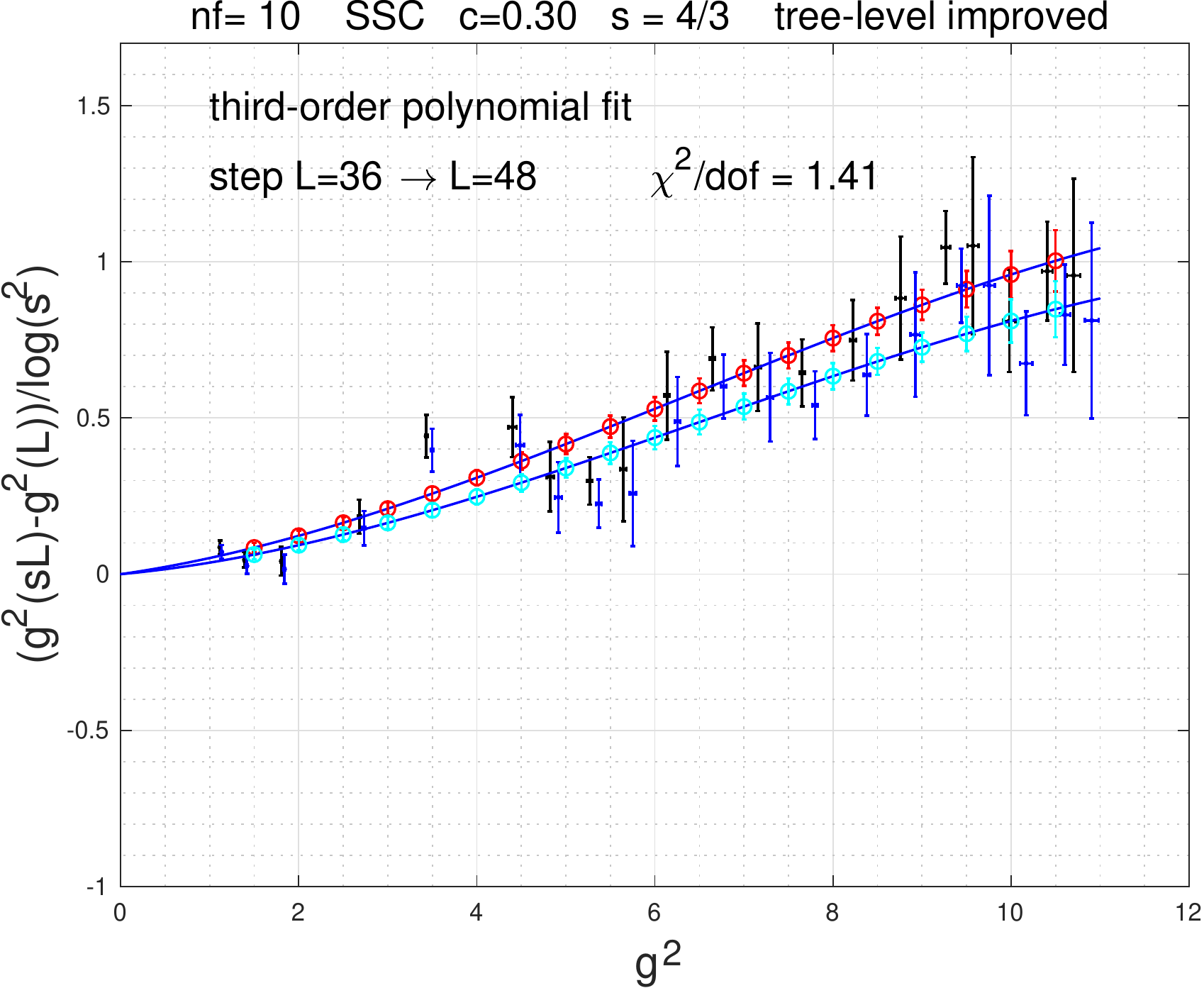}\\
\includegraphics[width=0.33\textwidth]{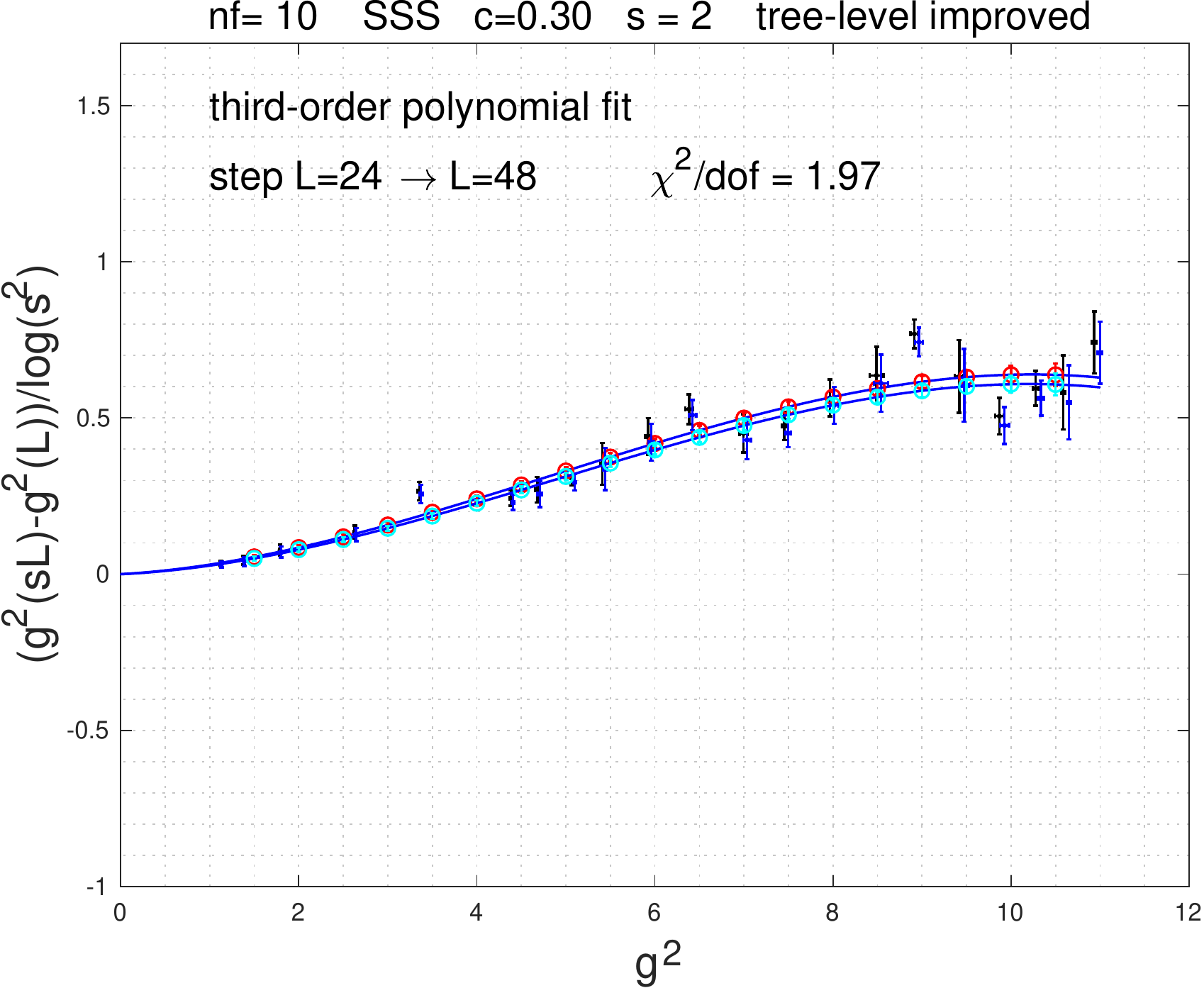}&
\includegraphics[width=0.33\textwidth]{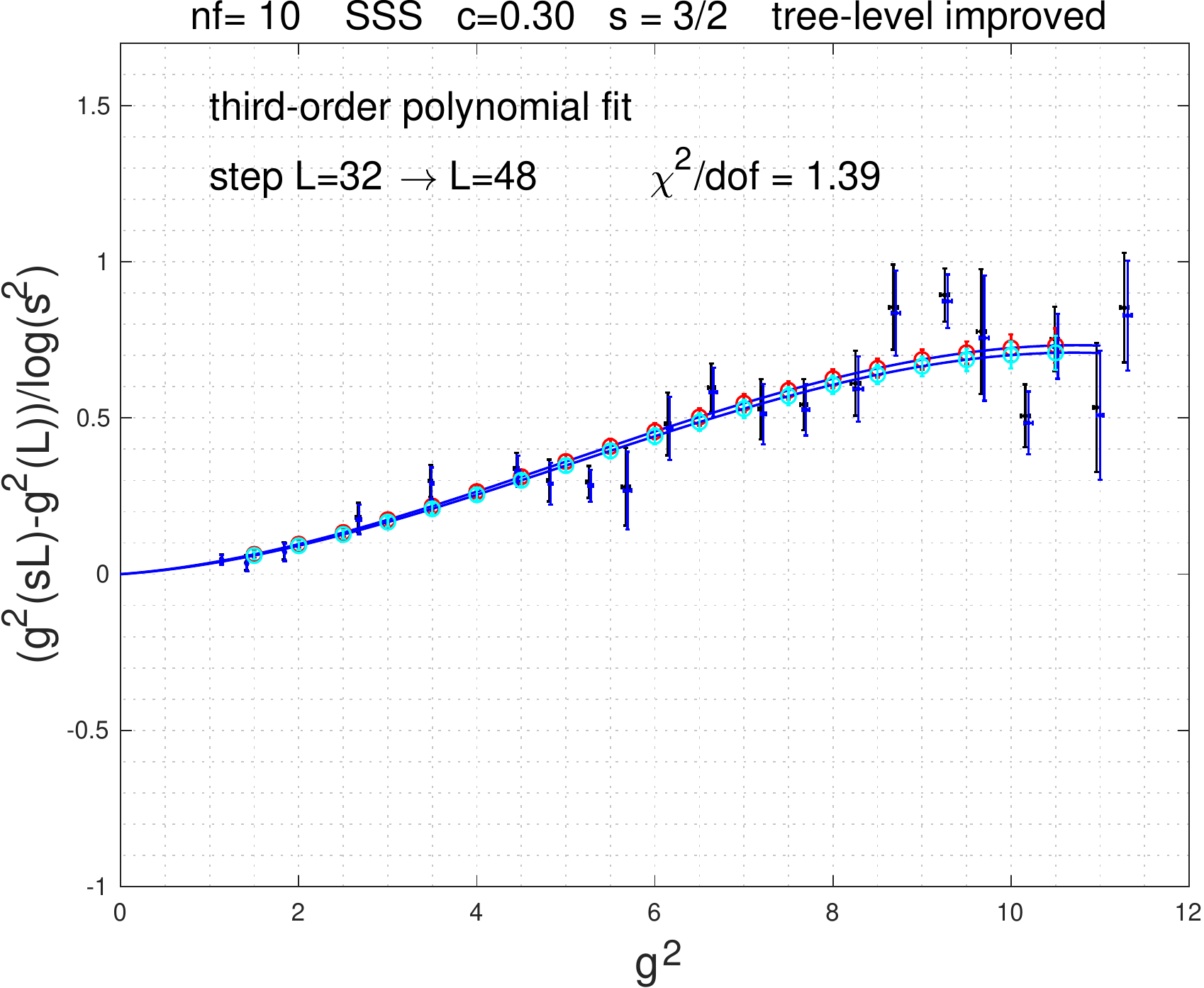}&
\includegraphics[width=0.33\textwidth]{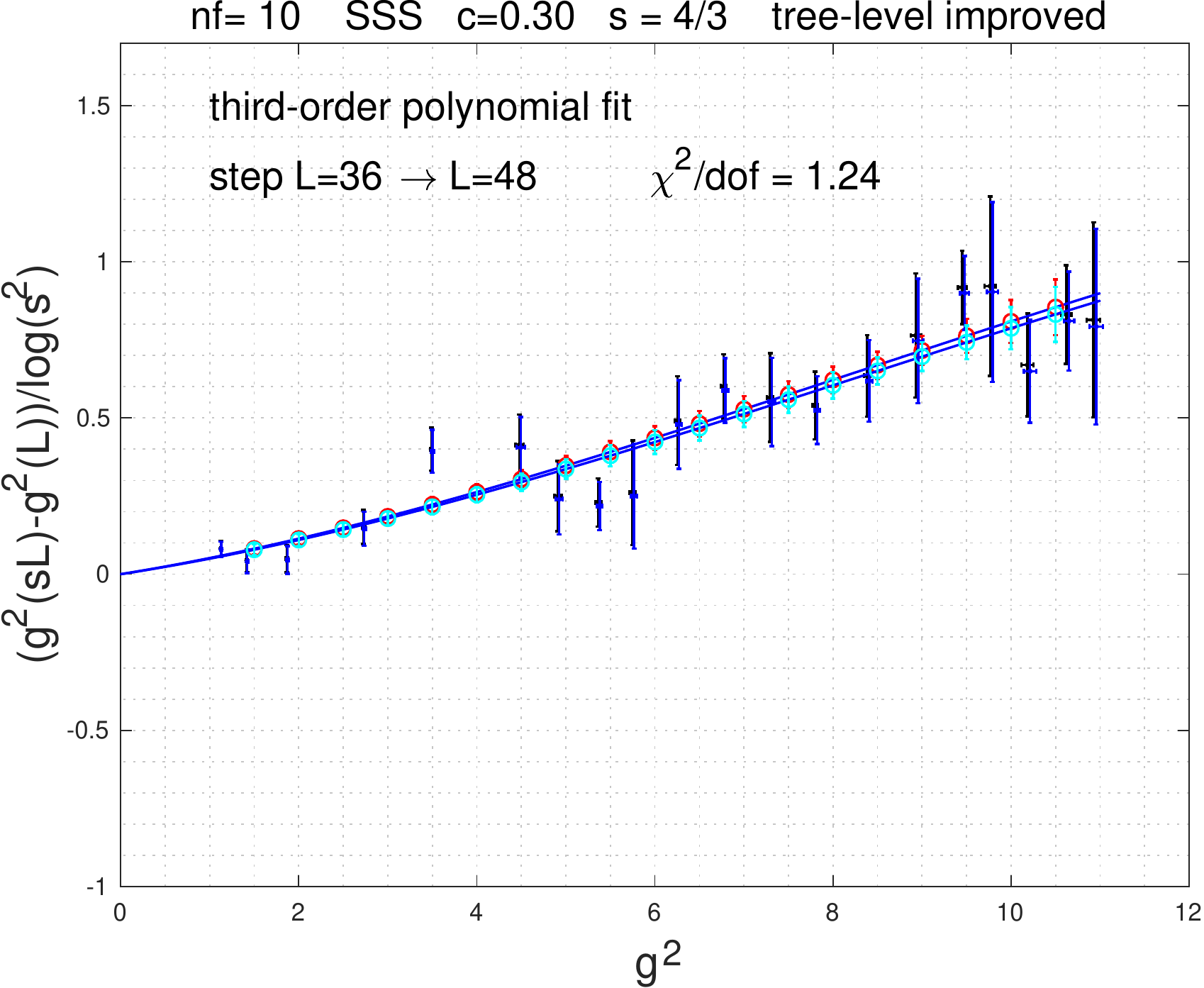}\\			
\includegraphics[width=0.33\textwidth]{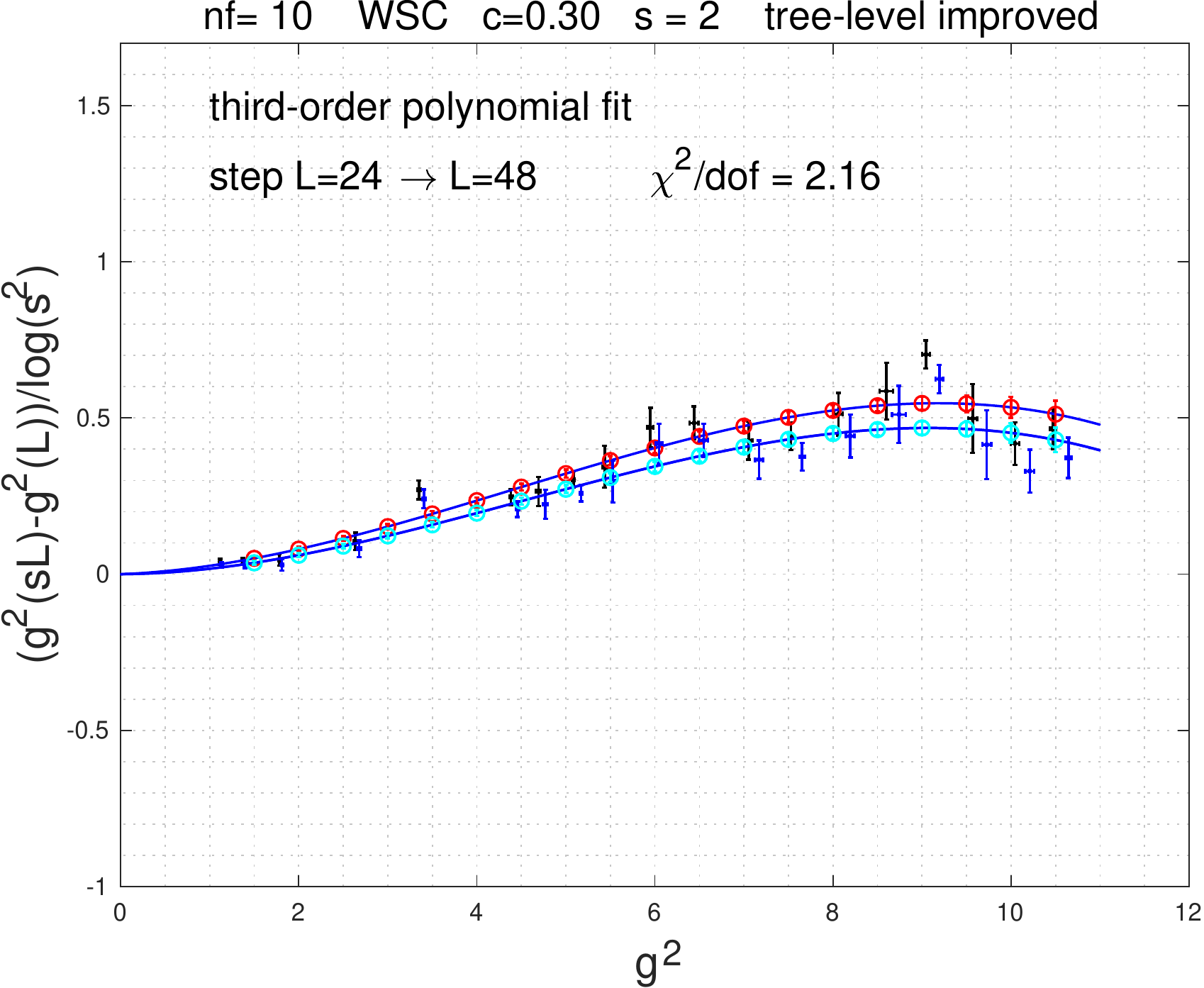}&
\includegraphics[width=0.33\textwidth]{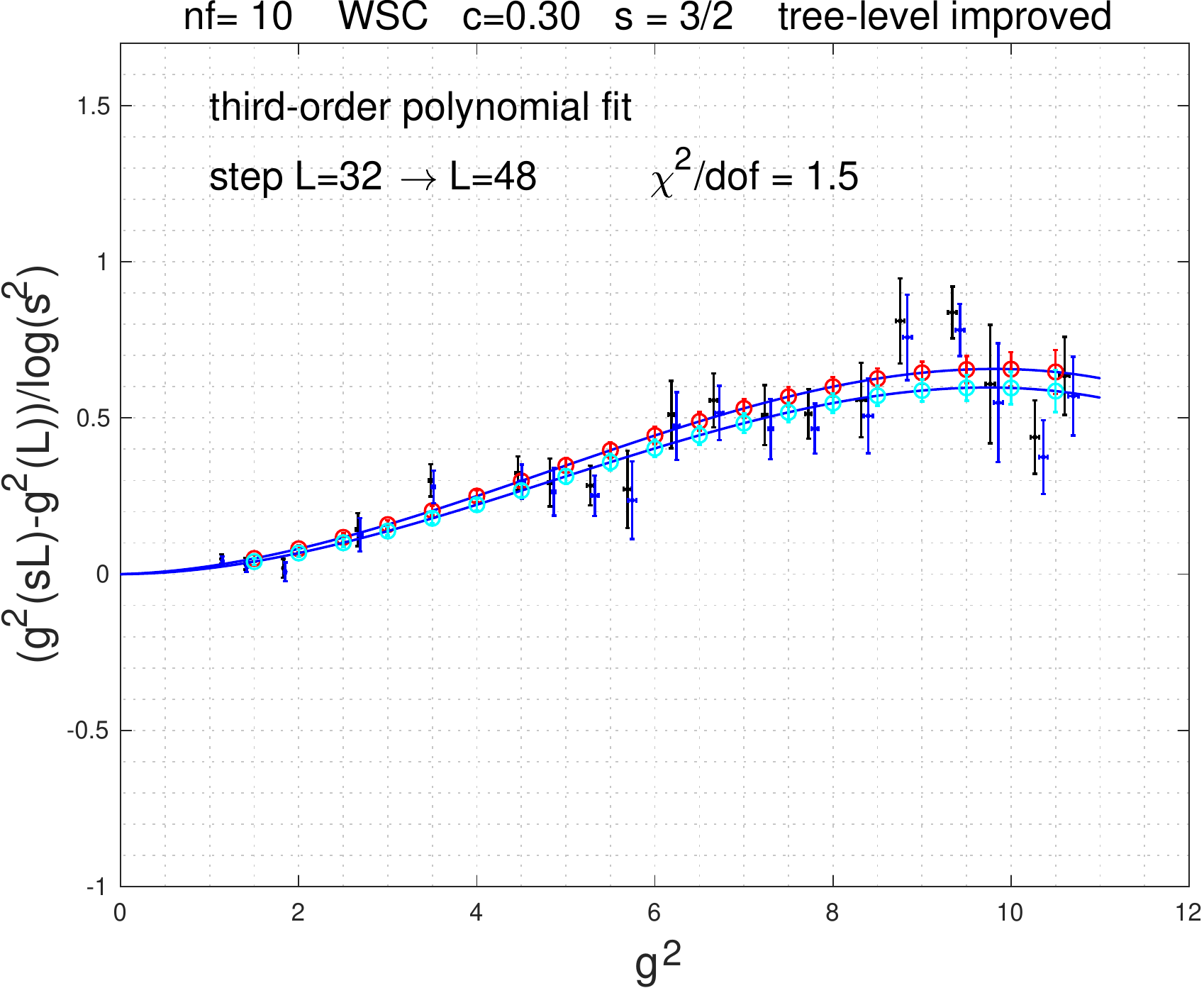}&
\includegraphics[width=0.33\textwidth]{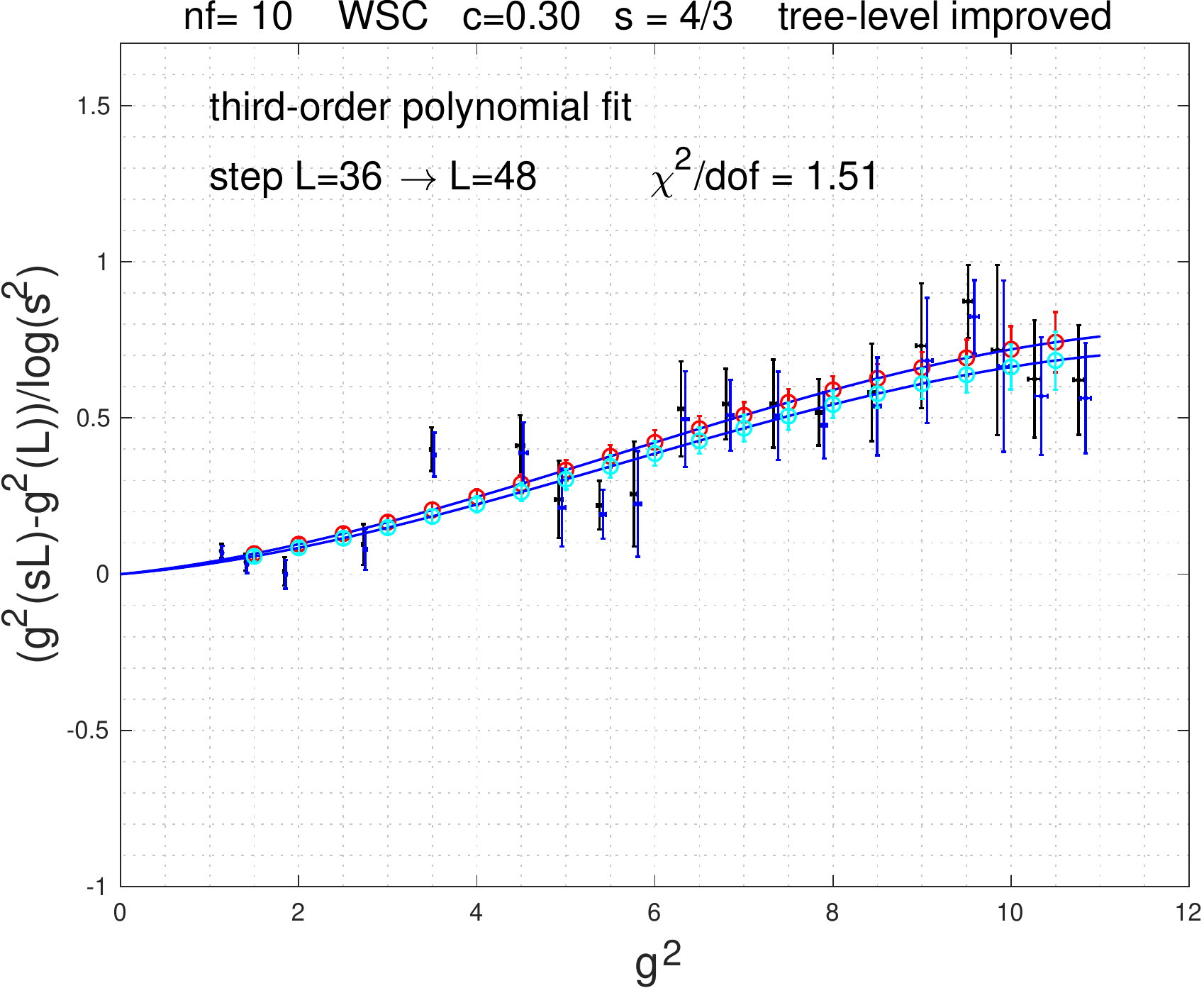}\\
\includegraphics[width=0.33\textwidth]{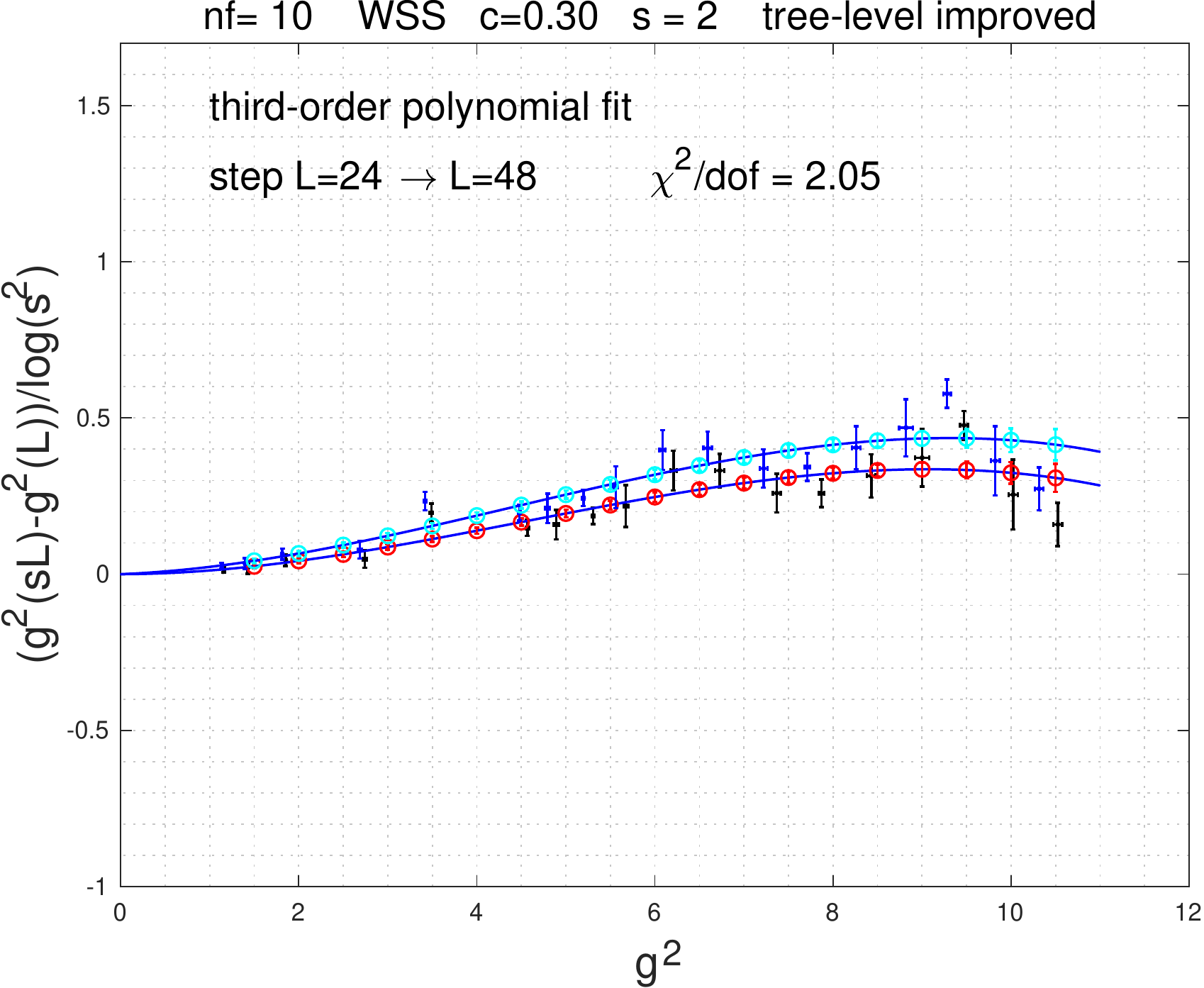}&
\includegraphics[width=0.33\textwidth]{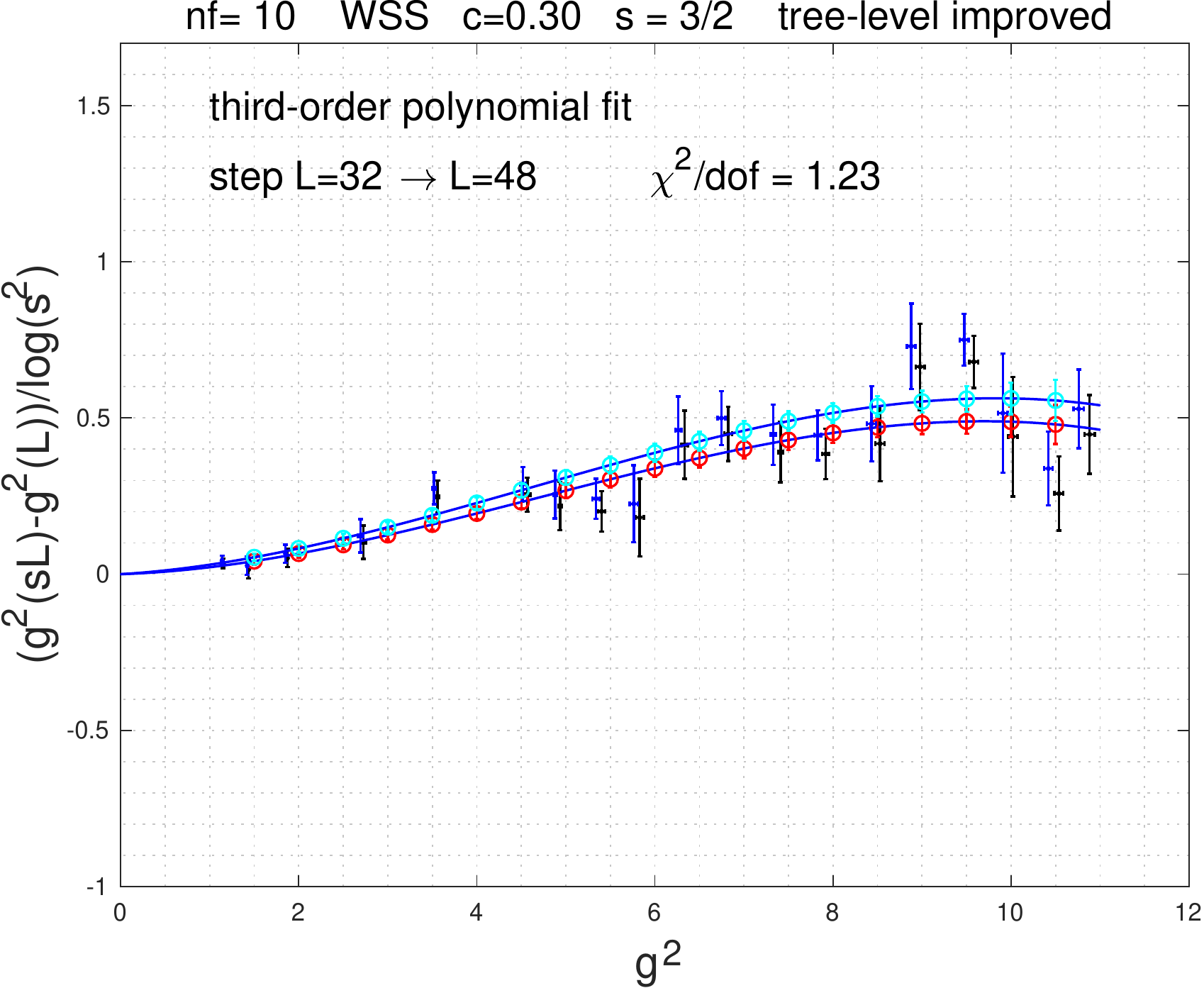}&
\includegraphics[width=0.33\textwidth]{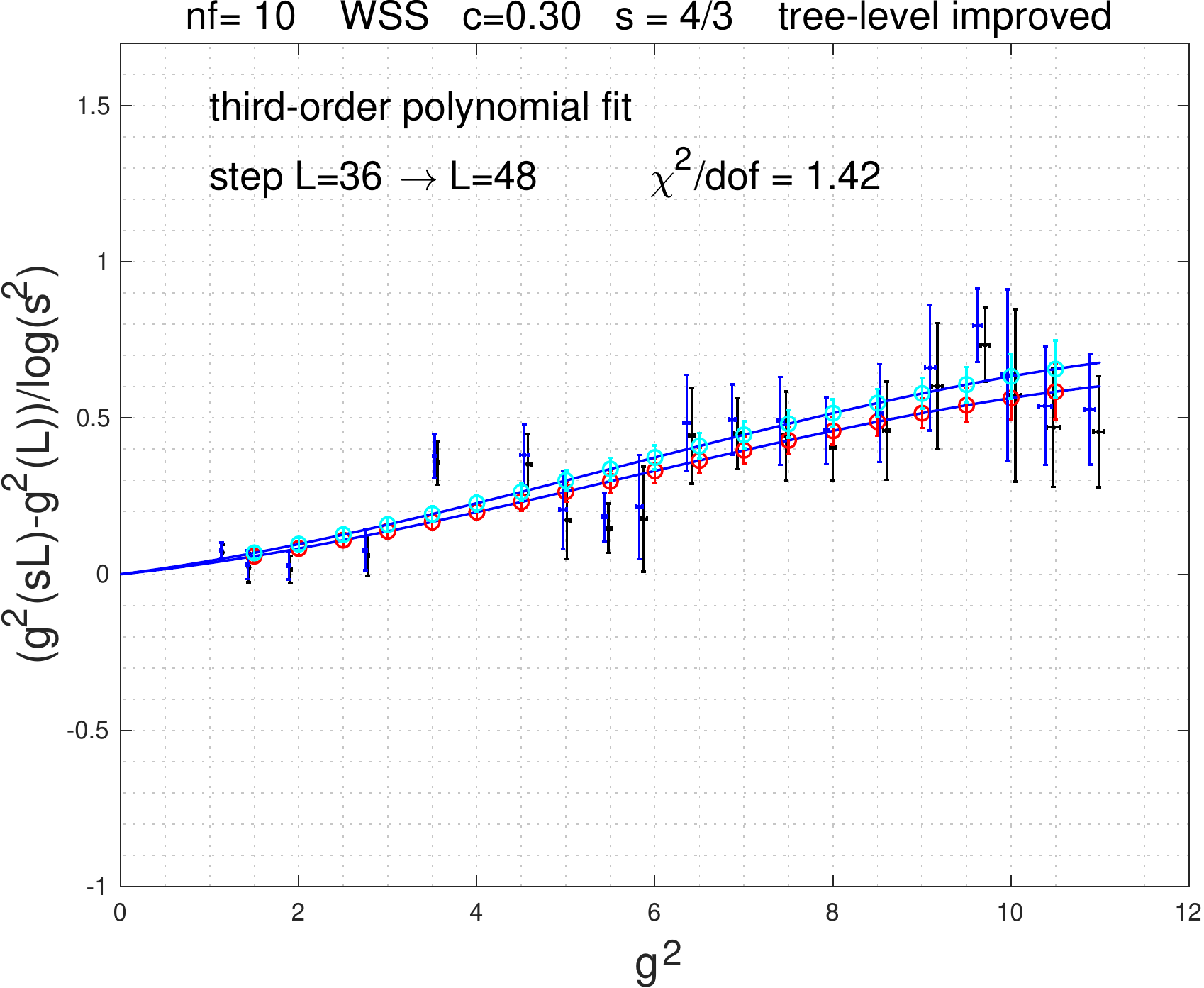}
		\end{tabular}
	\end{center}		
	\caption{\label{c300App} {\small  The fitting procedure is similar to what is described in Fig.~\ref{c250App} and in the main text.}}
\end{figure}

\bibliographystyle{elsarticle-num}
\bibliography{jk2021beta,nf10_2020,nf12_2018,jkNf12}

\begin{thebibliography}{10}
\expandafter\ifx\csname url\endcsname\relax
  \def\url#1{\texttt{#1}}\fi
\expandafter\ifx\csname urlprefix\endcsname\relax\def\urlprefix{URL }\fi
\expandafter\ifx\csname href\endcsname\relax
  \def\href#1#2{#2} \def\path#1{#1}\fi

\bibitem{Hasenfratz:2020ess}
A.~Hasenfratz, C.~Rebbi, O.~Witzel, {Gradient flow step-scaling function for
  SU(3) with ten fundamental flavors}, Phys. Rev. D 101~(11) (2020) 114508.
\newblock \href {http://arxiv.org/abs/2004.00754} {\path{arXiv:2004.00754}},
  \href {https://doi.org/10.1103/PhysRevD.101.114508}
  {\path{doi:10.1103/PhysRevD.101.114508}}.

\bibitem{LatticeStrongDynamics:2020uwo}
T.~Appelquist, et~al., {Near-conformal dynamics in a chirally broken system},
  Phys. Rev. D 103~(1) (2021) 014504.
\newblock \href {http://arxiv.org/abs/2007.01810} {\path{arXiv:2007.01810}},
  \href {https://doi.org/10.1103/PhysRevD.103.014504}
  {\path{doi:10.1103/PhysRevD.103.014504}}.

\bibitem{Chiu:2017kza}
T.-W. Chiu, {Discrete $\beta$-function of the $SU(3)$ gauge theory with 10
  massless domain-wall fermions}, PoS LATTICE2016 (2017) 228.
\newblock \href {https://doi.org/10.22323/1.256.0228}
  {\path{doi:10.22323/1.256.0228}}.

\bibitem{Chiu:2018edw}
T.-W. Chiu, {Improved study of the $\beta$-function of $SU(3)$ gauge theory
  with $N_f = 10$ massless domain-wall fermions}, Phys. Rev. D 99~(1) (2019)
  014507.
\newblock \href {http://arxiv.org/abs/1811.01729} {\path{arXiv:1811.01729}},
  \href {https://doi.org/10.1103/PhysRevD.99.014507}
  {\path{doi:10.1103/PhysRevD.99.014507}}.

\bibitem{Fodor:2018tdg}
Z.~Fodor, K.~Holland, J.~Kuti, D.~Nogradi, C.~H. Wong, {Fate of a recent
  conformal fixed point and $\beta$-function in the SU(3) BSM gauge theory with
  ten massless flavors}, PoS LATTICE2018 (2018) 199.
\newblock \href {http://arxiv.org/abs/1812.03972} {\path{arXiv:1812.03972}},
  \href {https://doi.org/10.22323/1.334.0199} {\path{doi:10.22323/1.334.0199}}.

\bibitem{Fodor:2019ypi}
Z.~Fodor, K.~Holland, J.~Kuti, D.~Nogradi, C.~H. Wong, {Case studies of
  near-conformal $\beta$-functions}, PoS LATTICE2019 (2019) 121.
\newblock \href {http://arxiv.org/abs/1912.07653} {\path{arXiv:1912.07653}},
  \href {https://doi.org/10.22323/1.363.0121} {\path{doi:10.22323/1.363.0121}}.

\bibitem{Fodor:2017die}
Z.~Fodor, K.~Holland, J.~Kuti, D.~Nogradi, C.~H. Wong, {A new method for the
  beta function in the chiral symmetry broken phase}, EPJ Web Conf. 175 (2018)
  08027.
\newblock \href {http://arxiv.org/abs/1711.04833} {\path{arXiv:1711.04833}},
  \href {https://doi.org/10.1051/epjconf/201817508027}
  {\path{doi:10.1051/epjconf/201817508027}}.

\bibitem{Narayanan:2006rf}
R.~Narayanan, H.~Neuberger, {Infinite N phase transitions in continuum Wilson
  loop operators}, JHEP 03 (2006) 064.
\newblock \href {http://arxiv.org/abs/hep-th/0601210}
  {\path{arXiv:hep-th/0601210}}, \href
  {https://doi.org/10.1088/1126-6708/2006/03/064}
  {\path{doi:10.1088/1126-6708/2006/03/064}}.

\bibitem{Luscher:2010iy}
M.~Lüscher, {Properties and uses of the Wilson flow in lattice QCD}, JHEP 08
  (2010) 071.
\newblock \href {https://doi.org/10.1007/JHEP08(2010)071,
  10.1007/JHEP03(2014)092} {\path{doi:10.1007/JHEP08(2010)071,
  10.1007/JHEP03(2014)092}}.

\bibitem{Luscher:2010we}
M.~Luscher, {Topology, the Wilson flow and the HMC algorithm}, PoS LATTICE2010
  (2010) 015.
\newblock \href {http://arxiv.org/abs/1009.5877} {\path{arXiv:1009.5877}}.

\bibitem{Harlander:2016vzb}
R.~V. Harlander, T.~Neumann, {The perturbative QCD gradient flow to three
  loops}, JHEP 06 (2016) 161.
\newblock \href {http://arxiv.org/abs/1606.03756} {\path{arXiv:1606.03756}},
  \href {https://doi.org/10.1007/JHEP06(2016)161}
  {\path{doi:10.1007/JHEP06(2016)161}}.

\bibitem{Borsanyi}
{Work in collaboration with Szabolcs Bors\'anyi}.

\bibitem{Luscher:2009eq}
M.~Luscher, {Trivializing maps, the Wilson flow and the HMC algorithm}, Commun.
  Math. Phys. 293 (2010) 899--919.
\newblock \href {http://arxiv.org/abs/0907.5491} {\path{arXiv:0907.5491}},
  \href {https://doi.org/10.1007/s00220-009-0953-7}
  {\path{doi:10.1007/s00220-009-0953-7}}.

\bibitem{Luscher:2011bx}
M.~Luscher, P.~Weisz, {Perturbative analysis of the gradient flow in
  non-abelian gauge theories}, JHEP 02 (2011) 051.
\newblock \href {https://doi.org/10.1007/JHEP02(2011)051}
  {\path{doi:10.1007/JHEP02(2011)051}}.

\bibitem{Lohmayer:2011si}
R.~Lohmayer, H.~Neuberger, {Continuous smearing of Wilson Loops}, PoS
  LATTICE2011 (2011) 249.
\newblock \href {http://arxiv.org/abs/1110.3522} {\path{arXiv:1110.3522}}.

\bibitem{Luscher:2013vga}
M.~L\"uscher, {Future applications of the Yang-Mills gradient flow in lattice
  QCD}, PoS LATTICE2013 (2014) 016.
\newblock \href {http://arxiv.org/abs/1308.5598} {\path{arXiv:1308.5598}},
  \href {https://doi.org/10.22323/1.187.0016} {\path{doi:10.22323/1.187.0016}}.

\bibitem{Borsanyi:2012zs}
S.~Borsanyi, et~al., {High-precision scale setting in lattice QCD}, JHEP 09
  (2012) 010.
\newblock \href {http://arxiv.org/abs/1203.4469} {\path{arXiv:1203.4469}},
  \href {https://doi.org/10.1007/JHEP09(2012)010}
  {\path{doi:10.1007/JHEP09(2012)010}}.

\bibitem{Fodor:2012td}
Z.~Fodor, K.~Holland, J.~Kuti, D.~Nogradi, C.~H. Wong, {The Yang-Mills gradient
  flow in finite volume}, JHEP 11 (2012) 007.
\newblock \href {http://arxiv.org/abs/1208.1051} {\path{arXiv:1208.1051}},
  \href {https://doi.org/10.1007/JHEP11(2012)007}
  {\path{doi:10.1007/JHEP11(2012)007}}.

\bibitem{Luscher:1992an}
M.~Luscher, R.~Narayanan, P.~Weisz, U.~Wolff, {The Schrodinger functional: A
  Renormalizable probe for nonAbelian gauge theories}, Nucl. Phys. B384 (1992)
  168--228.
\newblock \href {http://arxiv.org/abs/hep-lat/9207009}
  {\path{arXiv:hep-lat/9207009}}, \href
  {https://doi.org/10.1016/0550-3213(92)90466-O}
  {\path{doi:10.1016/0550-3213(92)90466-O}}.

\bibitem{Fodor:2016zil}
Z.~Fodor, K.~Holland, J.~Kuti, S.~Mondal, D.~Nogradi, C.~H. Wong, {Fate of the
  conformal fixed point with twelve massless fermions and SU(3) gauge group},
  Phys. Rev. D94~(9) (2016) 091501.
\newblock \href {http://arxiv.org/abs/1607.06121} {\path{arXiv:1607.06121}},
  \href {https://doi.org/10.1103/PhysRevD.94.091501}
  {\path{doi:10.1103/PhysRevD.94.091501}}.

\bibitem{Fodor:2014cpa}
Z.~Fodor, K.~Holland, J.~Kuti, S.~Mondal, D.~Nogradi, C.~H. Wong, {The lattice
  gradient flow at tree-level and its improvement}, JHEP 09 (2014) 018.
\newblock \href {http://arxiv.org/abs/1406.0827} {\path{arXiv:1406.0827}},
  \href {https://doi.org/10.1007/JHEP09(2014)018}
  {\path{doi:10.1007/JHEP09(2014)018}}.

\bibitem{Artz:2019bpr}
J.~Artz, R.~V. Harlander, F.~Lange, T.~Neumann, M.~Prausa, {Results and
  techniques for higher order calculations within the gradient-flow formalism},
  JHEP 06 (2019) 121, [Erratum: JHEP 10, 032 (2019)].
\newblock \href {http://arxiv.org/abs/1905.00882} {\path{arXiv:1905.00882}},
  \href {https://doi.org/10.1007/JHEP06(2019)121}
  {\path{doi:10.1007/JHEP06(2019)121}}.

\bibitem{Fodor:2015zna}
Z.~Fodor, K.~Holland, J.~Kuti, S.~Mondal, D.~Nogradi, C.~H. Wong, {The running
  coupling of the minimal sextet composite Higgs model}, JHEP 09 (2015) 039.
\newblock \href {http://arxiv.org/abs/1506.06599} {\path{arXiv:1506.06599}},
  \href {https://doi.org/10.1007/JHEP09(2015)039}
  {\path{doi:10.1007/JHEP09(2015)039}}.

\bibitem{Hasenfratz:2019hpg}
A.~Hasenfratz, O.~Witzel, {Continuous renormalization group $\beta$ function
  from lattice simulations}, Phys. Rev. D 101~(3) (2020) 034514.
\newblock \href {http://arxiv.org/abs/1910.06408} {\path{arXiv:1910.06408}},
  \href {https://doi.org/10.1103/PhysRevD.101.034514}
  {\path{doi:10.1103/PhysRevD.101.034514}}.

\bibitem{DallaBrida:2018rfy}
M.~Dalla~Brida, P.~Fritzsch, T.~Korzec, A.~Ramos, S.~Sint, R.~Sommer, {A
  non-perturbative exploration of the high energy regime in $N_{\mathrm{f}}=3$
  QCD}, Eur. Phys. J. C 78~(5) (2018) 372.
\newblock \href {http://arxiv.org/abs/1803.10230} {\path{arXiv:1803.10230}},
  \href {https://doi.org/10.1140/epjc/s10052-018-5838-5}
  {\path{doi:10.1140/epjc/s10052-018-5838-5}}.

\end{thebibliography}

\end{document}